\documentclass[allclo,onecollarge,natbib]{svjour2}

\usepackage[english]{babel}
\usepackage[utf8x]{inputenc}

\usepackage{amsmath}
\usepackage{amsfonts}
\usepackage{amssymb}
\usepackage{bm}
\usepackage{pzccal}

\usepackage{epsfig}
\usepackage{graphicx}

\usepackage{dcolumn}
\usepackage{color}
\usepackage{soul}

\usepackage{color}
\definecolor{purple}{rgb}{0.5,0,0.5}
\definecolor{blue}{rgb}{0.0,0,0.9}

\def\tstrut{\vrule height2.25ex depth0pt width0pt} 

\bibpunct{[}{]}{,}{n}{,}{,} 

\begin{document}

\title{Elastic and transition form factors of the $\Delta(1232)$}

\authorrunning{Jorge Segovia \emph{et al}.}
\titlerunning{Form factors of the $\Delta(1232)$}

\author{Jorge Segovia
\and
        Chen Chen
\and
        Ian C.~Clo\"et
\and
        Craig~D.~Roberts
\and
        Sebastian M. Schmidt
\and
        Shaolong~Wan
}

\institute{Jorge Segovia \and Ian C.~Clo\"et \and Craig D.~Roberts
\at
Physics Division, Argonne National Laboratory, Argonne
Illinois 60439, USA
\and
Chen Chen  \and Shaolong Wan  
\at
Institute for Theoretical Physics and Department of Modern Physics,\\
University of Science and Technology of China, Hefei 230026, P. R. China
\and
Sebastian M. Schmidt
\at
Institute for Advanced Simulation, Forschungszentrum J\"ulich and JARA, D-52425 J\"ulich, Germany
}

\date{23 August 2013}

\maketitle

\begin{abstract}
Predictions obtained with a confining, symmetry-preserving treatment of a vector$\,\otimes\,$vec\-tor contact interaction at leading-order in a widely used truncation of QCD's Dyson-Schwinger equations are presented for $\Delta$ and $\Omega$ baryon elastic form factors and the $\gamma N \to \Delta$ transition form factors.  This simple framework produces results that are practically indistinguishable from the best otherwise available, an outcome which highlights that the key to describing many features of baryons and unifying them with the properties of mesons is a veracious expression of dynamical chiral symmetry breaking in the hadron bound-state problem.
The following specific results are of particular interest.  The $\Delta$ elastic form factors are very sensitive to $m_\Delta$.  Hence, given that the parameters which define extant simulations of lattice-regularised QCD produce $\Delta$-resonance masses that are very large, the form factors obtained therewith are a poor guide to properties of the $\Delta(1232)$.  Considering the $\Delta$-baryon's quadrupole moment, whilst all computations produce a negative value, the conflict between theoretical predictions entails that it is currently impossible to reach a sound conclusion on the nature of the $\Delta$-baryon's deformation in the infinite momentum frame.  Results for analogous properties of the $\Omega$ baryon are less contentious.  In connection with the $N\to \Delta$ transition, the Ash-convention magnetic transition form factor falls faster than the neutron's magnetic form factor and nonzero values for the associated quadrupole ratios reveal the impact of quark orbital angular momentum within the nucleon and $\Delta$; and, furthermore, these quadrupole ratios do slowly approach their anticipated asymptotic limits.
\end{abstract}


\maketitle

\section{Introduction}
\label{sec:Introduction}

Given the challenges posed by nonperturbative quantum chromodynamics (QCD), it is insufficient to study hadron ground-states alone if one seeks a solution.  In order to chart the infrared behaviour of the quark-quark interaction via a collaborative effort between experiment and theory, every available tool must be exploited to its fullest.  In particular, the effort can benefit substantially by exposing the structure of nucleon excited states and measuring the associated transition form factors at large momentum transfers \cite{Aznauryan:2012ba}.  Large momenta are needed in order to pierce the meson-cloud that, often to a significant extent, screens the dressed-quark core of all baryons \cite{Suzuki:2009nj,Kamano:2013iva}; and, as one may infer from Refs.\,\cite{Cloet:2013gva,Chang:2013nia}, it is with the $Q^2$ evolution of form factors that one gains access to the running of QCD's coupling and masses from the infrared into the ultraviolet.

Motivated initially by a desire to understand the momentum dependence of the Ash-convention \cite{Ash1967165} $\gamma N \to \Delta(1232)$  magnetic dipole transition form factor, $G_{M,Ash}^{\ast}(Q^2)$, which was long held to be puzzling, we have employed the framework developed elsewhere \cite{GutierrezGuerrero:2010md,Roberts:2010rn,Roberts:2011cf,
Roberts:2011wy,Wilson:2011aa,Chen:2012qr,Chen:2012txa,Wang:2013wk,Segovia:2013rca} in a comprehensive study of $\Delta$- and $\Omega$-baryon elastic form factors.  The small-$Q^2$ behaviour of the $\Delta$ elastic form factors is a necessary element in computing the $\gamma N \to \Delta$ transition form factors \cite{Segovia:2013rca}; and once one has developed tools with which to calculate $\Delta$ elastic form factors, computing $\Omega$ elastic form factors is straightforward.  This article describes the outcomes of these studies and the insights they provide.

The $\Delta(1232)$-baryons were the first resonances discovered in $\pi N$
reactions \cite{Fermi:1952zz,Anderson:1952nw,Nagle:1984sg}.  They have since been
studied extensively, both experimentally and theoretically, so that their flavour
structure and total angular momentum are now well-known \cite{Beringer:1900zz}:
$\Delta(1232)$-baryons are positive-parity, total-spin $J=\frac{3}{2}$, total-isospin $I=\frac{3}{2}$ bound-states with no net strangeness.  As an $I=\frac{3}{2}$ multiplet, the $\Delta$-baryon occurs in four charge states: $\Delta^{++}$, $\Delta^{+}$, $\Delta^{0}$, $\Delta^{-}$.  These are the lightest baryon resonances, with a mass of $1.23\,$GeV, just 30\% heavier than the nucleon; and despite possessing a width of $0.12\,$GeV, the $\Delta$-resonances are well isolated from other nucleon excitations.  Much of the interest in $\Delta$-resonances finds its origin in the fact that, at the simplest level, the $\Delta^{+}$ and $\Delta^{0}$ can respectively be viewed as spin- and isospin-flip excitations of the proton and neutron.\footnote{As we shall see, however, this apparently elementary connection obscures a deeper truth; namely, the structure of the $\Delta$-baryon's dressed-quark-core is actually far simpler than that of the nucleons.}  In addition, the strong $\Delta \to \pi N$ coupling entails that the $\Delta(1232)$-resonance is an important platform for developing and honing an understanding of the role a meson cloud plays in baryon physics \cite{Sato:2000jf,JuliaDiaz:2006xt} so that this may be separated from effects more properly attributable to a baryon's dressed-quark core \cite{Roberts:2011cf,Chen:2012qr,Eichmann:2008ae,Eichmann:2008ef,Cloet:2008re}.

Since the $\Delta(1232)$ is a $J=\frac{3}{2}$ state, a complete description of its electromagnetic structure requires four form factors \cite{Scadron:1968zz}: electric charge, $G_{E0}$; magnetic dipole $G_{M1}$; electric quadrupole, $G_{E2}$; and magnetic octupole $G_{M3}$.  The first two listed here are, respectively, the analogues of those form factors which describe the momentum-space distribution of the nucleon's charge and magnetisation.  The remaining two may be associated with shape deformation of the $\Delta$-baryon because they are identically zero within any constituent-quark model framework in which $SU(6)$ spin-flavour symmetry is only broken by electromagnetism and the associated current transforms according to the adjoint representation of the symmetry group \cite{Beg:1964nm,Buchmann:1996bd}.

At zero momentum transfer the form factors can be used to define dimensionless $\Delta$ multipole moments: electric charge, $e_{\Delta}=G_{E0}(0)$; magnetic moment $\hat\mu_{\Delta}=G_{M1}(0)$; electric quadrupole moment $\hat{\cal Q}_{\Delta}=G_{E2}(0)$; and magnetic octupole moment $\hat{\cal O}_{\Delta}=G_{M2}(0)$.  Within the framework of $\mathpzc{N}=2$ SUperGRAvity with an elementary $J=3/2$ graviton coupled consistently to electromagnetism, it has been argued that a point particle with the positron's charge would posses the following values for the multipole moments \cite{Alexandrou:2009hs}: $e_{\mathpzc S}=1$; $\hat\mu_{\mathpzc S}=3$; $\hat{\cal Q}_{\mathpzc S}=-3$; and $\hat{\cal O}_{\mathpzc S}=-1$.  For comparison, an analogous $J=1$ particle would have three moments, with the values \cite{Brodsky:1992px,Ferrara:1992yc}: $e_{J=1}=1$; $\hat\mu_{J=1}=2$; $\hat{\cal Q}_{J=1}=-1$.  One might contend that deviations from these natural values measure the impact of compositeness.  Reviewing these analyses, it appears to us that the argument for natural values of $\hat\mu$ and $\hat{\cal Q}$ is strongest.

Whilst it is relatively straightforward to extend a theoretical framework applicable to the nucleon so that it may be employed to describe the $\Delta(1232)$, the state thus obtained is commonly stable; i.e., one obtains the zero-width dressed-quark-core of the $\Delta$.  Given that the width-to-mass ratio is small, this is a reasonable approximation, when interpreted judiciously, just as it is for the $\rho$-meson \cite{Pichowsky:1999mu,Maris:1999nt,Jarecke:2002xd}.  Empirically, on the other hand, one must deal with the very short $\Delta$-lifetime: $\tau_\Delta \sim 10^{-16} \tau_{\pi^+}$, and therefore little is experimentally known about the electromagnetic properties of $\Delta(1232)$-baryons.  Information has been obtained through analysis of the $\pi p \to \pi p \gamma$ and $\gamma p \to p \pi^0 \gamma^\prime$ reactions, so that Ref.\,\cite{Beringer:1900zz} reports $\mu_{\Delta^{++}} = 3.7\,$--$\,7.5\,\mu_N$ and
$\mu_{\Delta^{+}}=2.7^{+1.0}_{-1.3}{\rm (stat)}\pm1.5{\rm (syst)}\pm3.0{\rm (theor)}\,\mu_{N}$, where $\mu_N$ is the nuclear magneton.

In recent times, owing to the appearance of intense, energetic electron-beam facilities, it has become possible to sidestep some of the difficulties associated with the small value of $\tau_\Delta$ and learn about the $\Delta(1232)$ by studying the $\gamma^\ast p \to \Delta^+$ transition: data are now available on $0 \leq Q^2 \lesssim 8\,$GeV$^2$ \cite{Aznauryan:2011ub,Aznauryan:2011qj}.  This transition is described by three form factors \cite{Jones:1972ky}: magnetic-dipole, $G_{M}^{\ast}$; electric quadrupole, $G_{E}^{\ast}$; and Coulomb (longitudinal) quadrupole, $G_{C}^{\ast}$.  A qualitative and semiquantitative dynamical explanation of their behaviour may be found in Ref.\,\cite{Segovia:2013rca}, some elements of which we will recapitulate upon herein.  Here we will simply note that the magnetic dipole form factor is dominant and the quadrupole form factors are small but nonzero.  A careful analysis of the $\gamma N \to \pi N$ reaction in combination with relations valid at leading-order in a $1/N_c$ expansion has enabled an inference of the $\Delta^+$ quadrupole moment \cite{Alexandrou:2009hs}: $\hat{\cal Q}_{\,\Delta^+}=-1.87 \pm 0.08$, which is 40\% smaller in magnitude than the ``natural'' value mentioned above.

Following this background it is appropriate to briefly review some aspects of the framework developed in Refs.\,\cite{GutierrezGuerrero:2010md,Roberts:2010rn,Roberts:2011cf,
Roberts:2011wy,Wilson:2011aa,Chen:2012qr,Chen:2012txa,Wang:2013wk,Segovia:2013rca}.  That approach exploits a continuum perspective on quantum field theory based upon QCD's Dyson-Schwinger equations (DSEs) \cite{Chang:2011vu,Roberts:2012sv,Bashir:2012fs}.  It employs a symmetry-preserving treatment of a vector$\,\otimes\,$vector contact interaction because that has proven to be a reliable tool in a wide range of applications, which include meson and baryon spectra, and their electroweak elastic and transition form factors.  It is apposite to remark that this interaction produces form factors which are typically too hard but, when interpreted carefully, can nevertheless be used to draw valuable insights.  The simplicity of the interaction and its capacity to provide a unified explanation of a diverse array of phenomena, many of which are currently unreachable with more sophisticated DSE kernels owing to weaknesses in the numerical algorithms employed, are features that continue to supply grounds for its further application.  It is within this context that we undertake the calculation of $\Delta$ and $\Omega$ elastic form factors and the $\gamma N \to \Delta$ transition form factors.

The report is organised as follows.
In Sect.~\ref{sec:overview} we present a short survey of our framework, both the Faddeev equation treatment of the baryon dressed-quark cores, and the currents that describe the interaction of a photon with a baryon composed from such consistently dressed constituents.  Additional material is expressed in appendices and referred to as necessary.
Following this, our results for the dressed-quark-core contributions to the $\Delta(1232)$ elastic form factors are discussed in Sect.\,\ref{sec:GammaDeltaDelta}; and those for the $\Omega^-$ are detailed in Sect.\,\ref{sec:GammaOmegaOmega}.
Building upon the foundation provided by that material, Sect.\,\ref{sec:GammaNucleonDelta} presents our results for the $\gamma^\ast N \to \Delta$ transition form factors and explains the insights they provide.
We provide a summary and perspective in Sect.\,\ref{sec:summary}.

\section{Composite baryons and their electromagnetic currents}
\label{sec:overview}
\subsection{Faddeev equations}
Baryon bound-states in quantum field theory are described by a Faddeev amplitude, $\Psi$, obtained from a Poincar\'e-covariant Faddeev equation \cite{Cahill:1988dx}, which sums all possible quantum field theoretical interactions that can take place between the three quarks that define its valence-quark content.  The appearance of nonpointlike colour-antitriplet diquark correlations within the proton is a dynamical prediction of Faddeev equation studies; and empirical evidence in support of the presence of diquarks in the proton is accumulating
\cite{Wilson:2011aa,Close:1988br,Cloet:2005pp,Cates:2011pz,Cloet:2012cy,Qattan:2012zf,Roberts:2013mja}.  Importantly, use of the Faddeev equation allows one to treat mesons and baryons on the same footing and, in particular, enables the impact of dynamical chiral symmetry breaking (DCSB), the origin of more than 98\% of the visible mass in the universe \cite{Chang:2011vu,Roberts:2012sv,Bashir:2012fs,national2012Nuclear}, to be expressed in the prediction of baryon properties.

Since the nucleon and $\Delta$ have positive parity, $J^P=0^+$ (scalar) and $J^P=1^+$ (axial-vector) diquarks are the dominant correlations within them \cite{Roberts:2011cf,Chen:2012qr}.  The presence of pseudoscalar and vector diquarks can be ignored because such correlations are characterised by much larger mass-scales and they have negative parity \cite{Roberts:2011cf,Chen:2012qr}.
Owing to Fermi-Dirac statistics, scalar diquarks are necessarily $I=0$ states, whilst axial-vector diquarks are $I=1$ \cite{Cahill:1987qr}.  The nucleon ground-state contains both $0^+$ and $1^+$ diquarks.  However, the $\Delta(1232)$-baryon contains only axial-vector diquark correlations because it is impossible to combine an $I=0$ diquark with an $I=1/2$ quark to obtain $I=3/2$.  Its internal structure is therefore far simpler than that of the nucleon: in the rest frame the dressed-quark-core of the $\Delta(1232)$ is predominantly a quark$\,+\,$axial-vector diquark in a relative $S$-wave \cite{Nicmorus:2010sd}.  This feature is captured herein [see, e.g., Eq.\,\eqref{FAmplitudesContact}].

The dynamically generated correlations described here should not be confused with the pointlike diquarks introduced \cite{Lichtenberg:1967zz,Lichtenberg:1968zz} in order to simplify the study of systems constituted from three constituent-quarks.  The modern dynamical diquark correlation is nonpointlike and interacting, with the charge radius of a given diquark being typically 10\% larger than its mesonic analogue \cite{Roberts:2011wy}.  Hence, diquarks are soft components within baryons.

\begin{figure}[!t]
\begin{center}
\includegraphics[clip,width=0.50\textwidth]{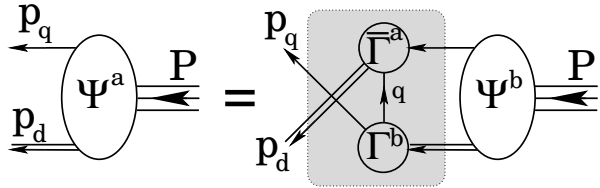}
\end{center}
\caption{\label{fig:FaddeevEquation} Poincar\'e covariant Faddeev equation
[Eq.\,(\ref{eq:NFaddeev}) for the Nucleon and Eq.\,(\ref{eq:DFaddeev}) for the
$\Delta$] employed herein to calculate baryon properties.  $\Psi$ in
Eq.\,(\ref{eq:psiFaddeev}) is the Faddeev amplitude for a baryon of total momentum
$P=p_{q}+p_{d}$.  It expresses the relative momentum correlation between the
dressed-quark and -diquarks within the baryon.  The shaded region demarcates the
kernel of the Faddeev equation (Apps.\,\ref{app:subsubsec:NKernel}
and~\ref{app:subsubsec:DKernel}), in which: the single line denotes the dressed-quark
propagator (App.\,\ref{app:subsec:GAP}); $\Gamma$ is the diquark Bethe-Salpeter
amplitude (App.\,\ref{app:subsec:MesonsDiquarks}); and the double line is the
diquark propagator [Eqs.\,(\ref{eq:0pprop}) and~(\ref{eq:1pprop})].  Quarks within a diquark are correlated via gluon exchange and the kernel in this Faddeev equation expresses additional binding within the baryon through diquark breakup and reformation, which is mediated by exchange of a dressed-quark with momentum $q$.}
\end{figure}

The Faddeev equation that describes the dressed-quark core of the nucleon and $\Delta(1232)$ is depicted in Fig.\,\ref{fig:FaddeevEquation}.  Details are provided in App.\,\ref{app:sec:ContactInteractionModel}, which also explains our treatment of the contact interaction.  The equations are completed with the quantities reported in Table~\ref{tab:mesonproperties}; and the computed values for the nucleon, $\Delta$ and $\Omega^-$ masses are \cite{Roberts:2011cf,Chen:2012qr}
\begin{equation}
\label{CompMasses}
m_{N} = 1.14\,{\rm GeV}, \quad m_{\Delta} = 1.39\,{\rm GeV}\,, \quad
m_\Omega = 1.76\,{\rm GeV} .
\end{equation}
The unit-normalised eigenvectors are: nucleon,
\begin{equation}
s= 0.88, \quad a_{1}^{\{uu\}}= -0.38, \quad a_{2}^{\{uu\}}= -0.065,
\end{equation}
with $a_{i}^{\{ud\}} = -a_{i}^{\{uu\}} /\sqrt{2} ,\; i=1,2$; and $\Delta^+$,
$d^{\{uu\}} =  0.58$, $d^{\{ud\}} =  \surd 2 d^{\{uu\}}$.  Having only one term in its amplitude, the $\Omega^-$ result is trivial, providing simply an overall canonical normalisation factor.

\subsection{Meson cloud}
\label{secMesonCloud}
The computed masses in Eq.\,\eqref{CompMasses} are greater than those determined empirically \cite{Beringer:1900zz}: $m_N^{\rm exp}=0.94\,$GeV; $m_{\Delta}^{\rm exp} = 1.23\,{\rm GeV}$; and $m_{\Omega}^{\rm exp} = 1.67\,{\rm GeV}$.  This is appropriate, given that the Bethe-Salpeter and Faddeev equation kernels omit resonant contributions; i.e., do not contain effects that may phenomenologically be associated with a meson cloud.  In practical calculations, meson-cloud effects divide into two distinct classes.  \label{page:pionloops}
The first (type-1) is within the gap equation, where pseudoscalar-meson loop corrections to the dressed-quark-gluon vertex act uniformly to reduce the infrared mass-scale associated with the mass-function of a dressed-quark \cite{Eichmann:2008ae,Blaschke:1995gr,Fischer:2007ze,Cloet:2008fw,Chang:2009ae}.  This effect can be pictured as a single quark emitting and reabsorbing a pseudoscalar meson.  It can be mocked-up by simply choosing the parameters in the gap equation's kernel so as to obtain a dressed-quark mass that is characterised by an energy-scale of approximately $400\,$MeV.  Such an approach has implicitly been widely employed with phenomenological success \cite{Chang:2011vu,Maris:2003vk,Roberts:2000aa,Roberts:2007jh} and we use it herein.

The second sort of correction (type-2) arises in connection with bound-states and may be likened to adding pseudoscalar meson exchange \emph{between} dressed-quarks within the bound-state \cite{Pichowsky:1999mu,Roberts:1988yz,Hollenberg:1992nj,Alkofer:1993gu,Mitchell:1996dn,Ishii:1998tw,%
Hecht:2002ej}, as opposed to the first type of effect; i.e., emission and absorption of a meson by the same quark.  The type-2 contribution, depicted explicitly in Fig.\,1 of Ref.\,\cite{Ishii:1998tw}, is that computed in typical evaluations of meson-loop corrections to hadron observables based on a point-hadron Lagrangian \cite{Hecht:2002ej}.  These are the corrections that should be added to the calculated results in Eq.\,\eqref{CompMasses}.  This is readily illustrated in connection with the $\Omega^-$ baryon.  With our value of the $s$-quark mass, the computed vector-meson dressed-quark-core mass is $m_\phi=1.13$, which is $110\,$MeV above the experimental value.  Pseudoscalar-meson loop corrections are estimated to reduce the core mass by $\simeq 100\,$MeV \cite{Eichmann:2008ae,Leinweber:2001ac}.  Furthermore, a similar analysis indicates that, at $m_{0^-}^2=0.5\,$GeV$^2$,
pseudoscalar-meson loop corrections in the $\Omega$ system produce a ($-100\,$MeV) shift in the mass of the baryon's dressed-quark core \cite{Young:2002cj}, a result which reconciles the value in Eq.\,\eqref{CompMasses} with experiment.

These observations underpin a view that bound-state kernels which omit type-2 meson-cloud corrections should produce dressed-quark-core masses for hadron ground-states that are larger than the empirical values.  This is certainly true in practice \cite{Roberts:2011cf,Chen:2012qr,Wang:2013wk}.  Moreover, as we shall again see herein, this perspective also has implications for the description of elastic and transition form factors \cite{Wilson:2011aa,Segovia:2013rca,Eichmann:2008ef,Cloet:2008re,Cloet:2008wg}.

\subsection{\mbox{\boldmath $\Delta$} elastic form factor}
The matrix element of the electromagnetic current operator between $J=3/2$ states can be expressed through four form factors: Coulomb monopole (E0); magnetic
dipole (M1); electric quadrupole (E2); and magnetic octupole (M3).  In order to construct those form factors, one may first write the $\Delta \gamma\Delta$ vertex as \cite{Nicmorus:2010sd}:
\begin{equation}
\Lambda_{\mu,\lambda\omega}(K,Q) = \Lambda_{+}(P_{f})R_{\lambda\alpha}(P_{f})
\Gamma_{\mu,\alpha\beta}(K,Q) \Lambda_{+}(P_{i})R_{\beta\omega}(P_{i}),
\label{eq:DEMcurrent}
\end{equation}
where the positive-energy projection matrix $\Lambda_{+}(P)$ and Rarita-Schwinger projection operator $R_{\mu\nu}(P)$ are defined, e.g., in Eqs.\,(A.13), (A.14) of Ref.\,\cite{Roberts:2011cf} and
\begin{equation}
\Gamma_{\mu,\alpha\beta}(K,Q) =
\left[(F_{1}^{\ast}+F_{2}^{\ast})i\gamma_{\mu}-\frac{F_{2}^{\ast}}{m_{\Delta}}K_{\mu}
\right]\delta_{\alpha\beta} -\left[(F_{3}^{\ast}+F_{4}^{\ast})i\gamma_{\mu}-\frac{F_{4}^{\ast}}{m_{\Delta}}K_{
\mu }\right]\frac{Q_{\alpha}Q_{\beta}}{4m_{\Delta}^{2}}.
\label{eq:Gammamualbe}
\end{equation}
This vertex involves two momenta, expressed through the ingoing, $P_{i}$, and
outgoing, $P_{f}$, baryon momenta, or by the average momentum $K=(P_{i}+P_{f})/2$ and the photon momentum $Q=P_{f}-P_{i}$.  Since the particle is on-shell, so that
$P_{i}^{2}=P_{f}^{2}=-m_{\Delta}^{2}$, one has
\begin{equation}
K^{2} = -m_{\Delta}^{2}(1+\tau), \quad K\cdot Q = 0,
\end{equation}
where $\tau=Q^{2}/(4m_{\Delta}^{2})$.  It follows that the Poincar\'e invariant form factors which constitute the vertex depend only on the photon momentum-transfer $Q^{2}$: $\{F_i^\ast=F_i^\ast(Q^2), i=1,2,3,4\}$.  The multipole form factors are constructed as follows:
\begin{subequations}
\label{DefsDeltaElastic}
\begin{eqnarray}
G_{E0}(Q^{2}) &=& \left(1+\frac{2\tau}{3}\right)(F_{1}^{\ast}-\tau F_{2}^{\ast}) -
\frac{\tau}{3} (1+\tau) (F_{3}^{\ast}-\tau F_{4}^{\ast}), \\
G_{M1}(Q^{2}) &=& \left(1+\frac{4\tau}{5}\right)(F_{1}^{\ast}+F_{2}^{\ast}) -
\frac{2\tau}{5} (1+\tau) (F_{3}^{\ast}+F_{4}^{\ast}), \\
G_{E2}(Q^{2}) &=& (F_{1}^{\ast}-\tau F_{2}^{\ast}) -
\frac{1}{2} (1+\tau) (F_{3}^{\ast}-\tau F_{4}^{\ast}), \\
G_{M3}(Q^{2}) &= &(F_{1}^{\ast}+F_{2}^{\ast}) -
\frac{1}{2} (1+\tau) (F_{3}^{\ast}+ F_{4}^{\ast})\,,
\end{eqnarray}
\end{subequations}
and, as stated in the Introduction, their $Q^2=0$ values define dimensionless multipole moments:
\begin{equation}
e_{\Delta} = G_{E0}(0)\,, \quad
\hat \mu_{\Delta} = G_{M1}(0)\,, \quad
\hat {\cal Q}_{\Delta} = G_{E2}(0) \,, \quad
\hat {\cal O}_{\Delta} = G_{M3}(0)\,.
\end{equation}

Given the vertex, $\Gamma_{\mu,\alpha\beta}(K,Q)$, one may obtain the multipole form factors using any four sensible projection operators; e.g., with \cite{Nicmorus:2010sd}
\begin{equation}
\mathpzc{p}_{1} = \check{P}_{\mu}\check{P}_{\lambda}\check{P}_{\omega}{\rm tr}_{\rm D} \Lambda_{\mu,\lambda\omega}\,, \quad
\mathpzc{p}_{2} = \check{P}_{\mu} {\rm tr}_{\rm D} \Lambda_{\mu,\lambda\lambda}\,, \quad
\mathpzc{p}_{3} = \check{P}_{\lambda}\check{P}_{\omega} {\rm
tr}_{\rm D} \Lambda_{\mu,\lambda\omega}\gamma^{\perp}_{\mu}\,, \quad
\mathpzc{p}_{4} = {\rm tr}_{\rm D} \Lambda_{\mu,\lambda\lambda}\gamma^{\perp}_{\mu}\,,
\label{eq:scalars}
\end{equation}
where the trace is over spinor indices, $P\cdot \gamma^{\perp}=0$ and $\check{P}^{2}=+1$, one has
\begin{equation}
\begin{array}{ll}
\displaystyle G_{E0} = \frac{\mathpzc{p}_{2}-2\mathpzc{p}_{1}}{4i\sqrt{1+\tau}}\,, &
\displaystyle G_{M1} = \frac{9i(\mathpzc{p}_{4}-2\mathpzc{p}_{3})}{40\tau}\,, \\
\displaystyle G_{E2} =
\frac{3\left[\mathpzc{p}_{1}\left(3+2\tau\right)
-\mathpzc{p}_{2}\tau\right]}{8i\tau^{2}\sqrt{1+\tau}}\,, &
\displaystyle G_{M3} = \frac{3i\left[\mathpzc{p}_{3}(5+4\tau)-2\mathpzc{p}_{4}\tau\right]}{32\tau^{3}}\,.
\end{array}
\end{equation}

\begin{figure}[t]
\begin{center}
\epsfig{figure=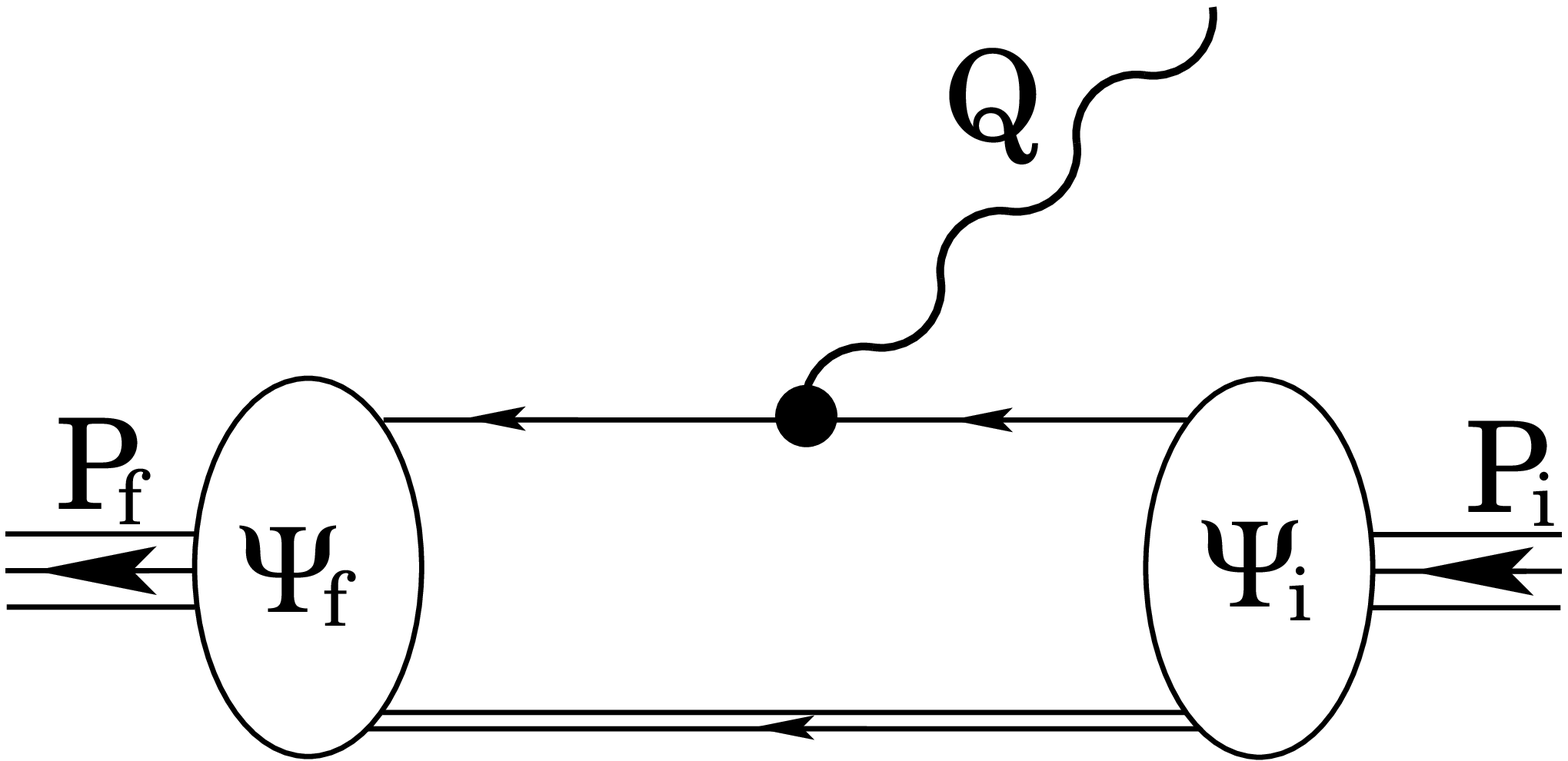,width=0.28\textwidth}\hspace*{\fill}
\epsfig{figure=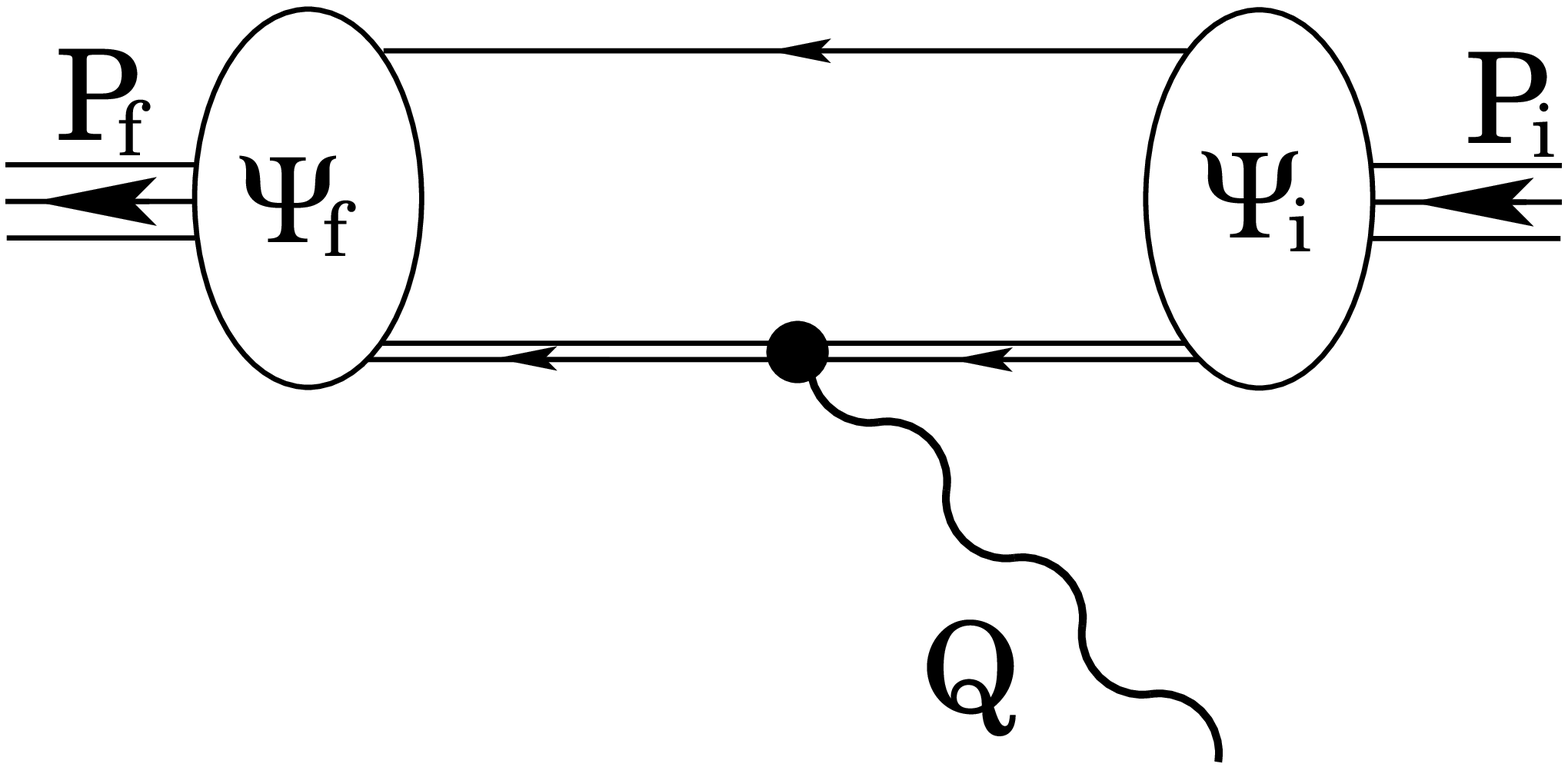,width=0.28\textwidth}\hspace*{\fill}
\epsfig{figure=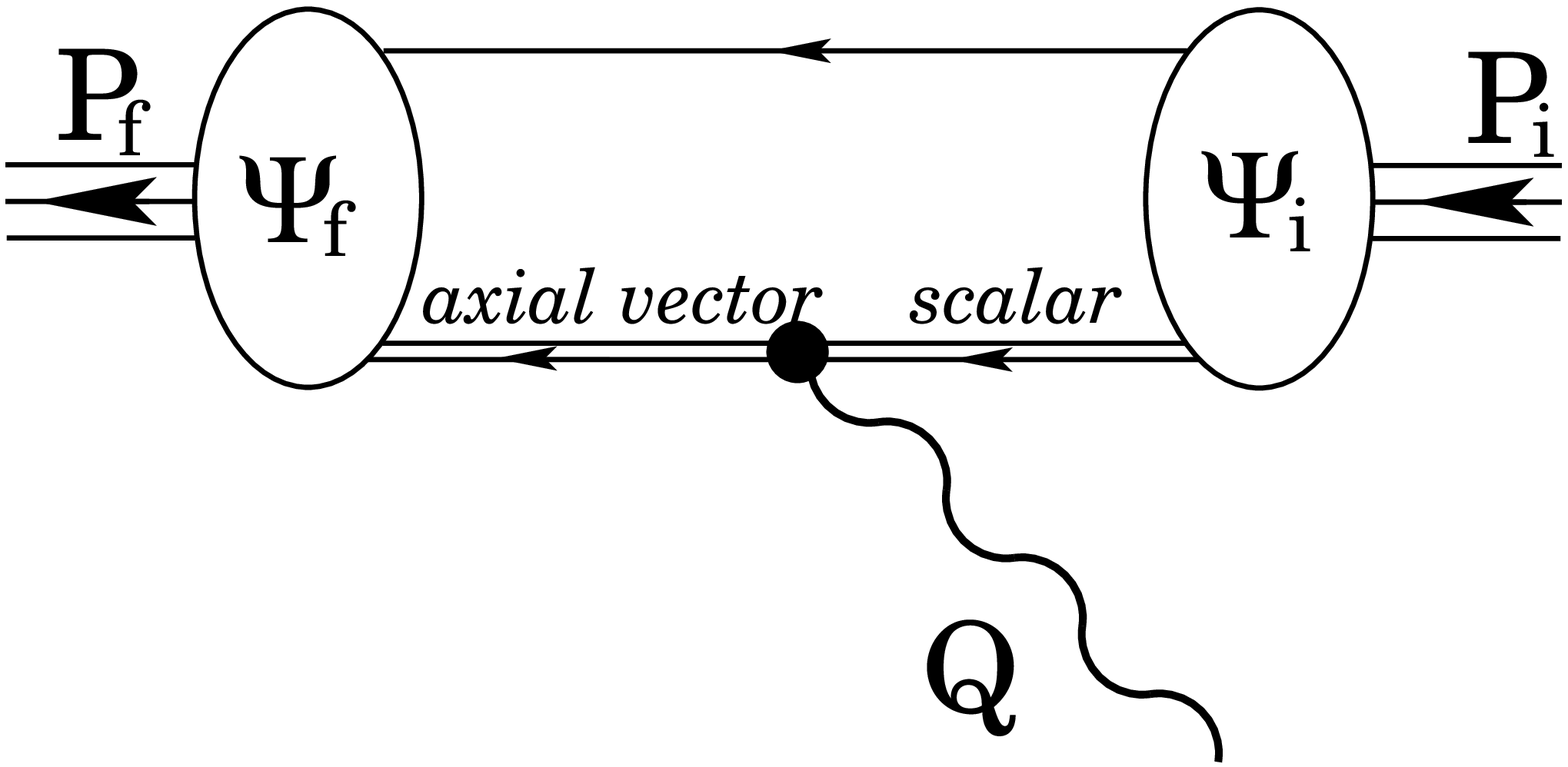,width=0.28\textwidth}
\caption{\label{fig:Transitioncurrent} One-loop diagrams in the $\Delta\gamma\Delta$, $\Omega\gamma\Omega$ and $N\gamma\Delta$ vertices.
The single line represents a dressed-quark propagator, $S(p)$; the double line, a diquark propagator; and the vertices are, respectively, the incoming and outgoing baryon, $\Psi_i$, $\Psi_f$.
From left to right, the diagrams describe the photon coupling: directly to a dressed-quark, Sect.\,\protect\ref{app:subsubsec:qgammaq}; to a diquark, in an elastic scattering event, Sect.\,\protect\ref{subsubsec:agammaa}; or inducing a transition between scalar and axial-vector diquarks, Sect.\,\protect\ref{subsubsec:agammas}.
In the computation of the elastic $J=3/2$ vertex, there is no contribution from the third diagram because such baryons contain only axial-vector diquark correlations (see Eqs.\,\protect\eqref{DeltaA}, \protect\eqref{eq:D31p}).
In the general case, there are three more diagrams, described in detail elsewhere \cite{Cloet:2008re}, which represent two-loop integrals.
}
\end{center}
\end{figure}

With our symmetry-preserving regularisation of the contact interaction and treatment of the Faddeev equation described in
App.\,\ref{app:sec:ContactInteractionModel}, there are two contributions to $\Gamma_{\mu,\alpha\beta}$.  They are illustrated in Fig.\,\ref{fig:Transitioncurrent} and detailed in App.\,\ref{app:sec:EMcurrentQuarkDiquark}.  The computation of the $\Delta$-baryon elastic form factors is straightforward following the procedures outlined therein.  One may obtain the analogous $\Omega^-$-baryon elastic form factors via uncomplicated modifications of these formulae.

\subsection{\mbox{\boldmath $\gamma+N\to \Delta$} transition form factor}
Data on the $\gamma^\ast p \to \Delta^+$ transition \cite{Aznauryan:2011ub,Aznauryan:2011qj} have stimulated a great deal of theoretical analysis, and speculation about, \emph{inter alia}:
the relevance of perturbative QCD (pQCD) to processes involving moderate momentum transfers \cite{Aznauryan:2011qj,Carlson:1985mm,Pascalutsa:2006up};
shape deformation of hadrons \cite{Alexandrou:2012da};
and, of course, the role that resonance electroproduction experiments can play in exposing nonperturbative features of QCD \cite{Aznauryan:2012ba}.

The $\gamma N \to \Delta$ transition is described by three Poincar\'e-invariant form factors~\cite{Jones:1972ky}: magnetic-dipole, $G_{M}^{\ast}$; electric quadrupole, $G_{E}^{\ast}$; and Coulomb (longitudinal) quadrupole, $G_{C}^{\ast}$.  They arise through consideration of the $N\to \Delta$ transition current:
\begin{equation}
J_{\mu\lambda}(K,Q) =
\Lambda_{+}(P_{f})R_{\lambda\alpha}(P_{f})i\gamma_{5}\Gamma_{\alpha\mu}(K,Q)\Lambda_
{+}(P_{i}),
\label{eq:JTransition}
\end{equation}
where: $P_{i}$, $P_{f}$ are, respectively, the incoming nucleon and outgoing $\Delta$
momenta, with $P_{i}^{2}=-m_{N}^{2}$, $P_{f}^{2}=-m_{\Delta}^{2}$; the incoming
photon momentum is $Q_\mu=(P_{f}-P_{i})_\mu$ and $K=(P_{i}+P_{f})/2$; and
$\Lambda_{+}(P_{i})$, $\Lambda_{+}(P_{f})$ are, respectively, positive-energy
projection operators for the nucleon and $\Delta$, with the Rarita-Schwinger tensor
projector $R_{\lambda\alpha}(P_f)$ arising in the latter connection.

In order to succinctly express $\Gamma_{\alpha\mu}(K,Q)$, we define
\begin{equation}
\check K_{\mu}^{\perp} = {\cal T}_{\mu\nu}^{Q} \check{K}_{\nu}
= (\delta_{\mu\nu} - \check{Q}_{\mu} \check{Q}_{\nu}) \check{K}_{\nu},
\end{equation}
with $\check{K}^{2} = 1 = \check{Q}^{2}$, in which case
\begin{equation}
\Gamma_{\alpha\mu}(K,Q) =
\mathpzc{k}
\left[\frac{\lambda_m}{2\lambda_{+}}(G_{M}^{\ast}-G_{E}^{\ast})\gamma_{5}
\varepsilon_{\alpha\mu\gamma\delta} \check K_{\gamma}\check{Q}_{\delta}  - G_{E}^{\ast} {\cal T}_{\alpha\gamma}^{Q} {\cal T}_{\gamma\mu}^{K}
- \frac{i\varsigma}{\lambda_m}G_{C}^{\ast}\check{Q}_{\alpha} \check K^\perp_{\mu}\right],
\label{eq:Gamma2Transition}
\end{equation}
where
$\mathpzc{k} = \sqrt{(3/2)}(1+m_\Delta/m_N)$,
$\varsigma = Q^{2}/[2\Sigma_{\Delta N}]$,
$\lambda_\pm = \varsigma + t_\pm/[2 \Sigma_{\Delta N}]$
with $t_\pm = (m_\Delta \pm m_N)^2$,
$\lambda_m = \sqrt{\lambda_+ \lambda_-}$,
$\Sigma_{\Delta N} = m_\Delta^2 + m_N^2$, $\Delta_{\Delta N} = m_\Delta^2 - m_N^2$.

With a concrete expression for the current in hand, one may obtain the form factors
using any three sensibly chosen projection operations; e.g., with \cite{Eichmann:2011aa}
\begin{equation}
\mathpzc{t}_{1} = \mathpzc{n}  \frac{\sqrt{\varsigma(1+2\mathpzc{d})}}{\mathpzc{d}-\varsigma}
{\cal T}^{K}_{\mu\nu}\check K^\perp_{\lambda} {\rm tr}_{\rm D}
\gamma_{5}J_{\mu\lambda}\gamma_{\nu}\,,
\mathpzc{t}_{2} = \mathpzc{n} \frac{\lambda_{+}}{\lambda_m} {\cal T}^{K}_{\mu\lambda} {\rm tr}_{\rm D} \gamma_{5} J_{\mu \lambda}\,,
\mathpzc{t}_{3} =  3 \mathpzc{n}
\frac{\lambda_+}{\lambda_m}\frac{(1+2\mathpzc{d})}{\mathpzc{d}-\varsigma} \check
K^\perp_{\mu}\check K^\perp_{\lambda} {\rm tr}_{\rm D}\gamma_{5}J_{\mu\lambda} \,,
\end{equation}
where $\mathpzc{d}=\Delta_{\Delta N}/[2 \Sigma_{\Delta N}]$,
$\mathpzc{n}= \sqrt{1-4\mathpzc{d}^{2}}/[4i\mathpzc{k}\lambda_m]$), then
\begin{equation}
\label{GMGEGC}
G_{M}^{\ast} = 3
\left[ \mathpzc{t}_{2}+\mathpzc{t}_{1}\right]\,, \;
G_{E}^{\ast} = \mathpzc{t}_{2}-\mathpzc{t}_{1}\,, \;
G_{C}^{\ast} = \mathpzc{t}_{3}.
\end{equation}

\section{Elastic form factors: \mbox{\boldmath $\Delta(1232)$}}
\label{sec:GammaDeltaDelta}
\begin{figure}[t]
\begin{center}
\begin{tabular}{cc}
\includegraphics[clip,width=0.45\linewidth]{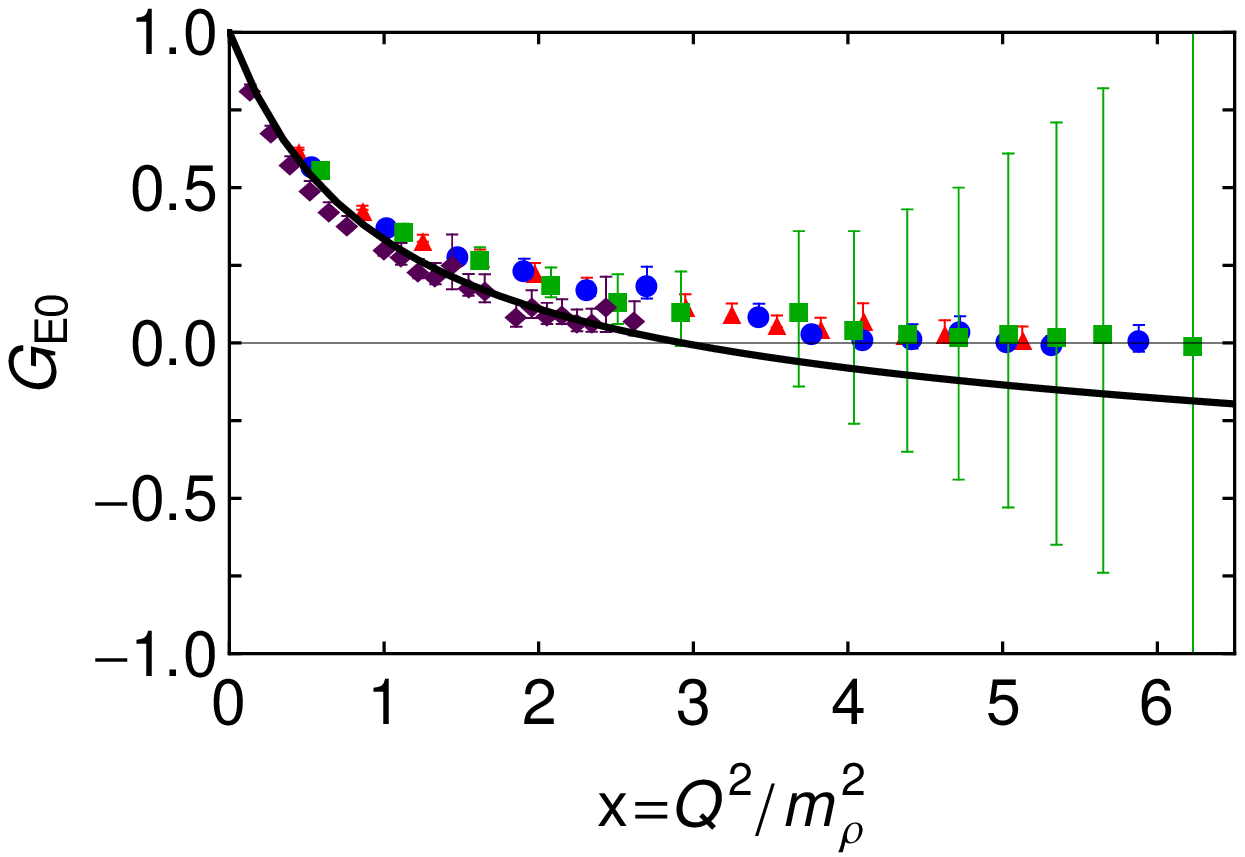}\vspace*{-1ex} &
\includegraphics[clip,width=0.43\linewidth]{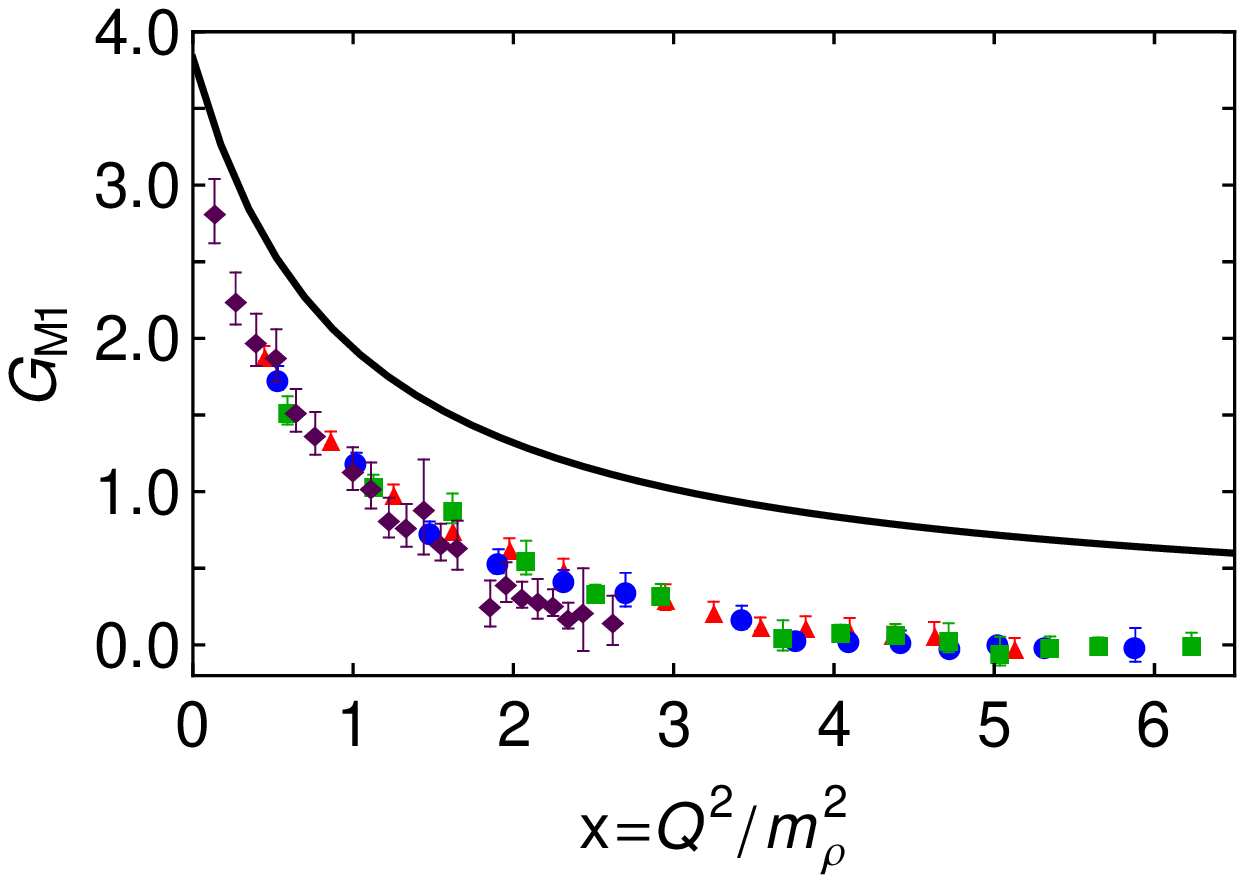}\vspace*{-1ex} \\
\includegraphics[clip,width=0.45\linewidth]{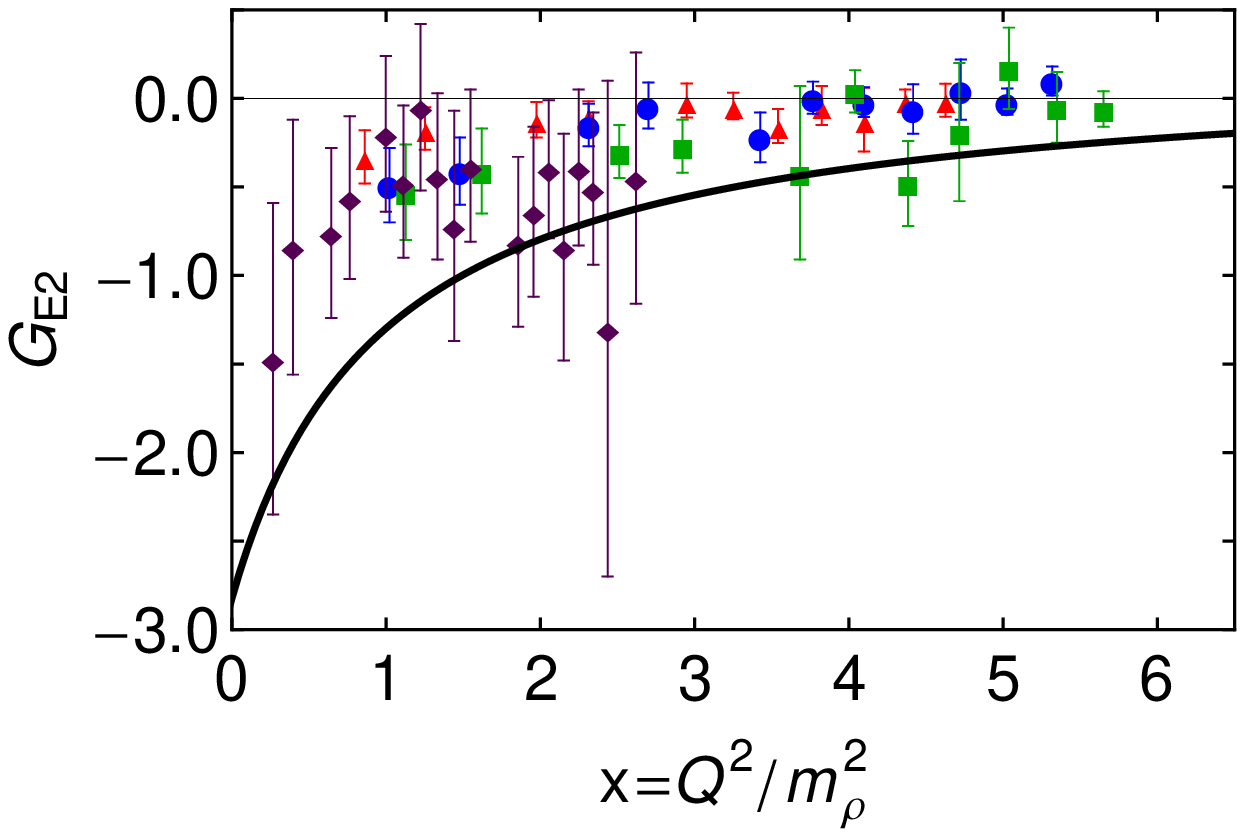}\vspace*{-1ex} &
\includegraphics[clip,width=0.43\linewidth]{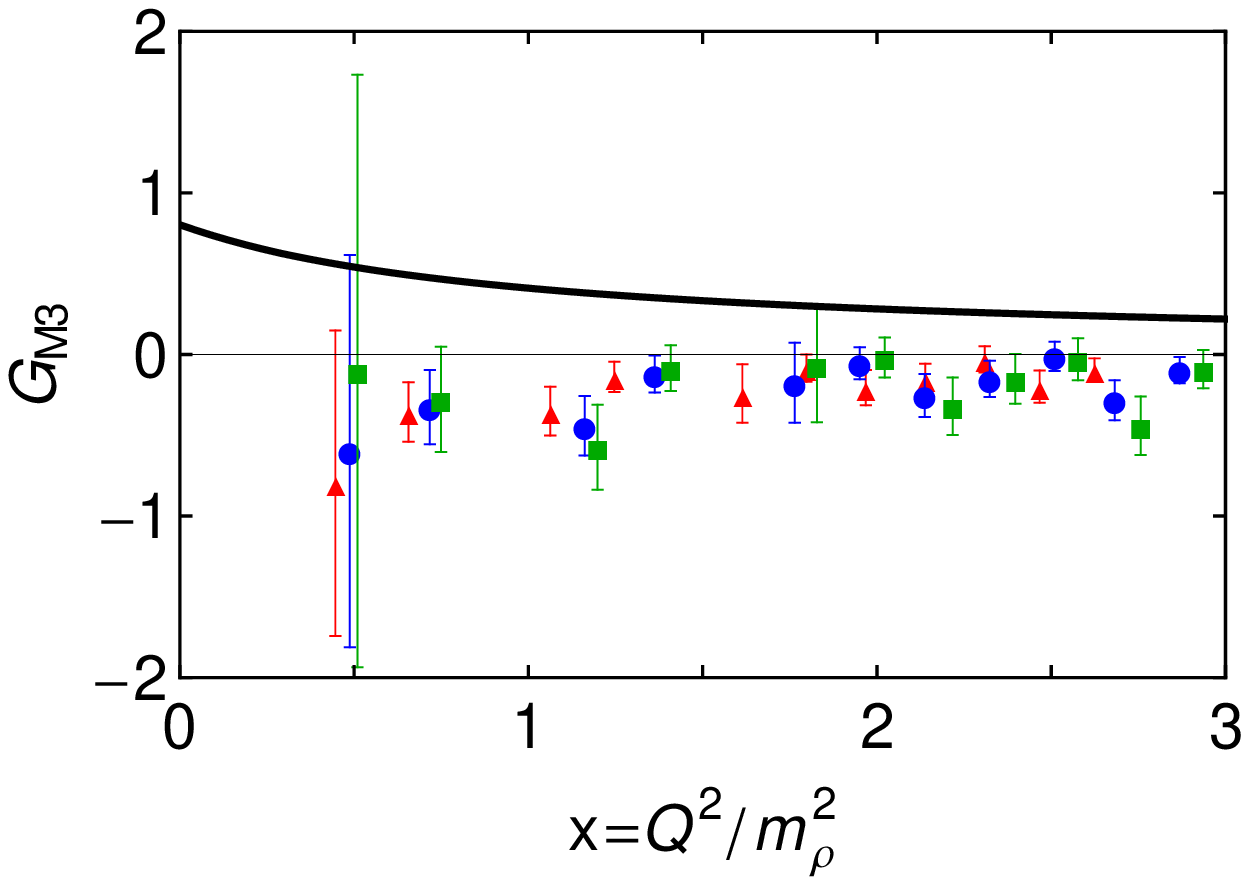}
\end{tabular}
\caption{\label{fig:elasticFFD} Dressed-quark-core contributions to the $\Delta^{+}$
electromagnetic form factors: $G_{E0}$, $G_{M1}$, $G_{E2}$ and $G_{M3}$.
%
%
In the absence of precise experimental data, the points depict results from numerical simulations of lattice-regularised QCD:
$G_{E0}$, $G_{M1}$ and $G_{E2}$, unquenched \protect\cite{Alexandrou:2009hs} (red triangles -- $m_{\pi}=691\,{\rm MeV}$, blue circles -- $m_{\pi}=509\,{\rm MeV}$, green squares -- $m_{\pi}=384\,{\rm MeV}$, and purple diamonds -- $m_{\pi}=353\,{\rm MeV}$);
and $G_{M3}$, quenched \protect\cite{Alexandrou:2007we} (red triangles -- $m_{\pi}=563\,{\rm MeV}$, blue circles -- $m_{\pi}=490\,{\rm MeV}$,
green squares -- $m_{\pi}=411\,{\rm MeV}$).
}
\end{center}
\end{figure}

In Fig.\,\ref{fig:elasticFFD} we depict our computed dressed-quark-core contributions to the $\Delta^{+}$ elastic electromagnetic form factors.  All our results are predictions; i.e., the values of our two parameters and the two current-quark masses were fixed elsewhere \cite{Roberts:2011cf} and are held at those values.  The behaviour of the form factors on $x\gtrsim 3$ highlights that contact interaction form factors are typically hard \cite{Wilson:2011aa,Chen:2012txa}.

Since there are no precise experimental data, the points depicted in Fig.\,\ref{fig:elasticFFD} are results obtained via numerical simulations of lattice-regularised QCD: $G_{E0}$, $G_{M1}$ and $G_{E2}$, unquenched \cite{Alexandrou:2009hs}; and $G_{M3}$, quenched only \cite{Alexandrou:2007we}.  The lattice simulation parameters produce the masses listed in Table~\ref{tab:pionmasses} and may be characterised as producing a stable $\Delta(1232)$-baryon with a root-mean-square mass of 1.55\,GeV (unquenched) or 1.43\,GeV (quenched).  Figure~\ref{fig:elasticFFD} reveals that our treatment of the contact interaction produces results which are qualitatively and semiquantitatively equivalent to those obtained in the best available lattice simulations.  There are some differences, however, and it is worth examining each panel separately.

First consider the top-left panel of Fig.\,\ref{fig:elasticFFD}, which displays the $\Delta^+$ electric monopole form factor: our computation is consistent with all lattice points.  In addition, our analysis predicts that $G_{E0}(Q^2)$ possesses a zero at $x=Q^2/m_\rho^2 \approx 3$.  For comparison, precisely the same framework predicts a zero in the $\rho$-meson's electric form factor at $x \approx 6$.
The available lattice data are not in conflict with this prediction.
It is worth noting that our treatment of the contact interaction also produces a zero in the proton's electric form factor \cite{Wilson:2011aa}.  It is located at $x \approx 4$, which, empirically, is too low in $Q^2$.  However, that underestimate comes about simply because the contact interaction produces form factors which are too hard.  Notably, the expression for the electric form factor in each case involves a destructive interference, with one or more negative contributions magnified by $Q^2$.  Naturally, this interference does not guarantee a zero in the electric form factor of a $J\geq 1/2$ bound-state but it does suggest both that a zero might be difficult to avoid, and that its appearance and location are a sensitive measure of the dynamics which underlies the bound-state's structure \cite{Cloet:2013gva}.

\begin{table}[t]
\begin{center}
\caption{\label{tab:pionmasses} Masses, in GeV, of the $\pi$, $\rho$ and $\Delta$, computed in the numerical simulations of lattice-regularised QCD that produced the points in Fig.\,\protect\ref{fig:elasticFFD}.}
\begin{tabular}{lccc}
\hline
Approach & $m_{\pi}$ & $m_{\rho}$ & $m_{\Delta}$ \\
\hline
Unquenched I   \protect\cite{Alexandrou:2009hs} & $0.691$ & $0.986$ & $1.687$ \\
Unquenched II  \protect\cite{Alexandrou:2009hs} & $0.509$ & $0.899$ & $1.559$ \\
Unquenched III \protect\cite{Alexandrou:2009hs}& $0.384$ & $0.848$ & $1.395$ \\
Hybrid         \protect\cite{Alexandrou:2009hs}& $0.353$ & $0.959$ & $1.533$ \\\hline
Quenched I    \protect\cite{Alexandrou:2007we} & $0.563$ & $0.873$ & $1.470$ \\
Quenched II   \protect\cite{Alexandrou:2007we} & $0.490$ & $0.835$ & $1.425$ \\
Quenched III  \protect\cite{Alexandrou:2007we} & $0.411$ & $0.817$ & $1.382$ \\
\hline
\end{tabular}
\end{center}
\end{table}

Given the electric form factor, one can readily compute a $\Delta^+$ charge radius:
\begin{equation}
\left\langle\right.\!\!r_{E0}^{2}\!\!\left.\right\rangle =
- 6 \left.\frac{dG_{E0}}{dQ^{2}}\right|_{Q^{2}=0} \approx
\frac{8}{m_\rho^2}=\frac{18}{m_\Delta^2}
\label{DeltaRadius}
\end{equation}
These values may be compared with the $\pi$ and $\rho$ meson radii computed in the same framework \cite{Roberts:2011wy}: $r_\pi^2 = 4.5/m_\rho^2$, $r_\rho^2 = 7.0/m_\rho^2$; and also the kindred proton radius \cite{Wilson:2011aa}: $r_p^2=6.7/m_\rho^2$.  Plainly, the electromagnetic size of the $\Delta^+$-baryon's dressed-quark-core is greater than that of these other light-quark systems.\footnote{N.B.\ The dimensionless product $m_\rho^2 r_{\Delta^+}^2$ computed using lattice-QCD is very sensitive to $m_\pi^2$: it grows rapidly as $m_\pi^2$ is decreased.  Therefore, in the absence of simulations at realistic masses, we choose not to report a lattice value for $r_{\Delta^+}$.}   To pursue this further, we considered the analysis in Refs.\,\cite{Buchmann:1996bd}, which leads us to provide the following comparison:
\begin{equation}
r_{\Delta^+}^2 = \frac{8}{m_\rho^2} \approx r_p^2-r_n^2 = \frac{7.7}{m_\rho^2};
\end{equation}
viz., numerically, at the level of 4\%, our computed $\Delta^+$ dressed-quark-core charge-radius-squared is approximately equal to the isovector combination of nucleon dressed-quark-core radii-squared computed in the same framework.  A precise equality between these two quantities is a prediction of the nonrelativistic chiral constituent-quark model and associated current constructed from numerous ingredients in Ref.\,\cite{Buchmann:1996bd}.  Whilst that result is model specific, it is curious that our Poincar\'e covariant framework, with its entirely different foundation and formulation, produces a numerically equivalent result.\footnote{Slightly modified estimates have been derived using a large-$N_c$ analysis \protect\cite{Buchmann:2000wf}.  Taking into account the isospin-symmetry identities derived in App.\,\ref{CurrentSymmetry}, it is evident that our framework is consistent with those relations.}


The top-right panel of Fig.\,\ref{fig:elasticFFD} depicts the $\Delta^+$ magnetic dipole form factor.  The computed dimensionless magnetic moment $\hat\mu_{\Delta^+}=G_{M1}(Q^2=0)$ is listed in Table~\ref{tab:comparative}.  Notably, the value of $\hat\mu_{\Delta^+}$ is dynamical; i.e., it is not constrained by any symmetry, and our results for $G_{M1}(Q^2)$ lie uniformly above the lattice output.  Critical in the latter connection, however, are the range of lattice-QCD masses for the pion, $\rho$-meson and $\Delta$-baryon in Table~\ref{tab:pionmasses}: they are too large.  We will return to this point.


The bottom-left panel in Fig.\,\ref{fig:elasticFFD} displays our calculated  $\Delta^+$ electric quadrupole form factor.  Once more, our computations agree with the trend of the lattice results but uniformly overestimate their magnitude.  The dimensionless quadrupole moment $\hat{\mathpzc{Q}}_{\,\Delta}=G_{E2}(Q^2=0)$ is listed in Table~\ref{tab:comparative}.  The value is negative; agrees with the quoted lattice result, within the large lattice error; and is slightly smaller in magnitude than the ``natural'' value.  Taken as a whole, on the other hand, it has been judged \cite{Alexandrou:2009hs} that the lattice calculations favour $|\hat{\mathpzc{Q}}_{\,\Delta}| \approx 1$ .  This is a significant difference between our analysis and estimates based on lattice-QCD.  Here, too, the large lattice masses (see Table~\ref{tab:pionmasses}) are playing a role.


The bottom-right panel in Fig.\,\ref{fig:elasticFFD} displays the $\Delta^+$ magnetic octupole form factor.  In this case, only quenched lattice results are available \cite{Alexandrou:2007we} and they are consistent with zero.  We, on the other hand, have a clear, positive result.  As we shall see, the question posed by this mismatch is again answered by the unrealistically large value of the lattice results for $m_\Delta$ in Table~\ref{tab:pionmasses}.  Another issue is not, however: our value of $\hat{\mathpzc{O}}_{\;\Delta^+} \approx 1$ has the same magnitude but opposite sign to the ``natural'' value proposed in Ref.\,\cite{Alexandrou:2009hs}.  This might be the clearest indication amongst the static observables that the $\Delta(1232)$ is far from an ideal point particle.  On the other hand, it could mean that the attempt to use SUperGRAvity to estimate the natural point-particle values has failed in this case.

\begin{table*}[t]
\begin{center}
\caption{\label{tab:comparative}
Static electromagnetic properties of the $\Delta^+(1232)$: the row labelled ``CI'' lists our contact interaction results. 
The experimental value for $G_{M1}(Q^{2}=0)$ is drawn from Ref.\,\protect\cite{Beringer:1900zz};
and the remaining rows report a representative selection of results from other calculations.  The $Q^2=0$ values associated with lattice simulations were obtained by fitting the available results and extrapolating (see Fig.\,\protect\ref{fig:elasticFFD}).
N.B.\ The symbol ``-'' in any location indicates that no result was reported for that quantity in the related reference.}

\begin{tabular}{lclll}\hline
Approach & Reference & $G_{M1}(0)$ & $G_{E2}(0)$ & $G_{M3}(0)$\\\hline
CI & & $+3.83$ & $-2.82$ &  $+0.80$ \rule{-0.6em}{0ex} \rule{0em}{2.2ex} \\\hline
%
``Natural'' point-particle values & \cite{Alexandrou:2009hs} & $+3$ & $-3$ & $-1$ \rule{-0.6em}{0ex} \rule{0em}{2.2ex}\\\hline
Exp & \cite{Beringer:1900zz} & $+3.6^{+1.3}_{-1.7}\pm2.0\pm4$ & - & - \rule{-0.6em}{0ex} \rule{0em}{2.2ex}\\\hline
Lattice-QCD (hybrid) & \cite{Alexandrou:2009hs}
& $+3.0\pm0.2$ & $-2.06^{+1.27}_{-2.35}$ & $~0.00$ \rule{-0.6em}{0ex} \rule{0em}{2.2ex}\\
$1/N_{c}+ N\to \Delta$ &\cite{Alexandrou:2009hs}
& - & $-1.87 \pm 0.08$ & - \\
%
Faddeev equation & \cite{Maris:1999nt,Sanchis-Alepuz:2013iia} & $+2.38$ & $-0.67$ & $>0$\\
Covariant $\chi$PT  & \cite{Geng:2009ys}
& $+3.74\pm0.03$ & $-0.9\pm0.6$ & $-0.9\pm2.1$ \\
$+\,$qLQCD & \cite{Boinepalli:2009sq} & & & \\
QCD-SR                     & \cite{Aliev:2009np} & $+4.2\pm1.1$ & $-0.6\pm0.2$ & $-0.7\pm0.2$ \\
$\chi$QSM                  & \cite{Ledwig:2008es} & $+3.1$ & $-2.0$ & - \\
General Param. Method  & \cite{Buchmann:2002xq,Buchmann:2008zza}
& - & $-4.4$ & $-2.6$ \\
QM$+$exchange-currents & \cite{Buchmann:1996bd} & $+4.6$ & $-4.6$ & - \\
$1/N_{c}+m_s$-expansion              & \cite{Luty:1994ub} & $+3.8\pm0.3$ & - & - \\
RQM                        & \cite{Schlumpf:1993rm} & $+3.1$ & - & - \\
HB$\chi$PT                 & \cite{Butler:1993ej} & $+2.8\pm0.3$ & $-1.2\pm0.8$ & - \\
nrCQM                        & \cite{Krivoruchenko:1991pm} & $+3.6$ & $-1.8$ & - \\
\hline
\end{tabular}
\end{center}
\end{table*}

It is now appropriate to address the $m_\Delta$-dependence of the $\Delta$ elastic form factors.  The pattern of pion, $\rho$-meson and $\Delta$-baryon masses in Table~\ref{tab:pionmasses} matches that explained in Refs.\,\cite{Eichmann:2008ae,Nicmorus:2010sd}.  The momentum-dependent interaction therein, based on Refs.\,\cite{Eichmann:2008ae,Maris:1999nt}, produces $m_\Delta \approx 1.8\,$GeV; i.e., both type-1 and type-2 meson-cloud corrections are omitted.  We have therefore recomputed the $\Delta$ elastic form factors by using $m_\Delta =1.8\,$GeV in Eqs.\eqref{DefsDeltaElastic}.  In doing this, we alter the kinematics but assume that the internal structure of the $\Delta(1232)$-baryon's dressed-quark-core is only weakly sensitive to changes in current-quark mass.  This assumption is suggested by the predominantly $S$-wave character of the $\Delta(1232)$ Faddeev amplitude and supported by the analysis in Ref.\,\cite{Roberts:2011cf}, which shows that the $\Delta$-baryon's dressed-quark-core is accurately described as an almost non-interacting system of a dressed-quark and axial-vector diquark over a large range of current-quark masses.
%

\begin{figure}[t]
\begin{center}
\begin{tabular}{cc}
\includegraphics[clip,width=0.46\linewidth]{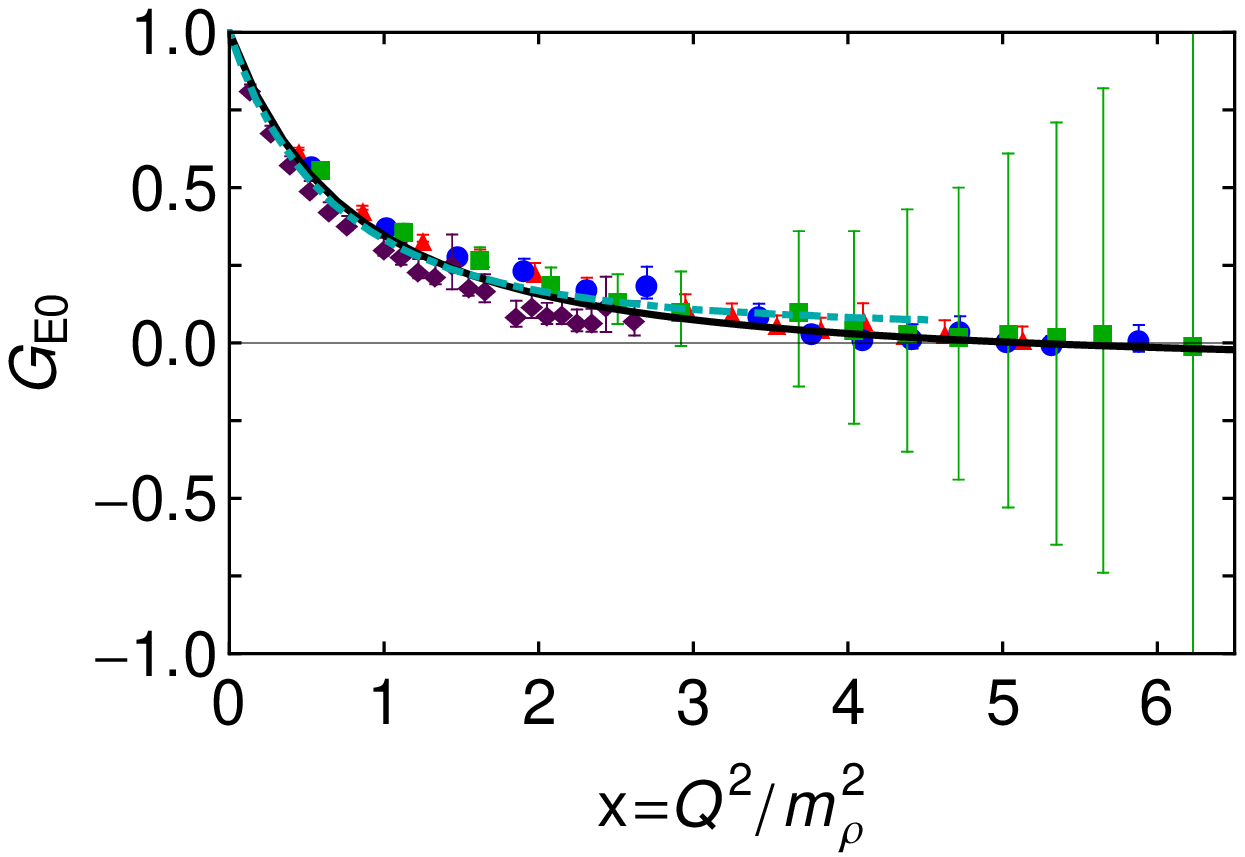}\vspace*{
-1ex }&
\includegraphics[clip,width=0.43\linewidth]{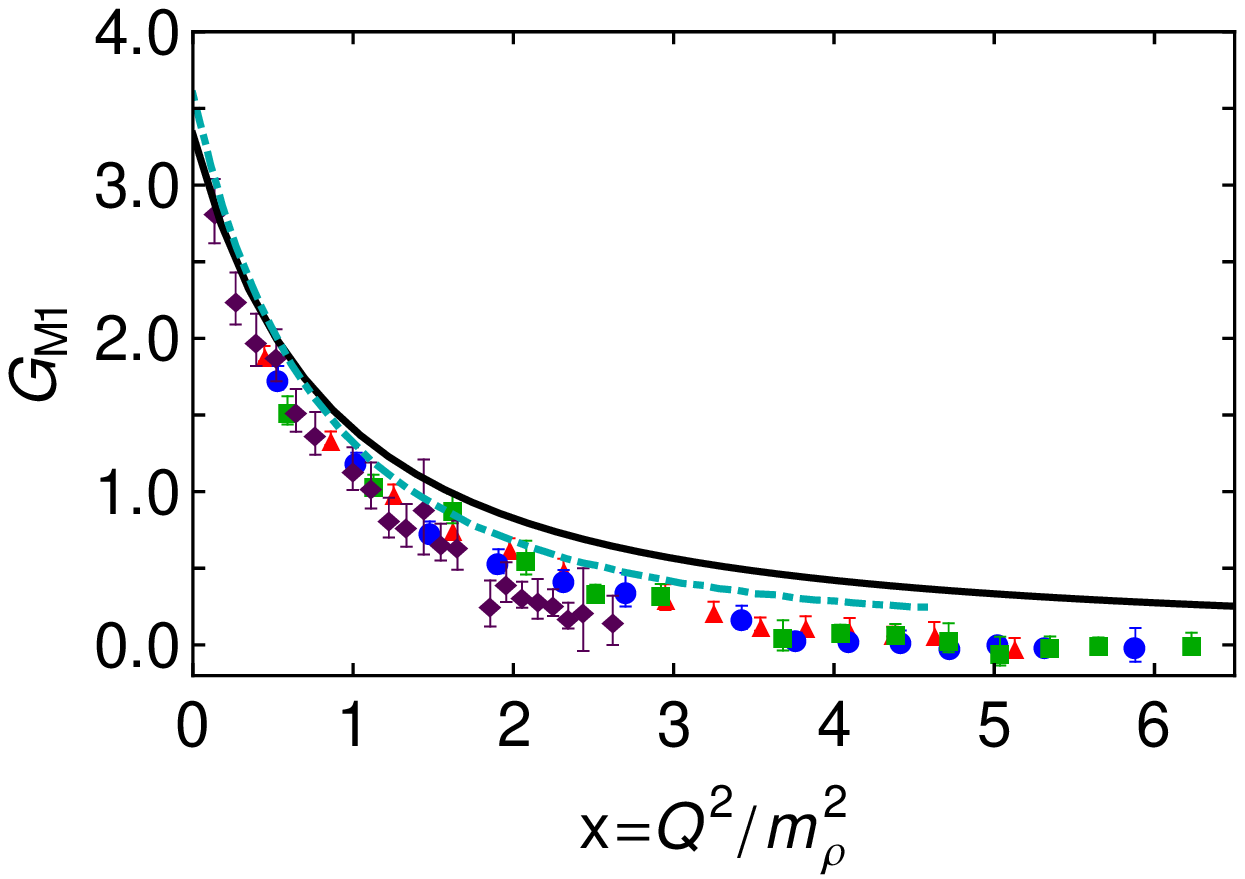}\vspace*{
-1ex} \\
\includegraphics[clip,width=0.43\linewidth]{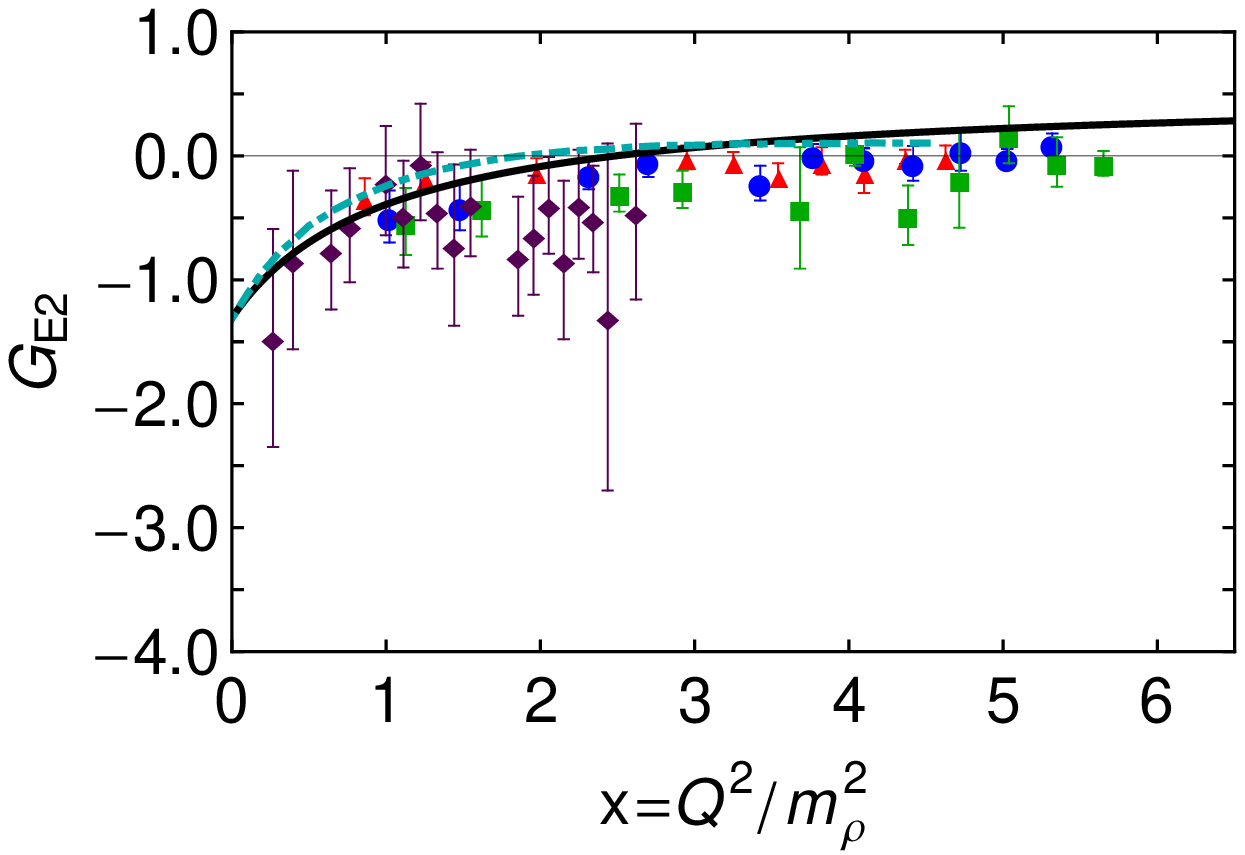}\vspace*{
-1ex}&
\includegraphics[clip,width=0.45\linewidth]{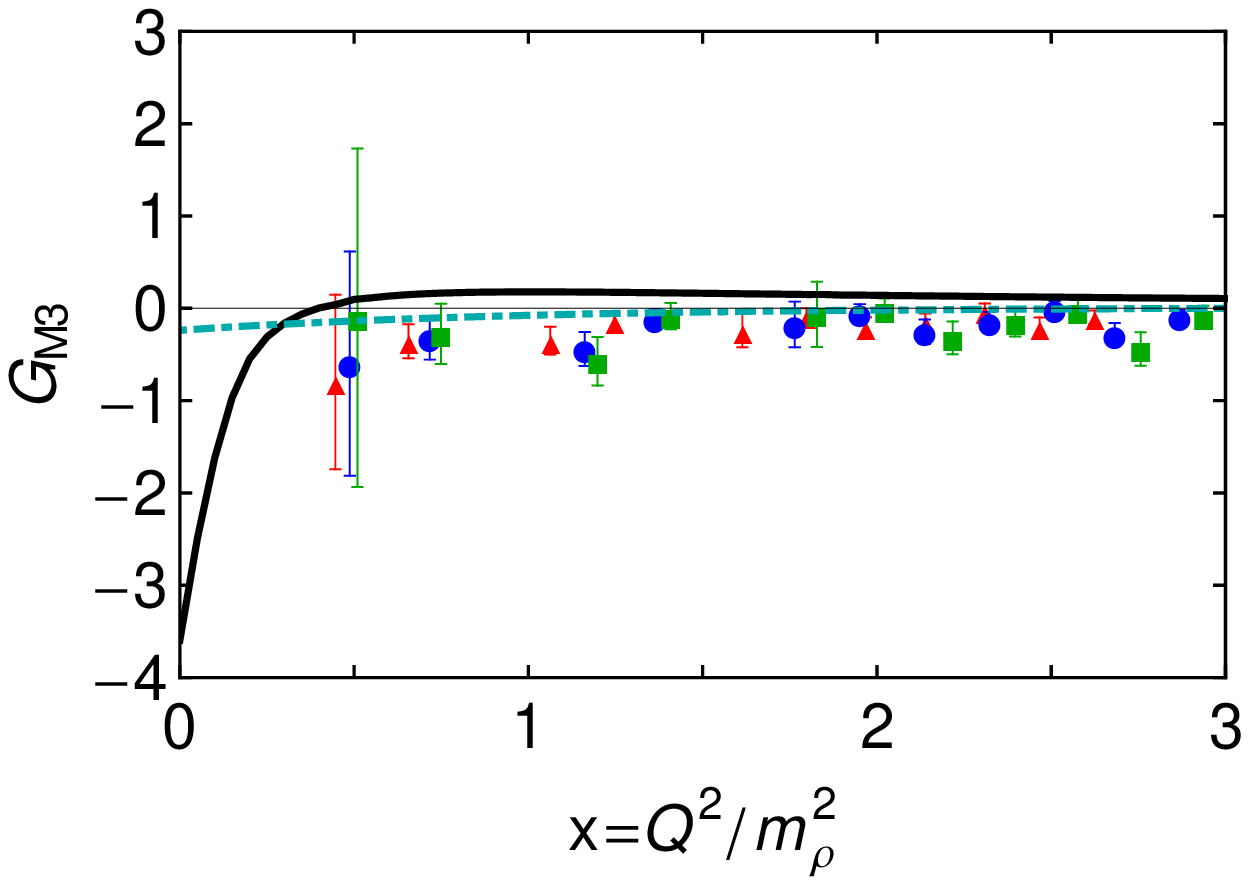}
\end{tabular}
\caption{\label{fig:hugemassFFD}
\emph{Solid curve} -- Contact-interaction dressed-quark-core contributions to the $\Delta^{+}$ electromagnetic form factors: $G_{E0}$, $G_{M1}$, $G_{E2}$ and $G_{M3}$, obtained with $m_\Delta = 1.8\,$GeV in Eqs.\,\protect\eqref{DefsDeltaElastic}.
\emph{Dash-dot curve} -- Analogous results from Ref.\,\protect\cite{Nicmorus:2010sd}: the numerical algorithm employed therein limits the calculations to $x \lesssim 4.5$.
In the absence of precise experimental data, the points depict results from numerical simulations of lattice-regularised QCD:
$G_{E0}$, $G_{M1}$ and $G_{E2}$, unquenched \protect\cite{Alexandrou:2009hs} (red triangles -- $m_{\pi}=691\,{\rm MeV}$, blue circles -- $m_{\pi}=509\,{\rm MeV}$, green squares -- $m_{\pi}=384\,{\rm MeV}$, and purple diamonds -- $m_{\pi}=353\,{\rm MeV}$);
and $G_{M3}$, quenched \protect\cite{Alexandrou:2007we} (red triangles -- $m_{\pi}=563\,{\rm MeV}$, blue circles -- $m_{\pi}=490\,{\rm MeV}$,
green squares -- $m_{\pi}=411\,{\rm MeV}$).}
\end{center}
\end{figure}

The results obtained by following this procedure are presented in Fig.\,\ref{fig:hugemassFFD}: the effects of the kinematic modification are dramatic.  As apparent in the figure, the magnitude and momentum-dependence of the form factors is now practically indistinguishable from the results in Ref.\,\cite{Nicmorus:2010sd}, Fig.\,6, even unto the appearance of zeros in the electric quadrupole and magnetic octupole form factors.   Furthermore, accounting for the hardness of contact-interaction form factors, the $m_\Delta=1.8\,$GeV results agree with the points produced by the numerical simulations of lattice-regularised QCD.  We judge, therefore, that the lattice results are grossly affected by the kinematic impact of an unrealistically large mass for the $\Delta(1232)$ so that it is misleading to infer too much about the empirical $\Delta(1232)$ resonance from existing lattice studies.

An examination of Table~\ref{tab:comparative} is instructive.  Omitting our computations for the present, results from a diverse array of analyses are presented.  If they are weighted equally, then one obtains a mean and median value of $\hat \mu_{\Delta^+} = 3.4$ with a standard deviation of $0.7$.  Including our result, then one has a mean of $3.5$, a median of $3.6$ and a standard deviation of $0.7$.  So, there is a modicum of agreement between the theoretical predictions.  With $\hat{\mathpzc{Q}}_{\;\Delta^+}$, on the other hand, one obtains a median value of $(-1.9)$ and a mean value of $(-2.1)$ with a standard deviation of $1.4$, so this quantity must be called uncertain.  There is plainly no consensus on the octupole moment but our results indicate that a positive value should be associated with the dressed-quark-core at a realistic $\Delta$ mass.

It is common to attempt to interpret a nonzero electric quadrupole moment with a deformation of the bound-state's charge distribution.  The robust indication from Table~\ref{tab:comparative} is $\hat{\mathpzc{Q}}_{\;\Delta^+}<0$: analyses with and without a meson-cloud agree on this sign.  If one supposes that for the $\Delta$-resonance the Fourier transform of a Breit-frame momentum-space form factor is, at least for small momentum transfers, a reasonable approximation to the configuration space charge distribution, then the negative value indicates an oblate deformation of the $\Delta^+$.  On the other hand, some would argue that it is the difference $(\hat{\mathpzc{Q}}_{\;\Delta^+}-\hat{\mathpzc{Q}}_{\;\Delta^+}^{\rm natural})$ which is the better measure, especially when drawing a correspondence with transverse densities in the infinite momentum frame \cite{Alexandrou:2009hs}.  However, there is no agreement on the sign of this difference and hence, as yet, no model-independent answer to the question of oblate versus prolate in that frame.

\section{Elastic form factors: \mbox{\boldmath $\Omega^-$}}
\label{sec:GammaOmegaOmega}
Using a $s$-quark current-mass of $0.17\,$GeV, we obtain a dressed-mass for the $s$-quark of $0.53\,$GeV.  The $\Omega^-$ baryon is then described by a Faddeev equation that involves a single axial-vector diquark correlation: $\{ss\}$, with an associated mass-scale of 1.42\,GeV.  (See Tables~\ref{tab:CQM}, \ref{tab:mesonproperties}.)  Since the $\Omega^{-}$ consists of three valence $s$-quarks, it can only decay via the weak interaction and is thus significantly more stable than other members of the baryon decuplet, with a lifetime $\tau_{\Omega^-} \sim 10^{13} \tau_{\Delta} \sim 10^{-3} \tau_{\pi^+}$.  It follows that the $\Omega^-$-baryon's electromagnetic properties must be more amenable to measurement and, indeed, its magnetic moment is gauged with some precision \cite{Beringer:1900zz}: $\mu_{\Omega^{-}}=-(2.02\pm0.05)\,{\rm \mu_{N}}$.  This makes the calculation of $\Omega^-$ electromagnetic properties considerably more interesting.
In order to compute the dressed-quark-core contributions to the $\Omega^-$ elastic form factors, one need only make obvious adjustments to the inputs and kinematics in the codes that produced the $\Delta$ elastic form factors.  

As remarked in Sec.\,\ref{secMesonCloud}, our symmetry-preserving treatment of the contact interaction produces an $\Omega^{-}$ with mass $1.76\,{\rm GeV}$.  This result may be compared with the empirical value \cite{Beringer:1900zz} $ 1.6725 \pm 0.0007\,$GeV: it is $\sim 0.1\,$GeV higher because we omit type-2 meson cloud corrections.  Owing to the OZI rule, those corrections are dominated by loops involving kaons and, since kaons are heavier than pions, the effect of meson loops is smaller for the $\Omega^-$ than for the $\Delta$-baryons.  It follows that the dressed-quark-core should be a good approximation for the $\Omega^-$.

\begin{table*}[t]
\begin{center}
\caption{\label{tab:comparative2}
Static electromagnetic properties of the $\Omega^-$-baryon: the row labelled ``CI'' lists our contact interaction results.
The experimental value for $G_{M1}(Q^{2}=0)$ is drawn from Ref.\,\protect\cite{Beringer:1900zz};
and the remaining rows report a representative selection of results from other calculations.  The $Q^2=0$ values associated with lattice simulations were obtained by fitting the available results and extrapolating (see Fig.\,\protect\ref{fig:FFomega}).
N.B.\ The symbol ``-'' in any location indicates that no result was reported for that quantity in the related reference.}
\begin{tabular}{lclll} \hline
\tstrut
Approach & & $G_{M1}(0)$ & $G_{E2}(0)$ & $G_{M3}(0)$\\
\hline
CI & & $-3.67$ & $+1.80$ & $-0.72$ \rule{-0.6em}{0ex} \rule{0em}{2.2ex} \\\hline
Exp & \cite{Beringer:1900zz} & $-3.61\pm0.09$ & - & - \rule{-0.6em}{0ex} \rule{0em}{2.2ex} \\\hline
Lattice-QCD & \cite{Alexandrou:2010jv}  & $-3.17$ & $+0.84$ & - \\
%
Faddeev equation & \cite{Nicmorus:2010sd} & $-3.75 \pm 0.17$ & $+1.18 \pm 0.16$ & $-0.22 \pm 0.03$\\
%
Faddeev equation & \cite{Maris:1999nt,Sanchis-Alepuz:2013iia} & $-2.42$ & $+0.54$ & $<-0.5$ \\
Covariant $\chi$PT & \cite{Geng:2009ys} & $-3.61$ & $+0.62$ & $+0.2\pm1.8$ \\
+ qlQCD & \cite{Boinepalli:2009sq} & & & \\
QCD-SR & \cite{Aliev:2009np} & - & $+7.19\pm2.16$ & $+3.3\pm1.1$ \\
$\chi$QSM & \cite{Ledwig:2008es} & $-4.09$ & - & - \\
General Param. Method  & \cite{Buchmann:2002xq,Buchmann:2008zza} & - & $+1.73$ & $+0.7$ \\
$1/N_{c}+m_s$-expansion & \cite{Luty:1994ub} & $-3.46$ & - & - \\
RQM & \cite{Schlumpf:1993rm} & $-4.20$ & - & - \\
HB$\chi$PT & \cite{Butler:1993ej} & $-3.46$ & $+0.65\pm0.36$ & - \\
nrCQM & \cite{Krivoruchenko:1991pm} & $-3.28$ & $+2.01$ & - \\ \hline
\end{tabular}
\end{center}
\end{table*}

Our predicted values for $\Omega^-$ static electromagnetic properties are presented in Table\,\ref{tab:comparative2}, along with results from an array of models.
Omitting our computations, one obtains a mean and median value of $\hat \mu_{\Omega^-} = -3.5 $ with a standard deviation of $0.5$.  Adding our result to the mix does not change these values.  Consequently, there is a degree of mutual agreement between the theoretical estimates and therewith confirmation of the experimental result.

Analogous to the quadrupole moment in Table~\ref{tab:comparative}, the case of $\hat{\mathpzc{Q}}_{\;\Omega^-}$ is completely different.  The QCD Sum Rule result \cite{Aliev:2009np} is plainly incompatible with all other computations.  If it is ignored, then one obtains a meaningful mean value of $\hat{\mathpzc{Q}}_{\;\Omega^-}=1.1 \pm 0.6$.

Regarding the octupole moment, $\hat{\mathpzc{O}}_{\;\Omega^-}$, once again, taken as a whole, the theory results do not even agree on the sign.  On the other hand, the Faddeev equation studies uniformly produce a negative value.  It will be observed that here, too, the theory predictions are not close to the so-called ``natural'' value, which is ``+1'' in this instance.
It should be noted that in the absence of experimental data, the results denoted $\chi$PT$+$qlQCD \cite{Geng:2009ys} and HB$\chi$PT \cite{Butler:1993ej} use lattice-QCD input in order to constrain low-energy constants.  Hence, they are not truly independent calculations and, naturally, when the lattice errors are large so, too, are those connected with these $\chi$PT estimates.

\begin{figure}[t]
\begin{center}
\begin{tabular}{cc}
\includegraphics[clip,width=0.46\linewidth]{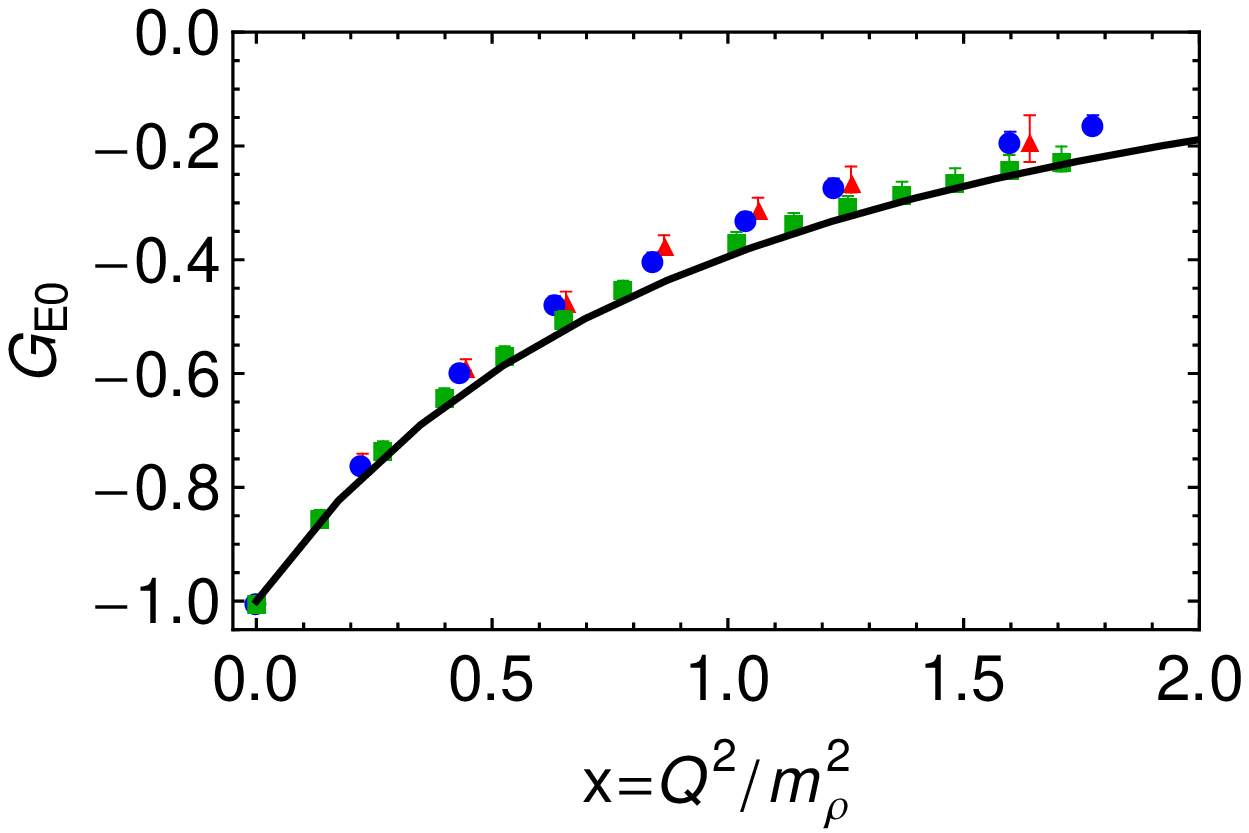}\vspace*{
-1ex } &
\includegraphics[clip,width=0.44\linewidth]{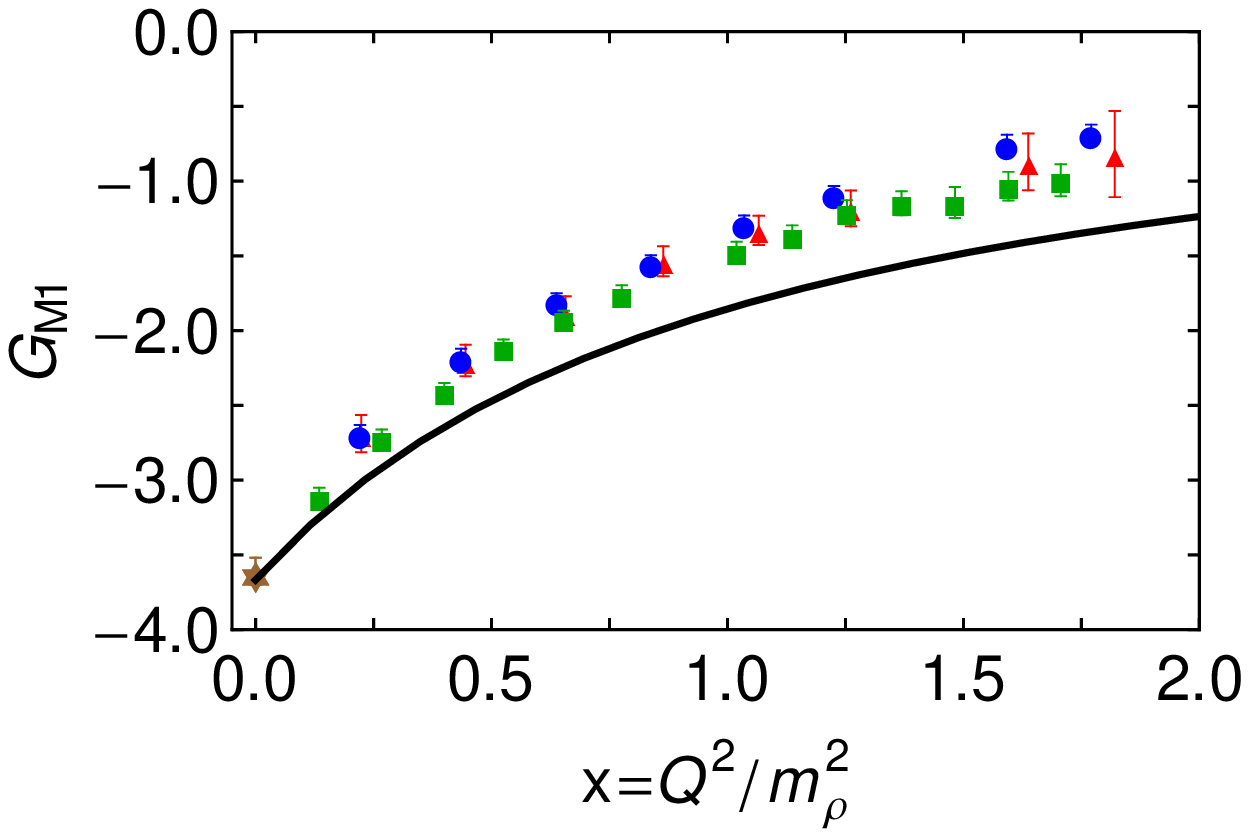}\vspace*{
-1ex} \\
\includegraphics[clip,width=0.44\linewidth]{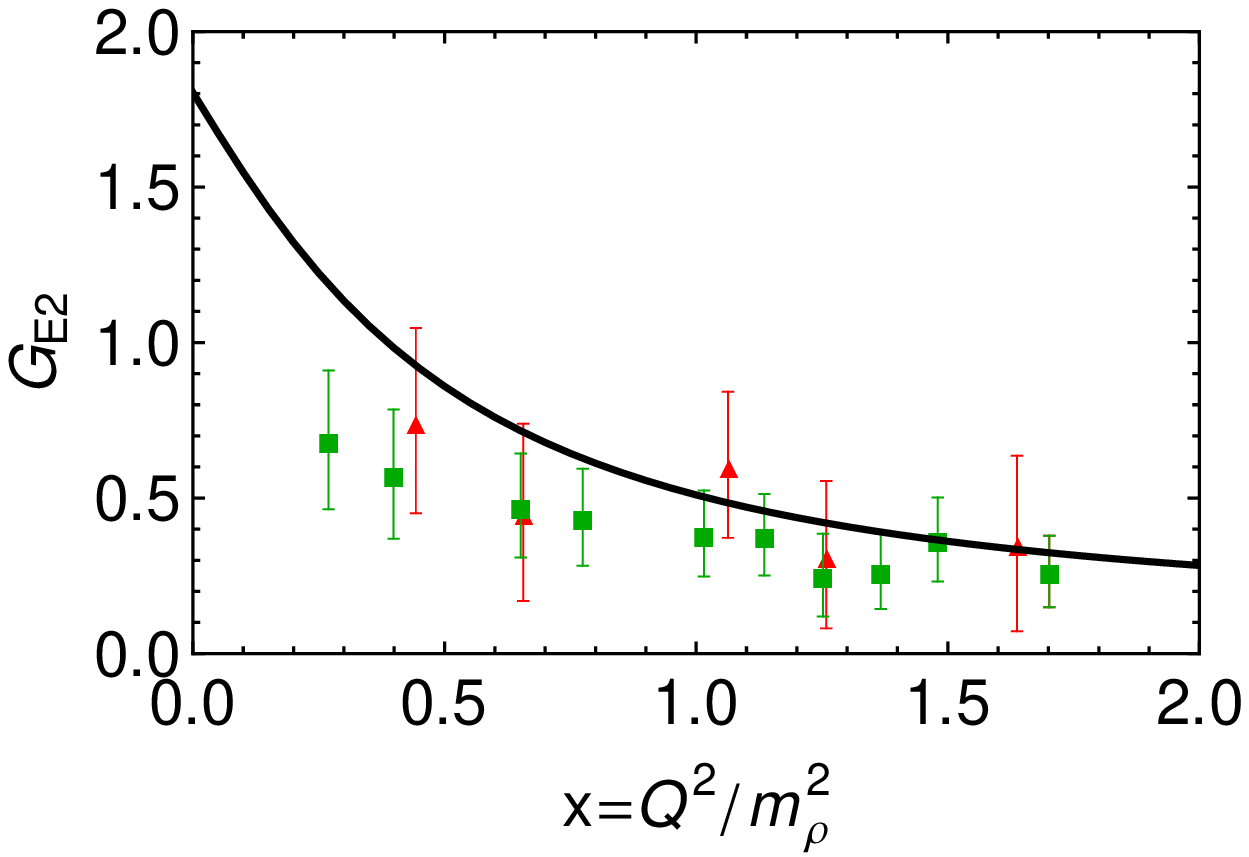}\vspace*{
-1ex} &
\includegraphics[clip,width=0.46\linewidth]{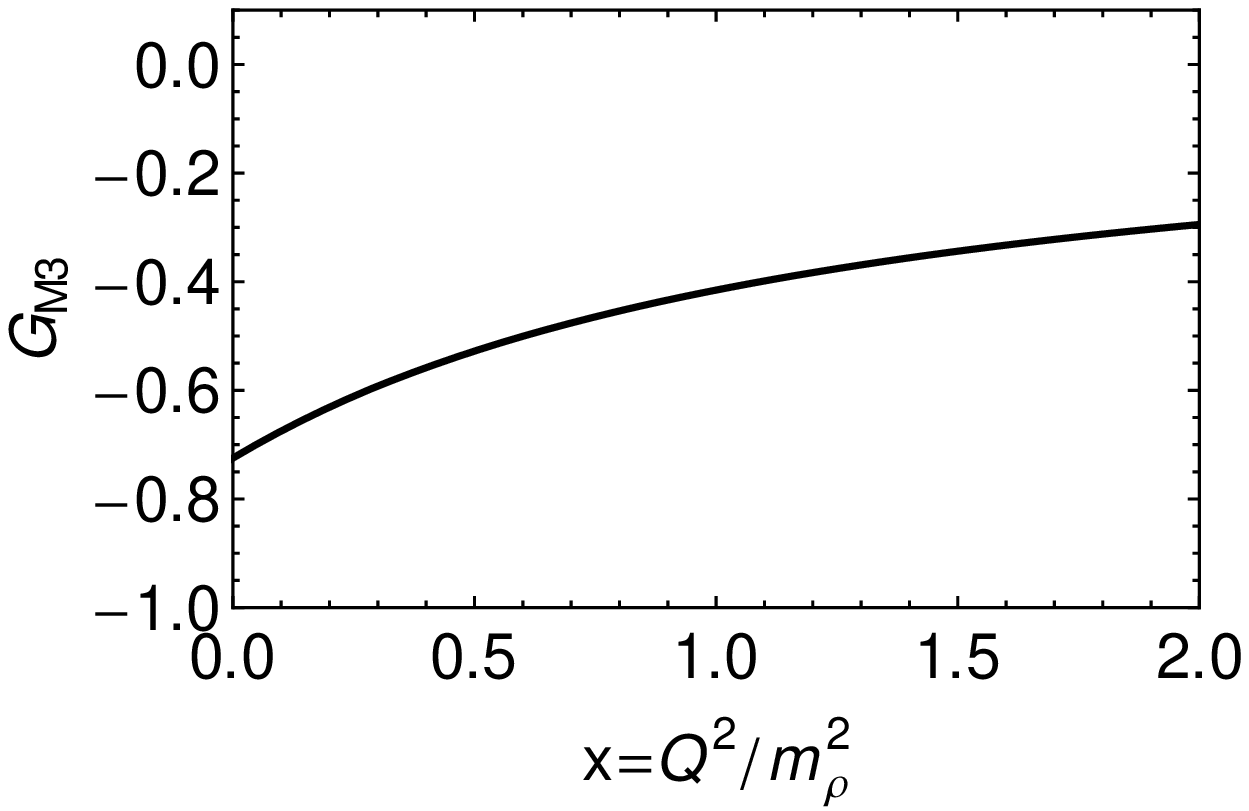}
\end{tabular}
\caption{\label{fig:FFomega} Contact interaction dressed-quark-core contributions to the $\Omega^{-}$ electromagnetic form factors: $G_{E0}$, $G_{M1}$, $G_{E2}$ and $G_{M3}$.
In the top-right panel, the experimental value of $G_{M1}(0)$ \protect\cite{Beringer:1900zz} is indicated by the star.
All other points are unquenched and hybrid results from numerical simulations of lattice-regularised QCD \protect\cite{Alexandrou:2010jv} at different pion mass (red triangles -- $m_{\pi}=297\,{\rm MeV}$, blue circles -- $m_{\pi}=330\,{\rm MeV}$, and green squares -- $m_{\pi}=353\,{\rm MeV}$).  The lattice results for $G_{M3}$ exhibit large statistical errors and are consistent with zero.}
\end{center}
\end{figure}

We depict our computed dressed-quark core contributions to the $\Omega^-$ elastic electromagnetic form factors in Fig.\,\ref{fig:FFomega}.  Since there is only one piece of experimental data; viz., $G_{M1}(0)=-3.61 \pm 0.09$, all but one of the points depicted are results obtained via numerical simulations of (unquenched) lattice-regularised QCD \cite{Alexandrou:2010jv}: no results are presented for the octupole form factor because the lattice points exhibit large statistical errors and are consistent with zero.  Allowing for the hardness of form factors obtained with our treatment of the contact interaction, the agreement between our results and the best available lattice simulations is striking: for practical purposes, they are indistinguishable.

One may obtain the electric radius of the $\Omega^{-}$-baryon's dressed-quark core from $G_{E0}(Q^2)$ in the top-left panel of Fig.\,\ref{fig:FFomega}:
\begin{equation}
\langle \; r_{E0_\Omega}^{2} \rangle =
6 \left. \frac{dG_{E0}}{dQ^{2}}\right|_{Q^{2}=0}
= \frac{7.1}{m_\rho^2} = \frac{16}{m_{\Delta}^2} = \frac{26}{m_{\Omega}^2}.
\end{equation}
A comparison with Eq.\,\eqref{DeltaRadius} shows that, not unexpectedly, this electric radius is smaller than that of the $\Delta^+$, which is comprised of lighter constituents.  On the other hand, the radius has decreased at a slower rate than that at which the bound-state's mass has grown.

\section{\mbox{\boldmath $\gamma^\ast N \to \Delta$} transition}
\label{sec:GammaNucleonDelta}
\subsection{General observations}
With the preceding material in hand we can now explore and explain the behaviour of the $\gamma^\ast N \to \Delta$ transition form factors, so let us begin by remarking that in analyses of baryon electromagnetic properties, using a quark model framework which implements a current that transforms according to the adjoint representation of
spin-flavor $SU(6)$, one finds simple relations between magnetic-transition matrix
elements~\cite{Beg:1964nm,Buchmann:2004ia}:
\begin{equation}
\label{eqBeg}
 \langle p | \mu | \Delta^+\rangle = -\langle n | \mu | \Delta^0\rangle\,,\quad
 \langle p | \mu | \Delta^+\rangle = - \surd 2 \langle n | \mu | n \rangle\,;
\end{equation}
i.e., the magnetic components of the $\gamma^\ast p \to \Delta^+$ and $\gamma^\ast n
\to \Delta^0$ are equal in magnitude and, moreover, simply proportional to the
neutron's magnetic form factor.  Furthermore, both the nucleon and $\Delta$ are
$S$-wave states (neither is deformed) and hence $G_{E}^{\ast} \equiv 0 \equiv
G_{C}^{\ast}$~\cite{Alexandrou:2012da}.

The second entry in Eq.\,\eqref{eqBeg} is consistent with pQCD \cite{Carlson:1985mm} in the following sense: both suggest that $G_{M}^{\ast p}(Q^2)$ should decay with $Q^2$ at the same rate as the neutron's magnetic form factor, which is dipole-like in QCD.  It is usually argued that this is not the case empirically
\cite{Aznauryan:2011ub,Aznauryan:2011qj}; and it is that claim which motivated us to undertake this entire study.

For baryons constituted as we have described herein, the $N\to \Delta$ transition current is represented by the three diagrams described in association with
Fig.~\ref{fig:Transitioncurrent}.  Plainly, with the presence of strong diquark
correlations, the assumption of $SU(6)$ symmetry for the associated state-vectors and
current is invalid.  Notably, too, since scalar diquarks are absent from the
$\Delta$, only axial-vector diquark correlations contribute in the left and middle
diagrams of Fig.~\ref{fig:Transitioncurrent}.

Each of the diagrams in Fig.~\ref{fig:Transitioncurrent} can be expressed like
Eq.~(\ref{eq:JTransition}), so that we may represent them as
\begin{equation}
\label{GammaTransition}
\Gamma_{\mu\lambda}^{m}(K,Q) = \Lambda_{+}(P_{f})R_{\lambda\alpha}(P_{f}) {\cal
J}_{\mu\alpha}^{m}(K,Q) \Lambda_{+}(P_{i})\,,
\end{equation}
where $m=1,2,\,$\ldots enumerates the diagrams, from left to right.

The left diagram describes a photon coupling directly to a dressed-quark with the
axial-vector diquark acting as a bystander.  If the initial-state is a proton, then
it contains two axial-vector diquark isospin states $(I,I_z) = (1,1)$, $(1,0)$, with
flavor content $\{uu\}$ and $\{ud\}$, respectively: in the isospin-symmetry limit,
they appear with relative weighting $(\sqrt{2/3})$:$(-\sqrt{1/3})$, which are just
the appropriate isospin-coupling Clebsch-Gordon coefficients. These axial-vector
diquarks also appear in the final-state $\Delta^{+}$ but with the orthogonal
weighting; i.e., $(\sqrt{1/3})$:$(\sqrt{2/3})$. For the process $\gamma^{\ast}p \to
\Delta^{+}$, Diagram~1 therefore represents a sum, which may be written
\begin{equation}
{\cal J}_{\mu\alpha}^{1 p} =
(\sqrt{2}/3) e_d {\mathcal I}_{\mu \alpha}^{1 \{uu\}}
-(\sqrt{2}/3) e_u {\mathcal I}_{\mu\alpha}^{1\{ud\}},
\label{eqJ1p}
\end{equation}
where we have extracted the isospin and charge factors associated with each
scattering.  Plainly, if the $\{uu\}$ diquark is a bystander, then the $d$-quark is
the active scatterer, and hence appears the factor $e_d=(-1/3)$.  Similarly,
$e_u=2/3$ appears with the $\{ud\}$ diquark bystander.

Now, having extracted the isospin and electric-charge factors, nothing remains to
distinguish between the $u$- and $d$-quarks in the isospin-symmetry limit.  Hence,
\begin{eqnarray}
\label{calIisospin}
&& {\mathcal I}_{\mu\alpha}^{1 \{uu\}}(K,Q) \equiv {\mathcal
I}_{\mu\alpha}^{1\{ud\}}(K,Q)
=: {\mathcal I}_{\mu\alpha}^{1 \{qq\}}(K,Q), \\
\label{J1pzero}
&\Rightarrow & {\cal J}_{\mu\alpha}^{1 p}(K,Q) = (-\sqrt{2}/3) {\mathcal
I}_{\mu\alpha}^{1 \{qq\}}(K,Q)\,.
\end{eqnarray}
It is known that diagrams with axial-vector diquark spectators do not contribute to
proton elastic form factors (Eq.~(C5) in Ref.~\cite{Wilson:2011aa}), so the
analogous contribution is absent from the proton's elastic form factors.  However,
this hard contribution is present in neutron elastic form factors.  In general, form
factors also receive a hard contribution from the two-loop diagrams omitted in
Fig.~\ref{fig:Transitioncurrent}.  In proton and neutron elastic magnetic form
factors, respectively, the large-$Q^2$ behaviour of this contribution matches that
produced by Diagram~1~\cite{Cloet:2008re}.

The remaining two diagrams in Fig.~\ref{fig:Transitioncurrent}; i.e., the central and
right images, describe a photon interacting with a composite object whose
electromagnetic radius is nonzero. (Indeed \cite{Roberts:2011wy}: $r_{1^+}
\gtrsim r_\pi$.)  They must therefore produce a softer contribution to the transition
form factors than anything obtained from the top diagram.

It follows from this discussion that the fall-off rate of $G_{M}^{\ast}(Q^2)$ in the
$\gamma^{\ast}p \to\Delta^{+}$ transition must match that of $G_M^n(Q^2)$. With
isospin symmetry, the first entry in Eq.\,\eqref{eqBeg} is valid, so the same is true of the $\gamma^{\ast}n \to\Delta^{0}$ magnetic form factor.  Note that these are statements about the dressed-quark-core contributions to the transitions. They will be valid empirically outside that domain upon which meson-cloud effects are important; i.e., for $Q^2\gtrsim 2\,$GeV$^2$~\cite{Sato:2000jf,JuliaDiaz:2006xt}.

\begin{figure}[t]
\centerline{\includegraphics[clip,width=0.50\linewidth]{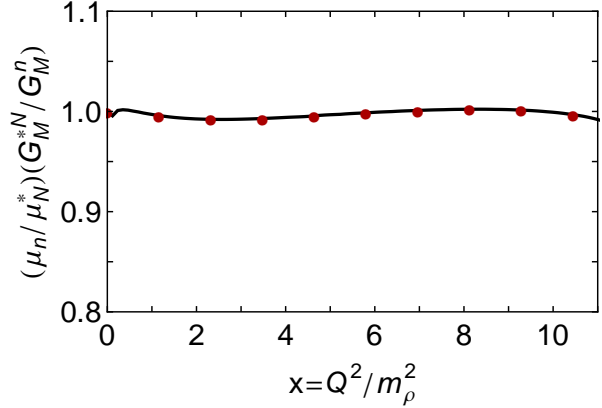}}
\caption{\label{figRatio} \emph{Solid curve} -- $\mu_n G_{M}^{\ast p}/\mu^\ast_p
G_{M}^{n}$ as a function of $x=Q^2/m_\rho^2$; and \emph{filled circles} -- $\mu_n
G_{M}^{\ast n}/\mu^\ast_n G_{M}^{n}$.
For $N=p,n$, $\mu^\ast_N = G_M^{\ast N}(Q^2=0)$; and $\mu_n = G_M^n(Q^2=0)$.  The
elastic form factor results are those presented in
Ref.\,\protect\cite{Wilson:2011aa}, so that the comparison is internally consistent.}
\end{figure}

\subsection{Quantitative illustrations and predictions}
In Fig.\,\ref{figRatio} we compare the momentum-dependence of the magnetic
$\gamma^{\ast}p \to\Delta^{+}$ and $\gamma^{\ast}n \to\Delta^{0}$ form factors with
$G_{M}^{n}(Q^{2})$. The prediction explained above is evident in a near identical
momentum dependence.

In connection with experiment, our formulation of a contact-interaction treatment of
the $N \to\Delta$ transition is quantitatively inadequate for two main reasons.
Namely, a contact interaction which produces Faddeev amplitudes that are independent
of relative momentum must underestimate the quark orbital angular momentum content of
the bound-state;\footnote{This is most serious for the nucleon, whose internal structure is far more complicated than that of the $\Delta$.} and the truncation which produces the momentum-independent amplitudes also suppresses the three two-loop diagrams in the current of Fig.~\ref{fig:Transitioncurrent}.  The detrimental effect can be illustrated via our computed values for the contributions to $G_M^\ast(0)$ that arise from the overlap axial-diquark($\Delta$)$\leftarrow$axial-diquark($N$) cf.\
axial-diquark($\Delta$)$\leftarrow$scalar-diquark($N$).  We find $0.85/0.18$, values
that may be compared with those in Table~3 of Ref.~\cite{Eichmann:2011aa}, which
uses momentum-dependent DSE kernels: $0.96/1.27$.  One may show algebraically that
the omitted two-loop diagrams facilitate a far greater contribution from
axial($\Delta$)-scalar($N$) mixing and the presence of additional orbital angular
momentum enhances both.

In recognition of both this defect and the general expectation that a comparison with
experiment should be sensible, we subsequently provide two sets of results.  Namely,
unameliorated predictions of the contact-interaction plus results obtained with two
corrections: we rescale the axial($\Delta$)-scalar($N$) diagram using the factor
\begin{equation}
\label{correction}
1+ \frac{g_{as/aa}}{1+Q^2/m_\rho^2}\,,
\end{equation}
with $g_{as/aa}=4.3$, so that its contribution to $G_M^{\ast p}(0)$ matches that of
the axial($\Delta$)-axial($N$) term; and incorporate the dressed-quark anomalous magnetic moment described in App.\,\ref{app:subsec:anomalousmagnetic}.

\begin{figure}[t]
\leftline{\includegraphics[clip,width=0.49\linewidth]{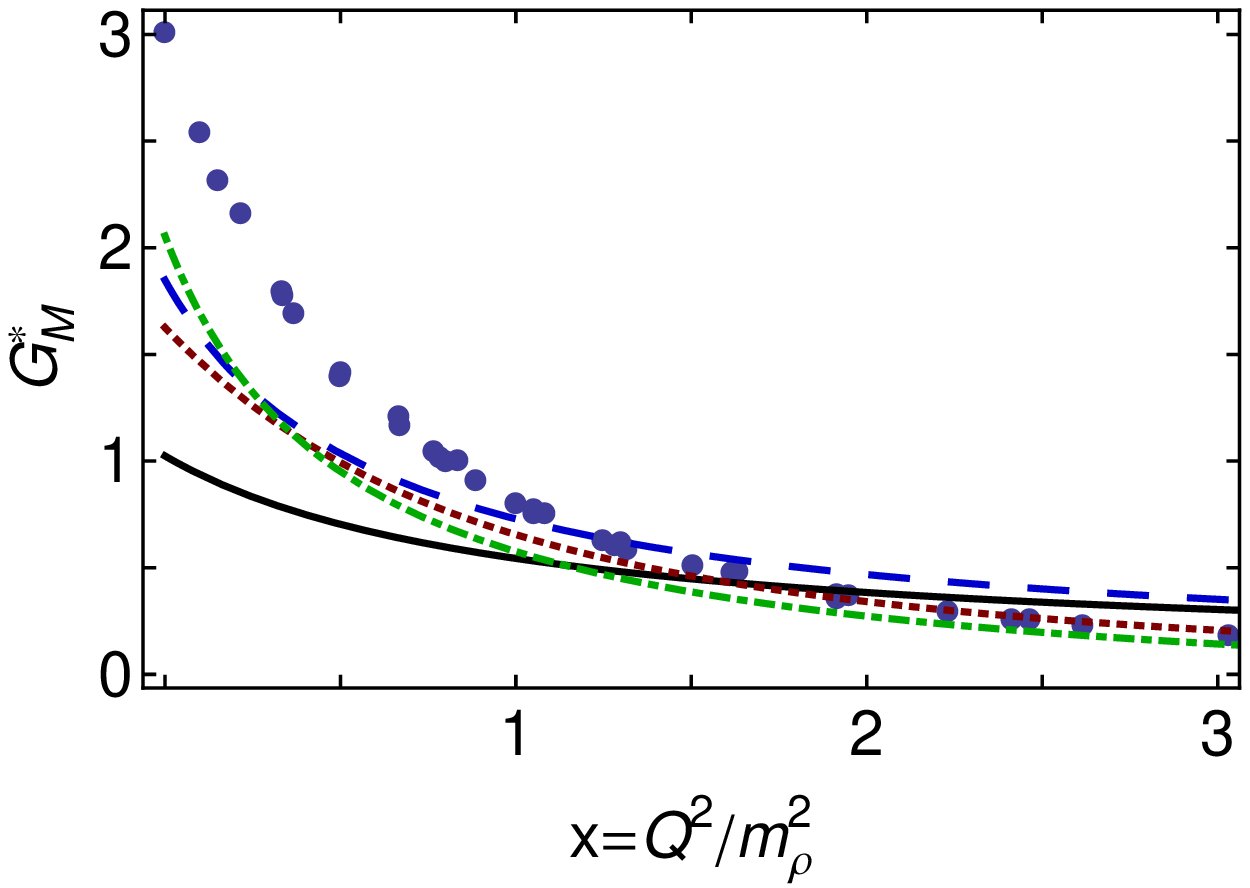}}
\vspace*{-43ex}

\rightline{\includegraphics[clip,width=0.49\linewidth]{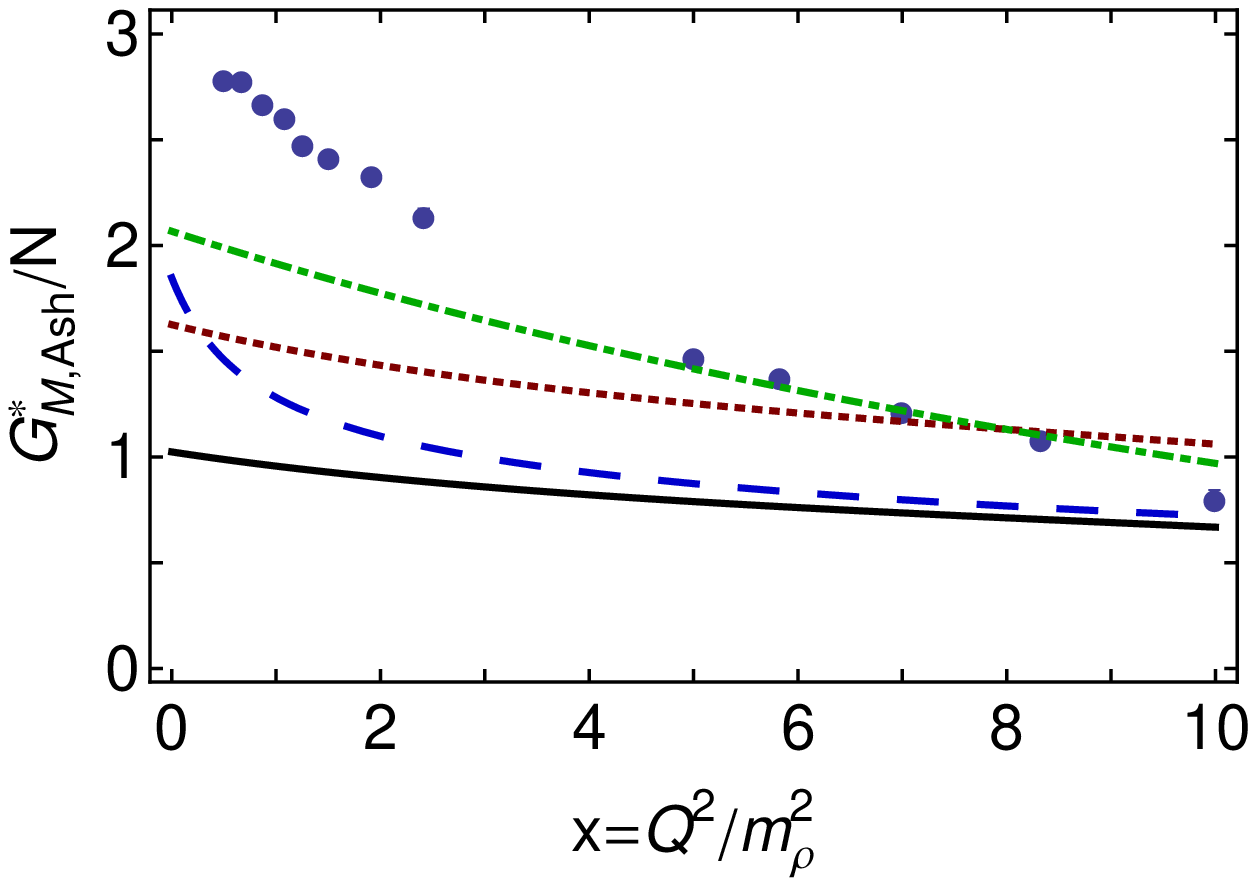}}
\caption{\label{figGMast}
\emph{Left panel}.  $G_{M}^{\ast}(Q^2)$:
contact-interaction result (solid curve);
ameliorated result (dashed curve), explained in connection with Eq.\,\protect\eqref{correction};
SL-model dressed-quark-core result \protect\cite{JuliaDiaz:2006xt} (dot-dashed curve);
and data from Refs.\,\protect\cite{Beringer:1900zz,Aznauryan:2009mx,Bartel:1968tw,Stein:1975yy,Sparveris:2004jn,Stave:2008aa}, whose errors are commensurate with the point size.
N.B.\ The contact interaction produces Faddeev amplitudes that are independent of relative momentum, hence $G_M^\ast(Q^2)$ is hard.  The dotted curve is an estimate of the result a realistic interaction would produce, obtained via multiplying $G_M^\ast(Q^2)$ by $G_M(Q^2)$-realistic$/G_M(Q^2)$-contact, where $G_M(Q^2)$-realistic is taken from Ref.\,\protect\cite{Cloet:2008re}.
\emph{Right panel}. $\mu_n G_{M,Ash}^{\ast}(Q^2)/N(Q^2)$: contact interaction (solid curve) and ameliorated result (dashed curve), both obtained with $N(Q^2)=G_M^n(Q^2)$.  (The dotted curve is the solid curve rescaled by $\mu_n$-realistic$/\mu_n$-contact.)  Also, empirical results \cite{Aznauryan:2009mx} for $G_{M,Ash}^{\ast}/N_D(Q^2)$, where $1/N_D(Q^2)=[1 + Q^2/\Lambda^2]^2$, $\Lambda=0.71\,$GeV, and SL-model's dressed-quark-core result for this ratio \protect\cite{JuliaDiaz:2006xt}.}
\end{figure}

The left panel of Fig.\,\ref{figGMast} displays the $\gamma^\ast p \to \Delta^+$ magnetic transition form factor.  (With $\tilde \mu_{N\Delta}^\ast:= (\sqrt{m_\Delta/m_N}) G_M^{\ast N}(0)$, we have a direct result of $\tilde \mu_{N\Delta}^\ast=1.13$ and an ameliorated value of $\tilde\mu_{N\Delta}^\ast=2.04$.)  Both computed curves are consistent with data for $x\gtrsim 2$ but, corrected or not, they are in marked disagreement at infrared momenta.  This is explained by the similarity between the ameliorated result (dashed curve) and the ``bare'' or dressed-quark-core result determined using the Sato-Lee (SL) dynamical meson-exchange model (dotted curve) \cite{JuliaDiaz:2006xt}.  The SL result supports a view that the discrepancy results from the omission of meson-cloud effects in the leading-order (rainbow-ladder) truncation of QCD's DSEs.

In contrast to the left panel of Fig.~\ref{figGMast}, presentations of experimental
data typically use the Ash form factor~\cite{Ash1967165}
\begin{equation}
\label{DefineAsh}
G_{M,Ash}^{\ast}(Q^2)= G_M^{\ast}(Q^2)/[1+Q^2/t_+ ]^{1/2}.
\end{equation}
This comparison is depicted in Fig.~\ref{figGMast}, right panel.  (Our dressed-quark
core result is quantitatively similar to Fig.~3 of Ref.~\cite{Aznauryan:2012ec}).
Plainly, $G_{M,Ash}^{\ast}(Q^2)$ falls faster than a dipole.  Historically, many have
viewed this as a conundrum.  However, as observed previously \cite{Carlson:1985mm}, elucidated elsewhere \cite{Segovia:2013rca}, and reiterated herein, there is no sound reason to expect $G_{M,Ash}^{\ast}(Q^2)/G_M^n(Q^2) \approx\,$constant.  Instead, the Jones-Scadron form factor should exhibit $G_{M}^{\ast}(Q^2)/G_M^n(Q^2) \approx\,$constant.  The empirical Ash form factor falls rapidly for two reasons. First: meson-cloud effects provide more than $30\%$ of the form factor for $x \lesssim 2$; these contributions are very soft; and hence they disappear rapidly.  Second: the additional kinematic factor $\sim 1/\sqrt{Q^2}$ in Eq.~\eqref{DefineAsh} provides material damping for $x\gtrsim 4$.

The dotted curves in Fig.\,\ref{figGMast} depict crude estimates of the behaviour to be expected of the associated form factors when they are computed with propagators and currents that exhibit QCD-like momentum-dependence, as employed, e.g., in Ref.\,\cite{Cloet:2008re}.  These curves hint that the leading-order treatment of a realistic interaction is capable of explaining the data, once meson cloud effects are removed.

\begin{figure}[t]
\leftline{\includegraphics[clip,width=0.49\linewidth]{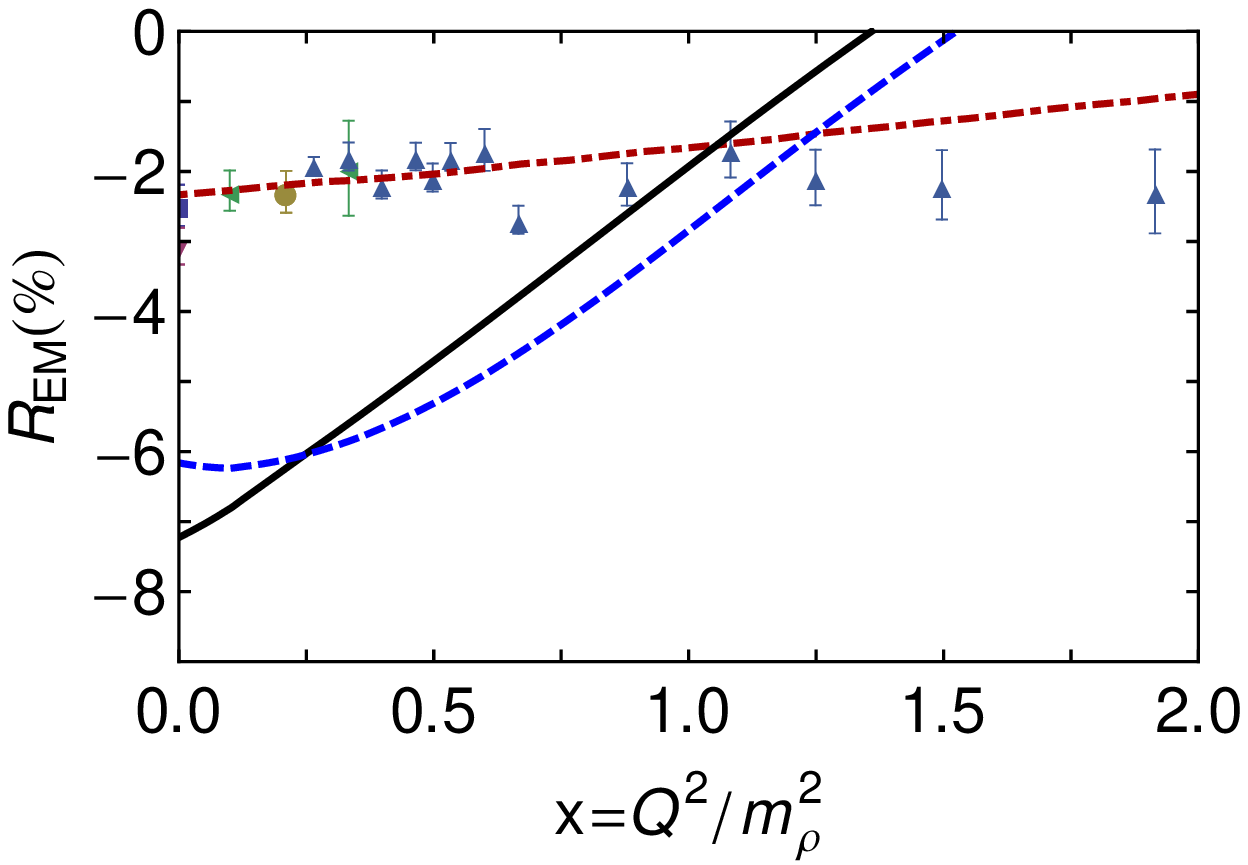}}
\vspace*{-41.5ex}

\rightline{\includegraphics[clip,width=0.50\linewidth]{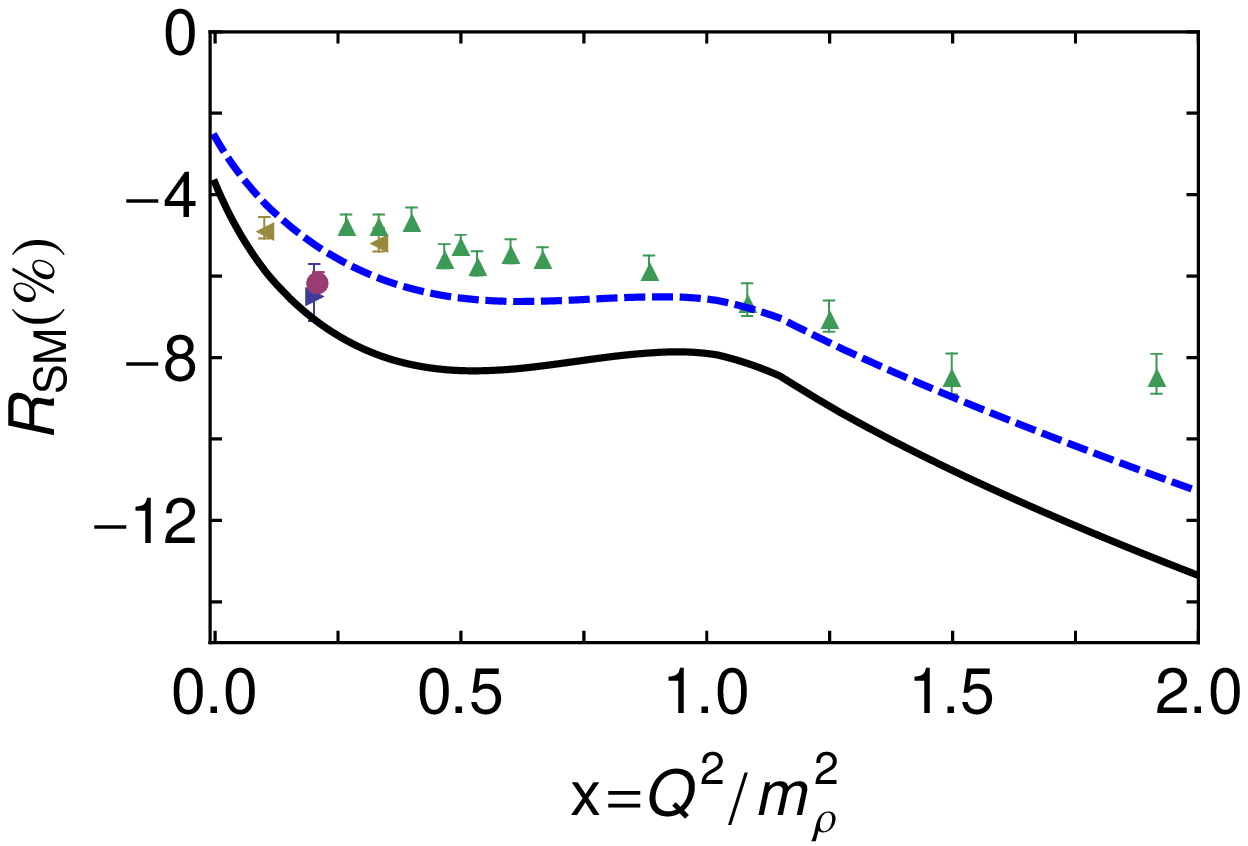}}
\caption{\label{figREMSM} Ratios in Eq.\,\protect\eqref{eqREMSM}.  Both panels:
\emph{solid curve} -- contact-interaction result; \emph{dashed curve} -- ameliorated
result, discussed in connection with Eq.\,\protect\eqref{correction};
and data
\protect\cite{Sparveris:2004jn,Stave:2008aa,Aznauryan:2009mx,Beck:1999ge,
Pospischil:2000ad,Blanpied:2001ae}.
The \emph{dash-dot curve} in the left panel is representative of the computation in
Ref.\,\protect\cite{Eichmann:2011aa}.
(N.B.\ $G_E^\ast$, $G_C^\ast$ are small, so EBAC could not reliably separate
meson-cloud and dressed-quark core contributions to these ratios.)}
\end{figure}

In Fig.\,\ref{figREMSM} we depict the ratios
\begin{equation}
\label{eqREMSM}
R_{\rm EM} = -\frac{G_E^{\ast}}{G_M^{\ast}}, \quad
R_{\rm SM} = - \frac{|\vec{Q}|}{2 m_\Delta} \frac{G_C^{\ast}}{G_M^{\ast}},
\end{equation}
which are commonly read as measures of deformation in one or both of the hadrons
involved because they are zero in $SU(6)$-symmetric constituent-quark models.
However, the ratios also measure the way in which such deformation influences the
structure of the transition current.

Our results show that even a contact-interaction produces correlations between
dressed-quarks within Faddeev wave-functions and related features in the current that
are comparable in size with those observed empirically.  They are actually too large
if axial($\Delta$)-axial($p$) contributions to the transition significantly outweigh
those from axial($\Delta$)-scalar($p$) processes.  This is highlighted effectively by the dash-dot curve in the left panel.  That result \cite{Eichmann:2011aa}, obtained in the same DSE truncation but with a QCD-motivated momentum-dependent interaction \cite{Maris:1999nt}, produces Faddeev amplitudes with a richer quark orbital angular momentum structure.  The left panel emphasises, therefore, that $R_{\rm EM}$ is a particularly sensitive measure of orbital angular momentum correlations, both within the hadrons involved and in the excitation current.  The simpler Coulomb quadrupole produces a ratio, $R_{\rm SM}$, that is more robust.

\begin{figure}[t]
\centerline{\includegraphics[clip,width=0.5\linewidth]{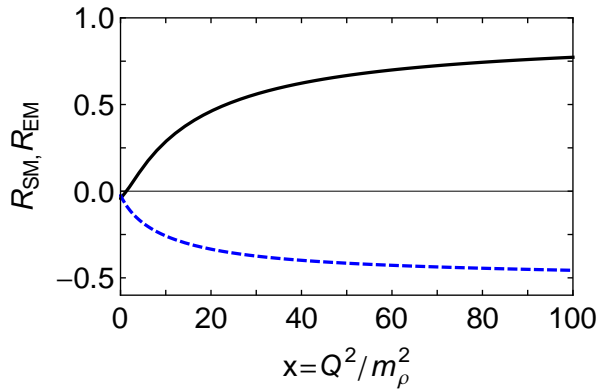}}
\caption{\label{UVREM}
$R_{\rm EM}$ (solid curve) and $R_{\rm SM}$ (dashed curve) in
Eq.\,\protect\eqref{eqREMSM}, computed using the ameliorated contact interaction,
discussed in connection with Eq.\,\protect\eqref{correction}.}
\end{figure}

Notwithstanding that the asymptotic power-law dependence of our computed form factors
is harder than that in QCD, one may readily show that the helicity conservation
arguments in Ref.\,\cite{Carlson:1985mm} should apply equally to an
internally-consistent symmetry-preserving treatment of a contact interaction.  As a
consequence, we have
\begin{equation}
\label{eqUVREMSM}
R_{EM} \stackrel{Q^2\to\infty}{=} 1 \,,\quad
R_{SM} \stackrel{Q^2\to\infty}{=} \,\mbox{\rm constant}\,.
\end{equation}
The validity of Eqs.~\eqref{eqUVREMSM} may be read from Fig.~\ref{UVREM}. On one
hand, it is plain that truly asymptotic $Q^2$ is required before the predictions are
realised. On the other hand, they \emph{are} apparent. Importantly, $G_E^\ast(Q^2)$
does possess a zero (at an empirically accessible momentum) and thereafter $R_{\rm
EM}\to 1$. Moreover, $R_{\rm SM}\to\,$constant.
(N.B.\ The curve we display contains the $\ln^2 Q^2$-growth expected in QCD
\cite{Idilbi:2003wj} but it is not a prominent feature). Since it is relative damping
associated with helicity flips that yields Eqs.~\eqref{eqUVREMSM}, with the
$Q^2$-dependence of the leading amplitude being less important, it is plausible that
the pattern evident herein is also that to be anticipated in QCD.

\section{Epilogue}
\label{sec:summary}
Using a symmetry-preserving treatment of a vector$\,\otimes\,$vector contact interaction, we have presented predictions for the dressed-quark-core contributions to the elastic form factors of the $\Delta(1232)$ and $\Omega^-$ baryons, and the $\gamma N \to \Delta(1232)$ transition form factors.  With judicious consideration given to the fact that our usage of the contact interaction typically produces hard form factors, this simple framework produces results that are practically indistinguishable from the best otherwise available.  Our analysis highlights that the key to describing a wide range of baryon features and unifying them with the properties of mesons is a faithful expression of dynamical chiral symmetry breaking in the hadron bound-state problem.

It is worth recapitulating here upon some of the important features of our Poincar\'e covariant analysis.  The basic element is the dressed-quark propagator, obtained as the dynamical solution to a gap equation.  The coupling between dressed-quarks and the photon is determined by an inhomogeneous Bethe-Salpeter equation, whose structure is symmetry-consistent with the gap equation.  That coupling is $Q^2$-dependent, where $Q$ is the photon momentum entering the vertex.  Within the baryon, dressed-quarks form diquark correlations, whose strength is described by the solution of homogeneous Bethe-Salpeter equations.  The structure of those equations is, again, symmetry consistent with the gap equation.  Moreover, the diquark correlations themselves possess nontrivial electromagnetic structure, which is calculated in a manner analogous to the computation of meson form factors.  A Faddeev equation describes how the dressed-quarks and -diquarks combine to form a baryon with nontrivial spin-flavour correlations.  Finally, the electromagnetic coupling of this baryon is determined by a current whose structure is completely determined by that of the elements from which it is composed.  It is crucial throughout that the diquark correlations are treated as dynamical.  It is unsound to consider them as inert and structureless.

It is appropriate now to highlight some of our findings.
The analysis shows that the $\Delta$ elastic form factors are very sensitive to $m_\Delta$.  Therefore, given that the parameters which define extant simulations of lattice-regularised QCD produce $\Delta$-resonance masses that are very large, the form factors obtained therewith should be interpreted carefully.
%
Moreover, at a realistic value of $m_\Delta$, the dressed-quark-core contribution to the octupole form factor is positive definite.
Considering the $\Delta$-baryon's quadrupole moment, the conflict between theoretical predictions indicates that it is currently impossible to judge whether the deformation of the $\Delta$ in the infinite momentum frame is oblate or prolate.  All predictions agree, however, that the quadrupole moment is negative.
Owing to the nature and relative stability of the $\Omega^-$, the dressed-quark-core provides a good approximation to the electromagnetic structure of this state, and there is agreement between our results and those available from lattice-regularised QCD.

Turning to the $N \to \Delta$ transition, we explained and illustrated that the Ash form factor connected with the $\gamma^\ast N \to \Delta$ transition should fall faster than the neutron's magnetic form factor, which is a dipole in QCD.  
In addition, we showed that the quadrupole ratios associated with this transition are a sensitive measure of quark orbital angular momentum within the nucleon and $\Delta$.  In Faddeev equation studies of baryons, this is commonly associated with the presence of strong diquark correlations.
Finally, direct calculation revealed that predictions for the asymptotic behaviour of these quadrupole ratios, which follow from considerations associated with helicity conservation, are valid, although only at truly large momentum transfers.

A natural next step is to replicate the calculations herein using a more realistic interaction \emph{plus} a numerical algorithm that enables one to compute the form factors to arbitrarily large momentum transfers, in particular those relating to the $N\to \Delta$ transition.  It is critical to overcome the weaknesses of earlier such computations, which employed simple algorithms that failed before reaching the interesting domain of momenta; viz., the domain whereupon QCD itself can be tested.  The approaches in Refs.\,\cite{Chang:2013nia,Cloet:2008re} are obvious candidates.  Progress in this direction will enable verification and improvement of both the predictions we have made and insights we have drawn using the contact interaction.

\section*{Acknowledgments}
We thank R.~Gothe, T.-S.\,H.~Lee and V.\,I.~Mokeev for helpful discussions.
Chen Chen acknowledges the support of the China Scholarship Council (file no.\ 2010634019); and CDR acknowledges support from an Helmholtz Association \emph{International Fellow Award}.
This work was otherwise supported by: U.\,S.\,Department of Energy, Office of Nuclear
Physics, contract no.~DE-AC02-06CH11357; and For\-schungs\-zentrum J\"ulich GmbH.

\appendix
\setcounter{equation}{0}
\renewcommand{\theequation}{\Alph{section}.\arabic{equation}}

\section{Contact interaction and Faddeev equations}
\label{app:sec:ContactInteractionModel}
The computations we perform require knowledge of the nucleon, $\Delta$ and $\Omega$ Faddeev amplitudes and in this appendix we explain how those amplitudes may be calculated using a symmetry-preserving regularisation of a vector$\,\otimes\,$vector contact interaction.  That interaction is specified by two parameters: an interaction strength $\alpha_{\rm IR}=0.93\pi$ and a momentum-space range $\Lambda_{\rm uv}=0.905\,$GeV, which were fixed elsewhere \cite{Roberts:2011wy}.  The $u=d$- and $s$-quark current-masses were chosen in order to obtain agreement with empirical values for the pion and kaon masses, since corrections to the rainbow-ladder truncation are small in the flavour-nonsinglet pseudoscalar meson channels \cite{Bender:2002as,Bhagwat:2004hn,Watson:2004kd,Chang:2009zb,Chang:2011ei}.

\subsection{Contact interaction}
\label{app:subsec:GAP}
For a given flavour of quark, associated with a current-quark mass $m_f$, the dressed-quark propagator in Fig.\,\ref{fig:FaddeevEquation} is obtained from the gap equation
\begin{equation}
 S_f^{-1}(p) =  i \gamma \cdot p + m_f +  \frac{16\pi}{3}\frac{\alpha_{\rm IR}}{m_G^2} \int\!\frac{d^4 q}{(2\pi)^4} \,
\gamma_{\mu} \, S_f(q) \, \gamma_{\mu}\,.
\label{gap-1}
\end{equation}
(Our Euclidean metric conventions are specified in Ref.\,\cite{Roberts:2011cf}, App.~A.)
In order to arrive at this expression from the general form we have written
\begin{equation}
g^{2} D_{\mu\nu}(p-q) = \delta_{\mu\nu} \frac{4\pi\alpha_{\rm IR}}{m_{\rm G}^{2}},
\label{eq:qqInteraction}
\end{equation}
where $m_G=0.8\,$GeV is a gluon mass-scale typical of the one-loop renormalisation-group-improved interaction detailed in Ref.\,\cite{Qin:2011dd}, and the fitted parameter $\alpha_{\rm IR} = 0.93 \pi$ is commensurate with contemporary estimates of the zero-momentum value of a running-coupling in QCD \cite{Aguilar:2009nf,Oliveira:2010xc,Aguilar:2010gm,Boucaud:2010gr,Pennington:2011xs,Wilson:2012em}.  As indicated above, we embed Eq.\,\eqref{eq:qqInteraction} in a rainbow-ladder truncation of the DSEs, which is the leading-order in the most widely used, global-symmetry-preserving truncation scheme \cite{Munczek:1994zz,Bender:1996bb}.  This means we use
\begin{equation}
\Gamma^{a}_{\nu}(q,p)=\frac{\lambda^{a}}{2} \gamma_{\nu}
\label{eq:VertexRL}
\end{equation}
in the gap equation and also in the subsequent construction of the Bethe-Salpeter kernels.

Equation~\eqref{gap-1} possesses a quadratic divergence, even in the chiral limit.  When the divergence is regularised in a Poincar\'e covariant manner, the solution is
\begin{equation}
\label{genS}
S_f(p)^{-1} = i \gamma\cdot p + M_f\,,
\end{equation}
where $M_f$ is momentum-independent and determined by
\begin{equation}
M_f = m_f + M_f\frac{4\alpha_{\rm IR}}{3\pi m_G^2} \int_0^\infty \!ds \, s\, \frac{1}{s+M_f^2}\,.
\label{eq:MGAP}
\end{equation}

Our regularisation procedure follows Ref.\,\cite{Ebert:1996vx}; i.e., we write
\begin{eqnarray}
\frac{1}{s+M^2} & = & \int_0^\infty d\tau\,{\rm e}^{-\tau (s+M^2)}  \rightarrow  \int_{\tau_{\rm uv}^2}^{\tau_{\rm ir}^2} d\tau\,{\rm e}^{-\tau (s+M^2)}
%
 =
\frac{{\rm e}^{- (s+M^2)\tau_{\rm uv}^2}-{\rm e}^{-(s+M^2) \tau_{\rm ir}^2}}{s+M^2} \,, \label{eq:Regularization}
\end{eqnarray}
where $\tau_{\rm ir,uv}$ are, respectively, infrared and ultraviolet regulators.  It is evident from the rightmost expression in Eq.\,(\ref{eq:Regularization}) that a finite value of $\tau_{\rm ir}=:1/\Lambda_{\rm ir}$ implements confinement by ensuring the absence of quark production thresholds \cite{Roberts:2012sv,Chang:2011vu}.  Since Eq.\,(\ref{eq:qqInteraction}) does not define a renormalisable theory, then $\Lambda_{\rm uv}:=1/\tau_{\rm uv}$ cannot be removed but instead plays a dynamical role, setting the scale of all dimensioned quantities.  Using Eq.\,\eqref{eq:Regularization}, the gap equation becomes
\begin{equation}
M = m + M \frac{4\alpha_{\rm IR}}{3\pi m_{\rm G}^{2}} \, {\cal C}^{\rm iu}(M^{2}),
\label{eq:SolutionGapEquation}
\end{equation}
where ${\cal C}^{\rm iu}(\sigma)/\sigma = \overline{\cal C}^{\rm iu}(\sigma) = \Gamma(-1,\sigma \tau_{\rm uv}^2) - \Gamma(-1,\sigma \tau_{\rm ir}^2)$, with $\Gamma(\alpha,y)$ being the incomplete gamma-function.

\begin{table}[t]
\caption{\label{tab:CQM}
Computed dressed-quark properties, required as input for the Bethe-Salpeter and Faddeev equations, and computed values for in-hadron condensates \protect\cite{Brodsky:2010xf,Chang:2011mu,Brodsky:2012ku}.  All results obtained with $\alpha_{\rm IR} =0.93 \pi$ and (in GeV) $\Lambda_{\rm ir} = 0.24\,$, $\Lambda_{\rm uv}=0.905$.  N.B.\ These parameters take the values determined in the spectrum calculation of Ref.\,\protect\cite{Roberts:2011cf}, which produces $m_\rho=0.928\,$GeV; and we assume isospin symmetry throughout.
(All dimensioned quantities are listed in GeV.)}
\begin{center}
\begin{tabular*}
{\hsize}
{
c@{\extracolsep{0ptplus1fil}}
c@{\extracolsep{0ptplus1fil}}
c@{\extracolsep{0ptplus1fil}}
c@{\extracolsep{0ptplus1fil}}
c@{\extracolsep{0ptplus1fil}}
c@{\extracolsep{0ptplus1fil}}
c@{\extracolsep{0ptplus1fil}}
c@{\extracolsep{0ptplus1fil}}
c@{\extracolsep{0ptplus1fil}}
c@{\extracolsep{0ptplus1fil}}}\hline
$m_u$ & $m_s$ & $m_s/m_u$ & $M_0$ &   $M_u$ & $M_s$ & $M_s/M_u$ & $\kappa_0^{1/3}$ & $\kappa_\pi^{1/3}$ & $\kappa_K^{1/3}$ \\\hline
0.007  & 0.17 & 24.3 & 0.36 & 0.37 & 0.53 & 1.43  & 0.241 & 0.243 & 0.246
\\\hline
\end{tabular*}
\end{center}
\end{table}

In Table~\ref{tab:CQM} we report values of $u$- and $s$-quark properties, computed from Eq.\,\eqref{eq:SolutionGapEquation}, that we use in bound-state calculations: the input ratio $m_s/\bar m$, where $\bar m = (m_u+m_d)/2$, is consistent with contemporary estimates \cite{Leutwyler:2009jg}.
N.B.\ It is a feature of Eq.\,\eqref{eq:SolutionGapEquation} that in the chiral limit, $m_f=m_0=0$, a nonzero solution for $M_0:= \lim_{m_f\to 0} M_f$ is obtained so long as $\alpha_{\rm IR}$ exceeds a minimum value.  With $\Lambda_{\rm ir,uv}$ as specified in the Table, that value is $\alpha_{\rm IR}^c\approx 0.4\pi$.  In the Table we also include chiral-limit and physical-mass values of the in-pseudoscalar-meson condensate \cite{Brodsky:2010xf,Chang:2011mu,Brodsky:2012ku}, $\kappa_H$, which is the dynamically generated mass-scale that characterises DCSB.  A growth with current-quark mass is anticipated in QCD \cite{Maris:1997tm,Roberts:2011ea}.

\subsection{Ward-Takahashi identities}
\label{app:subsec:WTIs}
In any study of hadron observables it is crucial to ensure that vector and axial-vector Ward-Green-Takahashi identities are satisfied.  The $m=0$ axial-vector identity states ($k_{+}=k+P$)
\begin{equation}
P_{\mu} \Gamma_{5\mu}(k_{+},k) = S^{-1}(k_{+}) i \gamma_{5} + i \gamma_{5} S^{-1}(k),
\label{eq:avWTI}
\end{equation}
where $\Gamma_{5\mu}(k_{+},k)$ is the axial-vector vertex, which is determined by
\begin{equation}
\Gamma_{5\mu}(k_{+},k) =\gamma_{5}\gamma_{\mu} - \frac{}{}\frac{16\pi\alpha_{\rm
IR}}{3m_{\rm G}^{2}} \int\frac{d^{4}q}{(2\pi)^4} \, \gamma_{\alpha}
\chi_{5\mu}(q_{+},q) \gamma_{\alpha},
\label{aveqn}
\end{equation}
with $\chi_{5\mu}(q_{+},q) = S(q+P) \Gamma_{5\mu} S(q)$.  One must implement a regularisation that maintains Eq.~(\ref{eq:avWTI}).  That amounts
to eliminating the quadratic and logarithmic divergences.  Their absence is just the
circumstance under which a shift in integration variables is permitted, an operation
required in order to prove Eq.~(\ref{eq:avWTI}).   It is guaranteed so long as one implements the constraint \cite{GutierrezGuerrero:2010md,Roberts:2010rn}
\begin{equation}
0 = \int_{0}^{1} d\alpha \,
\left[ {\cal C}^{\rm iu}(\omega(M^{2},\alpha,P^{2})) + {\cal C}^{\rm
iu}_{1}(\omega(M^{2},\alpha,P^{2}))\right],
\label{eq:avWTIP}
\end{equation}
with
\begin{equation}
\omega(M^{2},\alpha,P^{2}) = M^{2} + \alpha(1-\alpha) P^{2},
\end{equation}
and
\begin{equation}
{\cal C}^{\rm iu}_{1}(z) = - z (d/dz){\cal C}^{\rm iu}(z)
=z \left[ \Gamma(0,M^{2}\tau_{\rm uv}^{2})-\Gamma(0,M^{2}\tau_{\rm ir}^{2})\right].
\label{eq:C1}
\end{equation}

The vector Ward-Takahashi identity
\begin{equation}
P_{\mu} i\Gamma^{\gamma}_{\mu}(k_{+},k) = S^{-1}(k_{+}) - S^{-1}(k),
\label{eq:vWTI}
\end{equation}
wherein $\Gamma^{\gamma}_{\mu}$ is the dressed-quark-photon vertex, is crucial for a
sensible study of a bound-state's electromagnetic form factors~\cite{Roberts:1994hh}.
The vertex must be dressed at a level consistent with the truncation used to compute
the bound-state's Bethe-Salpeter or Faddeev amplitude. Herein this means the vertex
should be determined from the following inhomogeneous Bethe-Salpeter equation
\begin{equation}
\Gamma_{\mu}(Q) = \gamma_{\mu} - \frac{16\pi\alpha_{\rm IR}}{3m_{\rm G}^{2}}
\int \frac{d^{4}q}{(2\pi)^{4}} \, \gamma_{\alpha} \chi_{\mu}(q_{+},q)
\gamma_{\alpha},
\label{eq:GammaQeq}
\end{equation}
where $\chi_{\mu}(q_{+},q) = S(q+P) \Gamma_{\mu} S(q)$.  Owing to the
momentum-independent nature of the interaction kernel, the general form of the
solution is
\begin{equation}
\Gamma_{\mu}(Q) = \gamma^{\perp}_{\mu} P_{T}(Q^{2}) + \gamma_{\mu}^{\parallel} P_{L}(Q^{2})\,,
\label{eq:GammaQ}
\end{equation}
where $Q \cdot \gamma^{\perp}=0$ and $\gamma_{\mu}^{\parallel} + \gamma_\mu^{\perp}=\gamma_\mu$.

Inserting Eq.~(\ref{eq:GammaQ}) into Eq.~(\ref{eq:GammaQeq}), one readily obtains
\begin{equation}
P_{L}(Q^{2})= 1,
\label{eq:PL0}
\end{equation}
owing to corollaries of Eq.~(\ref{eq:avWTI}). Using these same identities, one finds
\cite{Roberts:2011wy}
\begin{equation}
P_{T}(Q^{2})= \frac{1}{1+K_{\gamma}(Q^{2})},
\label{eq:PTQ2}
\end{equation}
with $(\bar{\cal C}_{1}^{\rm iu}(z) = {\cal C}_{1}^{\rm iu}(z)/z)$
\begin{equation}
K_{\gamma}(Q^{2}) = \frac{4\alpha_{\rm IR}}{3\pi m_{\rm G}^{2}} \int_{0}^{1}
d\alpha\, \alpha(1-\alpha) Q^{2} \,  \bar{\cal C}^{\rm
iu}_{1}(\omega(M^{2},\alpha,Q^{2})). \label{eq:Kgamma}
\end{equation}

\subsection{Mesons and diquarks}
\label{app:subsec:MesonsDiquarks}
Since the rainbow-ladder truncation of the gap and Bethe-Salpeter equations provides a good approximation for ground-state vector and flavour-nonsinglet pseudoscalar mesons, it is sufficient herein to work with the following Bethe-Salpeter equation (BSE):
\begin{equation}
\Gamma_{q\bar{q}_{J^P}}(k;P) =  - \frac{16\pi\alpha_{\rm IR}}{3m_{\rm G}^{2}} \int \frac{d^{4}\ell}{(2\pi)^{4}} \gamma_{\mu}  S(\ell+P)\Gamma_{q\bar{q}_{J^P}}(\ell;P)S(\ell) \gamma_{\mu}.
\label{eq:BSEMeson}
\end{equation}
Plainly, the integrand does not depend on the external relative momentum, $k$.  Thus, a symmetry preserving regularisation of Eq.\,(\ref{eq:BSEMeson}) yields solutions
that are independent of $k$ so that the general solutions in the pseudoscalar and vector channels have the form:
\begin{eqnarray}
\label{eq:M0mBSE1}
\Gamma_{q\bar{q}_{0^{-}}}(P) &=& i\gamma_{5} E_{q\bar{q}_{0^{-}}}(P) + \frac{1}{M}
\gamma_{5} \gamma\cdot P F_{q\bar{q}_{0^{-}}}(P)\,,\\
\Gamma_{q\bar{q}_{1^{-}}}(P) &=&
\gamma_{\alpha}^{\perp} E_{q\bar{q}_{1^{-}}}(P)\,,
\label{eq:M0mBSE2}
\end{eqnarray}
where $M$ is the dressed light-quark mass in Table~\ref{tab:CQM}.

With the meson BSE in hand, one may readily infer the related equation for color-antitriplet quark-quark correlations (see, e.g., Ref.\,\cite{Roberts:2011cf}, Sec.\,2.1, for a derivation):
\begin{equation}
\Gamma^C_{qq_{J^P}}(k;P):=
\Gamma_{qq_{J^P}}(k;P)C^\dagger =  - \frac{8 \pi}{3} \frac{\alpha_{\rm IR}}{m_G^2}
\int \! \frac{d^4\ell}{(2\pi)^4} \gamma_\mu S_q(\ell+P) \Gamma^C_{qq_{J^P}}(\ell;P) S_q(\ell) \gamma_\mu \,,
\label{LBSEqq}
\end{equation}
where $C=\gamma_2\gamma_4$ is the charge-conjugation matrix.  Given the form of this equation, it will readily be understood that the solutions for scalar and axial-vector diquark correlations have the form:
\begin{eqnarray}
\label{eq:D0pBSE1}
\Gamma_{qq_{0^{+}}}(P) C^\dagger &=& i\gamma_{5} E_{qq_{0^{+}}}(P) + \frac{1}{M}
\gamma_{5} \gamma\cdot P F_{qq_{0^{+}}}(P)\,,\\
\Gamma_{qq_{1^{+}}}(P) C^\dagger &=&
\gamma_{\alpha}^{\perp} E_{qq_{1^{+}}}(P)\,.
\label{eq:D0pBSE2}
\end{eqnarray}

\begin{table}[t]
\begin{center}
\caption{\label{tab:mesonproperties} Computed diquark qualities that are required in order to complete the Faddeev equation kernels, obtained with $\alpha_{\rm IR}=0.93\pi$ and (in GeV): $m=0.007$, $m_{G}=0.8$, $\Lambda_{\rm ir} = 0.24$, $\Lambda_{\rm uv}=0.905$ \cite{Chen:2012qr}.}
\begin{tabular}{cccccccc}
\hline
$M$ & $m_{ud_{0^{+}}}$ & $E_{ud_{0^{+}}}$ & $F_{ud_{0^{+}}}$ & $m_{ud_{1^{+}}}$ &
$E_{ud_{1^{+}}}$ & $m_{ss_{1^{+}}}$ & $E_{ss_{1^{+}}}$\\
\hline
$0.368$ & $0.78$ & $2.74$ & $0.31$ & $1.06$ & $1.30$ & 1.26 & 1.42\\
\hline
\end{tabular}
\end{center}
\end{table}

In the following two subsections we present explicit forms of these BSEs for the ground-state $J^{P}=0^{-}$ and $1^{-}$ mesons and their respective $J^{P}=0^{+}$ and $1^{+}$ diquark partners.  The canonically normalised diquark solutions, which are inputs to the Faddeev equation kernels, are listed in Table~\ref{tab:mesonproperties}.

\subsubsection{Pseudoscalar mesons and scalar diquarks}
\label{app:subsubsec:pion}
With our symmetry preserving regularisation of the contact interaction, Eq.\,(\ref{eq:BSEMeson}) takes the following form in the pseudoscalar channel
\begin{equation}
\left[\begin{matrix} E_{q\bar{q}_{0^{-}}}(P) \\
F_{q\bar{q}_{0^{-}}}(P)\end{matrix}\right] = \frac{4\alpha_{\rm
IR}}{3\pi m_{\rm G}^{2}} \left[\begin{matrix} {\cal K}_{EE}^{\pi} & {\cal
K}_{EF}^{\pi} \\ {\cal K}_{FE}^{\pi} & {\cal K}_{FF}^{\pi}\end{matrix}\right]
\left[\begin{matrix} E_{q\bar{q}_{0^{-}}}(P) \\
F_{q\bar{q}_{0^{-}}}(P)\end{matrix}\right],
\label{eq:Eigenvalue0m}
\end{equation}
where
\begin{subequations}
\begin{eqnarray}
{\cal K}_{EE}^{\pi} &=& \int_{0}^{1} d\alpha \left[{\cal
C}^{\rm iu}(\omega(M^{2},\alpha,P^{2}))  -2\alpha(1-\alpha)P^{2}\bar{\cal
C}_{1}^{\rm iu}(\omega(M^{2},\alpha,P^{2}))\right], \\
{\cal K}_{EF}^{\pi} & =& P^{2} \int_{0}^{1} d\alpha \, \bar{\cal
C}_{1}^{\rm iu}(\omega(M^{2},\alpha,P^{2})), \\
{\cal K}_{FE}^{\pi} &=& \frac{1}{2} M^{2} \int_{0}^{1} d\alpha \, \bar{\cal
C}_{1}^{\rm iu}(\omega(M^{2},\alpha,P^{2})), \\
{\cal K}_{FF}^{\pi} & =& -2{\cal K}_{FE}^{\pi}.
\end{eqnarray}
\end{subequations}
It follows immediately that the explicit form of Eq.\,(\ref{eq:D0pBSE1}) is
\begin{equation}
\left[\begin{matrix} E_{qq_{0^{+}}}(P) \\
F_{qq_{0^{+}}}(P)\end{matrix}\right] = \frac{2\alpha_{\rm IR}}{3\pi m_{\rm G}^{2}}
\left[\begin{matrix} {\cal K}_{EE}^{\pi} & {\cal K}_{EF}^{\pi} \\ {\cal K}_{FE}^{\pi}
& {\cal K}_{FF}^{\pi}\end{matrix}\right] \left[\begin{matrix} E_{qq_{0^{+}}}(P) \\
F_{qq_{0^{+}}}(P)\end{matrix}\right].
\label{eq:Eigenvalue0p}
\end{equation}

Equations~(\ref{eq:Eigenvalue0m}) and (\ref{eq:Eigenvalue0p}) are eigenvalue problems: they each have a solution at isolated values of $P^{2}<0$, at which point the
eigenvector describes the associated on-shell Bethe-Salpeter amplitude.  That quantity must be normalised canonically before being used in the computation of observables;\footnote{This normalisation ensures that the meson's
electromagnetic form factor is unity at zero momentum transfer or, equivalently, the residue of the associated one-meson state in the quark-antiquark scattering matrix is one.}
i.e., one must use
$\Gamma_{q\bar{q}_{0^{-}}}^c= \Gamma_{q\bar{q}_{0^{-}}}/{\cal N}_{\,0^{-}}$, where
\begin{equation}
\label{eq:Normalisationp}
{\cal N}_{\,0^{-}}^2 =\left. \frac{d}{d P^2}\Pi_{0^-}(Q,P)\right|_{Q=P}^{P^2=-m^2_{q\bar q _{0^-}}},
\end{equation}
with (the remaining trace is over spinor indices)
\begin{equation}
\Pi_{0^-}(Q,P)= 2N_c {\rm tr}_{\rm D} \int\! \frac{d^4q}{(2\pi)^4}\Gamma_{q\bar{q}_{0^{-}}}(-Q)\,
 S_u(q+P) \, \Gamma_{q\bar{q}_{0^{-}}}(Q)\, S_d(q)\,
\end{equation}
and $N_{c}=3$ for a meson.
The canonical normalisation condition for the scalar diquark is almost identical: the only differences are that $N_{c}=3\to 2$ and the polarisation is evaluated at the diquark's mass.

\subsubsection{Vector mesons and axial-vector diquarks}
\label{app:subsubsec:rho}
The explicit form of Eq.~(\ref{eq:BSEMeson}) for the ground-state vector meson is
\begin{equation}
1 + K_{\gamma}(-m_{q\bar{q}_{1^{-}}}^{2}) = 0,
\end{equation}
with $K_{\gamma}$ given in Eq.\,\eqref{eq:Kgamma}.  The BSE for the axial-vector diquark again follows immediately; viz,
\begin{equation}
1+\frac{1}{2} K_{\gamma}(-m_{qq_{1^{+}}}^{2}) = 0.
\end{equation}
The canonical normalisation conditions are readily expressed; viz.,
\begin{equation}
\frac{1}{E_{q\bar{q}_{1^{-}}}^{2}} = \left. -\frac{9}{4\pi} \frac{m_{\rm
G}^{2}}{\alpha_{\rm IR}} \frac{d}{dP^{2}} K_\gamma(P^{2})
\right|_{P^{2}=-m_{q\bar{q}_{1^{-}}}^{2}} \!\!, \quad
\frac{1}{E_{qq_{1^{+}}}^{2}} = \left. -\frac{6}{4\pi} \frac{m_{\rm
G}^{2}}{\alpha_{\rm IR}} \frac{d}{dP^{2}} K_\gamma(P^{2})
\right|_{P^{2}=-m_{qq_{1^{+}}}^{2}} \!\!.
\end{equation}

\subsection{Nucleon and $\Delta$ Faddeev equations}
\label{app:subsec:Baryons}
The nucleon is represented by a Faddeev amplitude
\begin{equation}
\Psi_{N} = \Psi_{1} + \Psi_{2} + \Psi_{3},
\label{eq:psiFaddeev}
\end{equation}
where the subscript identifies the bystander quark and, e.g., $\Psi_{1,2}$ are
obtained from $\Psi_{3}$ by a cyclic permutation of all the quark labels.  The spin- and isospin-$1/2$ nucleon is a sum of scalar and axial-vector diquark correlations
\begin{equation}
\Psi_{3}(p_{i},\alpha_{i},\tau_{i}) = {\cal N}_{\,\Psi_{3}}^{0^{+}} + {\cal
N}_{\,\Psi_{3}}^{1^{+}},
\label{eq:Psi3}
\end{equation}
with $(p_{i},\alpha_{i},\tau_{i})$ the momentum, spin and isospin labels of the
quarks constituting the bound state, and $P=p_{1}+p_{2}+p_{3}$ the total
momentum of the system.

The scalar diquark piece in Eq.~(\ref{eq:Psi3}) is
\begin{equation}
{\cal N}_{\,\Psi_{3}}^{0^{+}} =
\sum_{[\tau_{1}\tau_{2}]\tau_{3}}\left[t^{[\tau_{1}\tau_{2}]}\Gamma^{0^{+}}_{
[\tau_{1}\tau_{2}]}(\frac{1}{2}p_{[12]};K)\right]_{\alpha_{1}
\alpha_{2}}^{\tau_{1}\tau_{2}} \Delta^{0^{+}}_{[\tau_{1}\tau_{2}]}(K)
[{\cal S}^{\Psi_{N}}(l;P)u^{\Psi_{N}}(P)]_{\alpha_{3}}^{\tau_{3}},
\label{eq:N30p}
\end{equation}
where the spinor satisfies Eq.\,(A.3) in Ref.\,\cite{Roberts:2011cf}, with $M$ the mass obtained by solving the Faddeev equation, and it is also a spinor in isospin space with $\varphi_{+}={\rm col}(1,0)$ for the proton and $\varphi_{-}={\rm col}(0,1)$ for the neutron; $K=p_{1}+p_{2}=p_{\{12\}}$, $p_{[12]}=p_{1}-p_{2}$,
$l=(-p_{\{12\}}+2p_{3})/3$;
\begin{equation}
\Delta^{0^{+}}_{[\tau_{1}\tau_{2}]}(K) =
\frac{1}{K^{2}+m_{[\tau_{1}\tau_{2}]_{0^{+}}}^{2}}
\label{eq:0pprop}
\end{equation}
is a propagator for the scalar diquark formed from quarks $1$ and $2$, with
$m_{[\tau_{1}\tau_{2}]_{0^{+}}}$ the mass-scale associated with this correlation,
$\Gamma^{0^{+}}_{[\tau_{1}\tau_{2}]}$ is the canonically-normalised Bethe-Salpeter
amplitude describing their relative momentum correlation and the flavor matrix in
this case is
\begin{equation}
t^{[ud]}=\left(\begin{matrix} 0 & 1 \\ -1 & 0 \end{matrix}\right);
\end{equation}
and ${\cal S}^{\Psi_{N}}$, a $4\times4$ Dirac matrix, describes the relative
quark-diquark momentum correlation. The color antisymmetry of $\Psi_{3}$ is
implicit in $\Gamma^{J^{P}}_{[\tau_{1}\tau_{2}]}$.

The axial-vector component in Eq.~(\ref{eq:Psi3}) is
\begin{equation}
{\cal N}^{1^{+}}_{\,\Psi_{3}} =
\sum_{\{\tau_{1}\tau_{2}\}\tau_{3}}\left[t^{\{\tau_{1}\tau_{2}\}}\Gamma^{1^{+}
}_{\mu,\{\tau_{1}\tau_{2}\}}(\frac{1}{2}p_{[12]};K)\right]_{\alpha_{1}\alpha_{2
}}^{\tau_{1}\tau_{2}} \Delta_{\mu\nu,\{\tau_{1}\tau_{2}\}}^{1^{+}}(K)
[{\cal A}_{\nu}^{\Psi_{N}}(l;P)u^{\Psi_{N}}(P)]_{\alpha_{3}}^{\tau_{3}},
\label{eq:N31p}
\end{equation}
where the symmetric isospin-triplet matrices are
\begin{equation}
t^{\{uu\}} = \left(\begin{matrix} \sqrt{2} & 0 \\ 0 & 0 \end{matrix}\right), \,
t^{\{ud\}} = \left(\begin{matrix} 0 & 1 \\ 1 & 0 \end{matrix}\right), \,
t^{\{dd\}} =\left(\begin{matrix} 0 & 0 \\ 0 & \sqrt{2} \end{matrix}\right),
\end{equation}
and the other elements in Eq.~(\ref{eq:N31p}) are straightforward generalisation of
those in Eq.~(\ref{eq:N30p}) with, e.g.,
\begin{equation}
\Delta_{\mu\nu,\{\tau_{1}\tau_{2}\}}^{1^{+}}(K) =
\frac{1}{K^{2}+m_{\{\tau_{1}\tau_{2}\}_{1^{+}}}^{2}}\left(\delta_{\mu\nu}+\frac{K_{
\mu}K_{\nu}}{m_{\{\tau_{1}\tau_{2}\}_{1^{+}}}^{2}}\right).
\label{eq:1pprop}
\end{equation}

Since it is not possible to combine an isospin-$0$ diquark with an isospin-$1/2$ to
obtain isospin-$3/2$, the spin- and isospin-$3/2$ $\Delta$-baryon contains only an
axial-vector diquark component
\begin{equation}
\label{DeltaA}
\Psi_{3}(p_{i},\alpha_{i},\tau_{i}) = {\cal D}_{\Psi_{3}}^{1^{+}}.
\end{equation}
where now the axial-vector component is
\begin{equation}
{\cal D}_{\Psi_{3}}^{1^{+}} = \sum_{\{\tau_{1}\tau_{2}\}\tau_{3}}
\left[t^{\{\tau_{1}\tau_{2}\}}\Gamma^{1^{+}}_{\mu,\{\tau_{1}\tau_{2}\}}(\frac{1}{2}p_
{[12]};K)\right]_{\alpha_{1}\alpha_{2}}^{\tau_{1}\tau_{2}}  \Delta_{\mu\nu,\{\tau_{1}\tau_{2}\}}^{1^{+}}(K) [{\cal
D}_{\nu\rho}^{\Psi_{\Delta}}(l;P)u_{\rho}^{\Psi_{\Delta}}(P)]_{\alpha_{3}}^{\tau_{3}}
,
\label{eq:D31p}
\end{equation}
where $u_{\rho}(P)$ is a Rarita-Schwinger spinor defined, e.g., via Eq.\,(A.11) in Ref.\,\cite{Roberts:2011cf}.

The general forms of the matrices ${\cal S}^{\Psi_{N}}(l;P)$, ${\cal
A}_{\nu}^{\Psi_{N}}(l;P)$ and ${\cal D}_{\nu\rho}^{\Psi_{\Delta}}(l;P)$, which
describe the momentum-space correlation between the quark and diquark in the nucleon
and $\Delta$-baryon are described in Refs.\,\cite{Oettel:1998bk,Cloet:2007pi}. The
requirement that ${\cal S}^{\Psi_{N}}(l;P)$ represents a positive energy nucleon
entails
\begin{equation}
{\cal S}^{\Psi_{N}}(l;P) = s_{1}(l;P) {\bf I_{D}} +
(i\gamma\cdot\hat{l}-\hat{l}\cdot\hat{P}\, {\bf I_{D}}) s_{2}(l;P),
\label{eq:SlP}
\end{equation}
where $({\bf I_{D}})_{rs} = \delta_{rs}$, $\hat{l}^{2}=1$, ${\hat P}^{2}=-1$. In the
nucleon rest frame, $s_{1,2}$ describe, respectively, the upper, lower component of
the bound-state nucleon's spinor. Placing the same constraint on the axial-vector
component, one has
\begin{equation}
{\cal A}_{\nu}^{i\, \Psi_{N}}(l;P) = \sum_{n=1}^{6} a^i_{n}(l;P)\gamma_{5}
A_{\nu}^{n}(l;P),\quad i=+,0,-\,,
\label{eq:AlP}
\end{equation}
where
\begin{equation}
\begin{array}{lll}
A_{\nu}^{1} = (\gamma\cdot\hat{l}^{\bot})\hat{P}_{\nu}, &
A_{\nu}^{2} = -i\hat{P}_{\nu}, &
A_{\nu}^{3} = (\gamma\cdot\hat{l}^{\bot})\hat{l}^{\perp}_{\nu},\\
A_{\nu}^{4} = i\hat{l}_{\nu}^{\perp}, &
A_{\nu}^{5} = \gamma_{\nu}^{\perp}-A_{\nu}^{3}, &
A_{\nu}^{6} = i\gamma_{\nu}^{\perp}(\gamma\cdot\hat{l}^{\bot})-A_{\nu}^{4}.
\end{array}
\end{equation}
Finally, requiring also that ${\cal D}_{\nu\rho}^{\Psi_{\!\Delta}}(l;P)$ be an
eigenfunction of the positive-energy projection operator, one obtains
\begin{equation}
{\cal D}_{\nu\rho}^{\Psi_{\!\Delta}}(l;P) = {\cal
S}^{\Psi_{\!\Delta}}(l;P)\delta_{\nu\rho} + \gamma_{5}{\cal
A}_{\nu}^{\Psi_{\!\Delta}}(l;P)l_{\rho}^{\bot},
\label{eq:DlP}
\end{equation}
with ${\cal S}^{\Psi_{\!\Delta}}$ and ${\cal A}_{\nu}^{\Psi_{\!\Delta}}$ given by obvious analogues of Eqs.~(\ref{eq:SlP}) and~(\ref{eq:AlP}), respectively.

At this point we have detailed forms for the dressed-quark propagators, the diquark Bethe-Salpeter amplitudes and the diquark propagators, so it is now possible to complete the kernels of the desired Faddeev equations.  As apparent in Fig.~\ref{fig:FaddeevEquation}, the kernels involve diquark breakup and reformation via exchange of a dressed-quark.  In proceeding we follow Ref.\,\cite{Roberts:2011cf} and make a severe simplification; namely, in the Faddeev equation for a baryon of type $B$, the quark exchange between the diquarks is truncated as follows
\begin{equation}
S^{T}(k) \to \frac{g_{B}^{2}}{M},
\label{eq:gB}
\end{equation}
where $g_{N}=1.18$ and $g_{\Delta,\Omega}=1.56$ are effective couplings \cite{Roberts:2011cf,Chen:2012qr} .  This is a variant of the so-called ``static approximation'', which itself was introduced in Ref.\,\cite{Buck:1992wz} and has subsequently been used in studies of a range of nucleon properties \cite{Bentz:2007zs}.  In combination with diquark correlations generated by Eq.\,(\ref{eq:qqInteraction}), whose Bethe-Salpeter amplitudes are momentum-independent, Eq.\,(\ref{eq:gB}) generates Faddeev equation kernels which themselves are momentum independent.  It follows that
Eqs.\, (\ref{eq:SlP}), (\ref{eq:AlP}) and (\ref{eq:DlP}) simplify dramatically, with
only those terms surviving that are independent of the relative momentum:
\begin{equation}
\label{FAmplitudesContact}
\begin{split}
{\cal S}^{\Psi_{N}}(l;P) &\to {\cal S}^{\Psi_{N}}(P) = s(P) {\bf I_{D}}, \\
{\cal A}_{\nu}^{i\,\Psi_{N}}(l;P) &\to {\cal A}_{\nu}^{i\,\Psi_{N}}(P) =
a_{1}^i(P)i\gamma_{5}\gamma_{\mu} + a_{2}^i(P)\gamma_{5}\hat{P}_{\mu}, \\
{\cal D}_{\nu\rho}^{\Psi_{\Delta}}(l;P) &\to {\cal D}_{\nu\rho}^{\Psi_{\Delta}}(P)
= d(P) \delta_{\nu\rho}
\end{split}
\end{equation}

One may now write the Faddeev equation appropriate to the contact interaction augmented by Eq.\,\eqref{eq:gB}:

\begin{equation}
 \left[ \begin{array}{r}
{\cal S}^{\Psi_{N}}(P)\, u(P)\\
{\cal A}^{i\,{\Psi_{N}}}_\mu(P)\, u(P)
\end{array}\right]\\
 = -\,4\,\int\frac{d^4\ell}{(2\pi)^4}\,{\cal M}(k,\ell;P)
\left[
\begin{array}{r}
{\cal S}^{\Psi_{N}}(P)\, u(P)\\
{\cal A}^{j\,{\Psi_{N}}}_\nu(P)\, u(P)
\end{array}\right] ,
\label{eq:NFaddeev}
\end{equation}
with the kernel detailed below.
%

The Faddeev amplitude for the charged or neutral $\Delta$-baryon may written:
\begin{equation}
{\cal D}_{\nu\mu}^{\Psi_{\Delta}}(P)u_{\mu}^{\Psi_{\Delta}}(P;s) =
\sum_{i=\{uu\},\{ud\}} d^{i}(P) \delta_{\nu\lambda} u_{\lambda}^{\Psi_{\Delta}}(P;s),
\end{equation}
so that the corresponding Faddeev equation has the form
\begin{equation}
\left[\begin{matrix} d^{\{uu\}}(P) \\ d^{\{ud\}}(P) \end{matrix}\right]
u_{\mu}^{\Psi_{\Delta}}(P;s) = -4 \int \frac{d^{4}l}{(2\pi)^{4}} \\
\left[\begin{matrix}
{\cal M}_{\mu\nu}^{\{uu\},\{uu\}}(l,P) & {\cal M}_{\mu\nu}^{\{uu\},\{ud\}}(l,P) \\
{\cal M}_{\mu\nu}^{\{ud\},\{uu\}}(l,P) & {\cal M}_{\mu\nu}^{\{ud\},\{ud\}}(l,P)
\end{matrix}\right]
\left[\begin{matrix}
d^{\{uu\}}(P) \\ d^{\{ud\}}(P) \end{matrix}\right] u_{\nu}^{\Psi_{\Delta}}(P;s).
\label{eq:DFaddeev}
\end{equation}
One could simplify this equation using isospin symmetry.  However, we choose not to, so that one may readily infer the equation for other decuplet baryons.  The simpler form, appropriate to $\Delta^{++}$, $\Delta^-$ and $\Omega^-$ baryons can be found, e.g., in App.\,C.1 of Ref.\,\cite{Chen:2012qr}.

\subsubsection{Nucleon kernel}
\label{app:subsubsec:NKernel}

The kernel in Eq.\,(\ref{eq:NFaddeev}) may be written in the following form
\begin{equation}
{\cal M}(l;P) = \left[\begin{matrix} {\cal M}_{00} & ({\cal M}_{01})_{\nu}^{j} \\
({\cal M}_{10})_{\mu}^{i} & ({\cal M}_{11})_{\mu\nu}^{ij} \end{matrix}\right],
\end{equation}
where $i,j$ range over $+=\{uu\}$, $0=\{ud\}$ in the case of the proton.  The explicit expressions for the entries are
\begin{subequations}
\begin{eqnarray}
{\cal M}_{00} &=& \frac{g_{N}^{2}}{M} \, t^{[ud]} \,
\Gamma_{[ud]}^{0^{+}}(k_{0^{+}}) \, (t^{[ud]})^{T} \,
\bar{\Gamma}_{[ud]}^{0^{+}}(-k_{0^{+}}) \, S(l) \,
\Delta_{[ud]}^{0^{+}}(k_{0^{+}}), \\
({\cal M}_{01})_{\nu}^{j} &= &\frac{g_{N}^{2}}{M} \, t^{j} \,
\Gamma_{\mu,j}^{1^{+}}(k_{1^{+}}) \,
(t^{[ud]})^{T} \, \bar{\Gamma}_{[ud]}^{0^{+}}(-k_{0}^{+}) \, S(l)
\, \Delta_{\mu\nu,j}^{1^{+}}(k_{1^{+}}), \\
({\cal M}_{10})_{\mu}^{i} &=& \frac{g_{N}^{2}}{M} \, t^{[ud]} \,
\Gamma_{[ud]}^{0^{+}}(k_{0^{+}}) \, (t^{i})^{T} \,
\bar{\Gamma}_{\mu,i}^{1^{+}}(-k_{1^{+}}) \, S(l) \,
\Delta_{[ud]}^{0^{+}}(k_{0^{+}}), \\
({\cal M}_{11})_{\mu\nu}^{ij} &=& \frac{g_{N}^{2}}{M} \, t^{i} \,
\Gamma_{\rho,i}^{1^{+}}(k_{1^{+}}) \, (t^{j})^{T} \,
\bar{\Gamma}_{\mu,j}^{1^{+}}(-k_{1}^{+}) \, S(l) \,
\Delta_{\rho\nu,i}^{1^{+}}(k_{1^{+}}).
\end{eqnarray}
\end{subequations}

The mass of the ground-state nucleon is then determined by a $5\times 5$ matrix
Faddeev equation; viz.,
\begin{equation}
K(P) =
\left[ \begin{matrix}
K^{00}_{ss} & -\sqrt{2} \, K^{01}_{sa_{1}} & K^{01}_{sa_{1}} & -\sqrt{2} \,
K^{01}_{sa_{2}} & K^{01}_{sa_{2}} \\[0.7ex]
-\sqrt{2} \, K^{10}_{a_{1}s} & 0 & \sqrt{2} \, K^{11}_{a_{1}a_{1}} & 0 & \sqrt{2} \,
K^{11}_{a_{1}a_{2}} \\[0.7ex]
K^{10}_{a_{1}s} & \sqrt{2} \, K^{11}_{a_{1}a_{1}} & K^{11}_{a_{1}a_{1}} &
\sqrt{2} \, K^{11}_{a_{1}a_{2}} & K^{11}_{a_{1}a_{2}} \\[0.7ex]
-\sqrt{2} \, K^{10}_{a_{2}s} & 0 & \sqrt{2} \, K^{11}_{a_{2}a_{1}} & 0 & \sqrt{2} \,
K^{11}_{a_{2}a_{2}} \\[0.7ex]
K^{10}_{a_{2}s} & \sqrt{2} \, K^{11}_{a_{2}a_{1}} & K^{11}_{a_{2}a_{1}} & \sqrt{2} \,
K^{11}_{a_{2}a_{2}} & K^{11}_{a_{2}a_{2}}
\end{matrix}\right],
\end{equation}
constructed using $c_{N}= g_{N}^{2}/(4\pi^{2}M)$,
\begin{equation}
\sigma_{N}^{0} = \sigma_{N}(\alpha,M,m_{qq_{0^+}},m_{N}) =
(1-\alpha)\,M^{2} + \alpha\,m_{qq_{0^+}}^{2} - \alpha (1-\alpha)
m_{N}^{2}, \quad
\sigma_{N}^{1} = \sigma_{N}(\alpha,M,m_{qq_{1^+}},m_N);
\end{equation}
and
\begin{subequations}
{\allowdisplaybreaks
\begin{eqnarray}
K^{00}_{ss} & = & K^{00}_{EE}+K^{00}_{EF}+K^{00}_{FF}\,,\\
K^{00}_{EE} & = & c_N E_{qq_{0^+}}^2 \!
\int_0^1 d\alpha \,\overline{\cal C}^{\rm iu}_1(\sigma_N^0)
(\alpha m_N + M)\,,\\
K^{00}_{EF} & = & - 2 c_N E_{qq_{0^+}} F_{qq_{0^+}}\frac{m_N}{M} \!
\int_0^1 d\alpha \,\overline{\cal C}^{\rm iu}_1(\sigma_N^0)
(1-\alpha) (\alpha m_N + M)\,,\\
K^{00}_{FF} & = & c_N F_{qq_{0^+}}^2\frac{m_{qq_{0^+}}^2}{M^2} \!
\int_0^1 d\alpha \,\overline{\cal C}^{\rm iu}_1(\sigma_N^0)(\alpha m_N + M)\,;\\
K^{01}_{s a_1} & = & K^{01}_{s_E a_1} + K^{01}_{s_F a_1}\,,\\
K^{01}_{s_E a_1} &=& c_N \frac{E_{qq_{0^+}}E_{qq_{1^+}}}{m_{qq_{1^+}}^2}\!
\int_0^1 d\alpha \,\overline{\cal C}^{\rm iu}_1(\sigma_N^1)
( m_{qq_{1^+}}^2 (3 M + \alpha m_N) + 2 \alpha (1-\alpha)^2 m_N^3 )\,, \\
K^{01}_{s_F a_1} &=& -c_N \frac{F_{qq_{0^+}}E_{qq_{1^+}}}{m_{qq_{1^+}}^2}
\frac{m_N}{M}\!
\int_0^1 d\alpha \,\overline{\cal C}^{\rm iu}_1(\sigma_N^1)
(1-\alpha)
(m_{qq_{1^+}}^2 (M + 3 \alpha m_N) + 2 (1-\alpha)^2 M m_N^2) \,; \\
K^{01}_{s a_2} & = & K^{01}_{s_E a_2} + K^{01}_{s_F a_2}\,,\\
K^{01}_{s_E a_2} & = & c_N \frac{E_{qq_{0^+}}E_{qq_{1^+}}}{m_{qq_{1^+}}^2}\!
\int_0^1 d\alpha \,\overline{\cal C}^{\rm iu}_1(\sigma_N^1)
(\alpha m_N - M) ((1-\alpha)^2 m_N^2-m_{qq_{1^+}}^2)\,,\\
K^{01}_{s_F a_2} & = & c_N \frac{F_{qq_{0^+}}E_{qq_{1^+}}}{m_{qq_{1^+}}^2}
\frac{m_N}{M} \!
\int_0^1 d\alpha \,\overline{\cal C}^{\rm iu}_1(\sigma_N^1)
(1-\alpha)(\alpha m_N - M) ((1-\alpha)^2 m_N^2-m_{qq_{1^+}}^2)\,; \\
K^{10}_{a_1 s} & = & K^{10}_{a_1 s_E} + K^{10}_{a_1 s_F}\,,\\
K^{10}_{a_1 s_E} & = & \frac{c_N}{3}\frac{E_{qq_{0^+}}E_{qq_{1^+}}}{m_{qq_{1^+}}^2}
\!
\int_0^1 d\alpha \,\overline{\cal C}^{\rm iu}_1(\sigma_N^0)
(\alpha m_N + M) (2 m_{qq_{1^+}}^2 + (1-\alpha)^2 m_N^2)\,,\\
K^{10}_{a_1 s_F} & = & -\frac{c_N}{3}\frac{F_{qq_{0^+}}E_{qq_{1^+}}}{m_{qq_{1^+}}^2}
\frac{m_N}{M} \!
\int_0^1 d\alpha \,\overline{\cal C}^{\rm iu}_1(\sigma_N^0)
(1-\alpha)(2 m_{qq_{1^+}}^2 + (1-\alpha)^2 m_N^2) (\alpha m_N + M)\,; \\
K^{10}_{a_2 s} & = & K^{10}_{a_2 s_E} + K^{10}_{a_2 s_F}\,,\\
K^{10}_{a_2 s_E} & = & \frac{c_N}{3} \frac{E_{qq_{0^+}}E_{qq_{1^+}}}{m_{qq_{1^+}}^2}
\!
\int_0^1 d\alpha \,\overline{\cal C}^{\rm iu}_1(\sigma_N^0)
(\alpha m_N + M) (m_{qq_{1^+}}^2 - 4 (1-\alpha)^2 m_N^2),\\
K^{10}_{a_2 s_F} & = & \frac{c_N}{3}
\frac{F_{qq_{0^+}}E_{qq_{1^+}}}{m_{qq_{1^+}}^2}\frac{m_N}{M} \!
\int_0^1 d\alpha \,\overline{\cal C}^{\rm iu}_1(\sigma_N^0)
(1-\alpha) (5 m_{qq_{1^+}}^2-2(1-\alpha)^2 m_N^2)(\alpha m_N + M)\,;\\
K^{11}_{a_1 a_1} & = & -\frac{c_N}{3}\frac{E_{qq_{1^+}}^2}{m_{qq_{1^+}}^2} \!
\int_0^1 d\alpha \,\overline{\cal C}^{\rm iu}_1(\sigma_N^1)
[ 2 m_{qq_{1^+}}^2 (M-\alpha m_N) + (1-\alpha)^2 m_N^2 (M+5 \alpha m_N)]\,;\\
K^{11}_{a_1 a_2} & = & -\frac{2 c_N}{3}\frac{E_{qq_{1^+}}^2}{m_{qq_{1^+}}^2} \!
\int_0^1 d\alpha \,\overline{\cal C}^{\rm iu}_1(\sigma_N^1)
(-m_{qq_{1^+}}^2+(1-\alpha)^2 m_N^2) (\alpha m_N - M)\,;\\
K^{11}_{a_2 a_1} & = & -\frac{c_N}{3}\frac{E_{qq_{1^+}}^2}{m_{qq_{1^+}}^2} \!
\int_0^1 d\alpha \,\overline{\cal C}^{\rm iu}_1(\sigma_N^1)
[m_{qq_{1^+}}^2(11 \alpha  m_N + M) - 2(1-\alpha)^2 m_N^2 (7\alpha m_N + 2 M)]\,;\\
K^{11}_{a_2 a_2}  & = & - \frac{5 c_N}{3} \frac{E_{qq_{1^+}}^2}{m_{qq_{1^+}^2}} \!
\int_0^1 d\alpha \,\overline{\cal C}^{\rm iu}_1(\sigma_N^1)
(m_{qq_{1^+}}^2 - (1-\alpha)^2 m_N^2) (\alpha m_N - M)\,.
\end{eqnarray}}
\end{subequations}

\subsubsection{The $\Delta$ kernel}
\label{app:subsubsec:DKernel}
The kernel in Eq.\,(\ref{eq:DFaddeev}) can be written in the following form
\begin{equation}
\begin{split}
{\cal M}_{\mu\nu}^{\{uu\},\{uu\}}(l,P) &= \frac{g_{\Delta}^{2}}{M} \, t^{\{uu\}} \,
\Gamma_{\rho,\{uu\}}^{1^{+}}(k_{1^{+}}) \, (t^{\{uu\}})^{T} \,
\bar{\Gamma}_{\mu,\{uu\}}^{1^{+}}(-k_{{1}^{+}}) \, S(l) \,
\Delta_{\rho\nu,\{uu\}}^{1^{+}}(k_{1^{+}}), \\
{\cal M}_{\mu\nu}^{\{uu\},\{ud\}}(l,P) &= \frac{g_{\Delta}^{2}}{M} \, t^{\{uu\}} \,
\Gamma_{\rho,\{uu\}}^{1^{+}}(k_{1^{+}}) \, (t^{\{ud\}})^{T} \,
\bar{\Gamma}_{\mu,\{ud\}}^{1^{+}}(-k_{{1}^{+}}) \, S(l) \,
\Delta_{\rho\nu,\{uu\}}^{1^{+}}(k_{1^{+}}), \\
{\cal M}_{\mu\nu}^{\{ud\},\{uu\}}(l,P) &= \frac{g_{\Delta}^{2}}{M} \,
t^{\{ud\}} \, \Gamma_{\rho}^{(\{ud\},1^{+})}(k_{1^{+}}) \, (t^{\{uu\}})^{T} \,
\bar{\Gamma}_{\mu}^{(\{uu\},1^{+})}(-k_{{1}^{+}}) \, S(l) \,
\Delta_{\rho\nu}^{(\{ud\},1^{+})}(k_{1^{+}}), \\
{\cal M}_{\mu\nu}^{\{ud\},\{ud\}}(l,P) &= \frac{g_{\Delta}^{2}}{M} \,
t^{\{ud\}} \, \Gamma_{\rho,\{ud\}}^{1^{+}}(k_{1^{+}}) \, (t^{\{ud\}})^{T} \,
\bar{\Gamma}_{\mu,\{ud\}}^{1^{+}}(-k_{{1}^{+}}) \, S(l) \,
\Delta_{\rho\nu,\{ud\}}^{1^{+}}(k_{1^{+}}).
\end{split}
\end{equation}
The mass of the ground-state $\Delta$ is then determined by a $2\times 2$ matrix
Faddeev equation; viz.,
\begin{equation}
K(P) = \left[\begin{matrix} 0 & \sqrt{2} \\ \sqrt{2} & 1 \end{matrix}\right]
K_{\Delta},
\end{equation}
where
\begin{equation}
K_{\Delta} = c_{\Delta} \frac{E_{qq_{1^{+}}}^{2}}{m_{qq_{1^{+}}}^{2}} \int_{0}^{1} d\alpha\, \bar{{\cal
C}}_{1}^{\rm iu}(\sigma_{\Delta}) (M+\alpha m_{\Delta})
(m_{qq_{1^{+}}}^{2}+m_{\Delta}^{2}(1-\alpha)^{2})\,,
\end{equation}
with
\begin{equation}
c_{\Delta} = \frac{g_{\Delta}^{2}}{4\pi^{2}M}, \quad
\sigma_{\Delta} = \sigma_{\Delta}(\alpha,M,m_{qq_{1^{+}}},m_{\Delta})
 =(1-\alpha) M^{2} + \alpha m_{qq_{1^{+}}}^{2} - \alpha(1-\alpha)m_{\Delta}^{2}.
\end{equation}

\setcounter{equation}{0}
\section{Baryon electromagnetic currents}
\label{app:sec:EMcurrentQuarkDiquark}
In order to compute the electromagnetic vertices one must specify how the photon
couples to the constituents within the composite hadrons.  In the present context this amounts to specifying the nature of the couplings of the photon to the dressed quarks and the diquark correlations, since the incoming and outgoing baryons are described by the quark-diquark Faddeev amplitudes.  As explained above, with our treatment of the contact interaction and Faddeev equation, there are three contributions to the currents, which are illustrated in Fig.~\ref{fig:Transitioncurrent} and detailed below.

\subsection{Contribution of the different diagrams}
\label{app:subsec:differentdiagrams}
The three contributions can be expressed in a form similar to those in Eqs.\,\eqref{eq:DEMcurrent}, (\ref{eq:JTransition}); viz., for the elastic interaction \begin{equation}
\Gamma_{\mu,\lambda\omega}^{m}(K,Q) = \Lambda_{+}(P_{f})R_{\lambda\alpha}(P_{f})
{\cal J}_{\mu,\alpha\beta}^{m}(K,Q) \Lambda_{+}(P_{i})R_{\beta\omega}(P_{i}),
\end{equation}
where $m=1,2,3$ counts the diagrams in Fig.\,\ref{fig:Transitioncurrent} from left-to-right, and for the transition
\begin{equation}
\Gamma_{\mu\lambda}^{m}(K,Q) = \Lambda_{+}(P_{f})R_{\lambda\alpha}(P_{f}) {\cal
J}_{\mu\alpha}^{m}(K,Q) \Lambda_{+}(P_{i}).
\end{equation}

\subsubsection{Photon on quark}
\label{app:subsubsec:qgammaq}
The leftmost diagram in Fig.~\ref{fig:Transitioncurrent} describes a photon coupling
directly to a dressed quark with the axial-vector diquark acting as a bystander.  For the elastic and transition processes, respectively, it corresponds to
\begin{equation}
{\cal J}_{\mu,\alpha\beta}^{1}(K,Q) = {\cal J}_{\mu,\alpha\beta}^{1\{uu\}}
+ {\cal J}_{\mu,\alpha\beta}^{1\{ud\}}\,, \quad
{\cal J}_{\mu\alpha}^{1}(K,Q) = {\cal J}_{\mu\alpha}^{1\{uu\}} +
{\cal J}_{\mu\alpha}^{1\{ud\}}. \\
\end{equation}
As we elucidate in App.\,\ref{CurrentSymmetry}, generality is not lost by specifying the flavour content of the $\Delta^+$ baryon.

Taking isospin symmetry into account, one has
\begin{equation}
\begin{array}{ll}
\displaystyle {\cal J}_{\mu,\alpha\beta}^{1\{uu\}} = d^{\{uu\}} \, d^{\{uu\}} \, e_{d} \,
{\cal I}_{\mu,\alpha\beta}^{1}, &
\displaystyle \quad {\cal J}_{\mu,\alpha\beta}^{1\{ud\}} = d^{\{ud\}} \, d^{\{ud\}} \, e_{u} \,
{\cal I}_{\mu,\alpha\beta}^{1}, \\
\displaystyle {\cal J}_{\mu\alpha}^{1\{uu\}} = d^{\{uu\}} \, a_{j}^{\{uu\}} \, e_{d} \,
{\cal I}_{\mu\alpha}^{1j},  &
\displaystyle \quad {\cal J}_{\mu\alpha}^{1\{ud\}} = d^{\{ud\}} \, a_{j}^{\{ud\}} \, e_{u}  \,
{\cal I}_{\mu\alpha}^{1j},
\end{array}
\end{equation}
where $d^{\{uu\}}$, $d^{\{ud\}}$ and $a_{j}^{\{uu\}}$, $a_{j}^{\{ud\}}$ $j=1,2$ are
the Faddeev components of the $\Delta^{+}$ and proton, respectively; $e_{u}=2/3$ and
$e_{d}=-1/3$ are the quark-charge in units of the positron charge; and
\begin{eqnarray}
{\cal I}_{\mu,\alpha\beta}^{1} &=& i \int \frac{d^{4}l}{(2\pi)^{4}} \, S(l_{+f})
\gamma_{\mu}^{\bot} P_{T}(Q^{2}) S(l_{+i}) \Delta_{\alpha\beta}^{1^{+}}(-l), \\
{\cal I}_{\mu,\alpha}^{1j} & =& i \int \frac{d^{4}l}{(2\pi)^{4}} \, S(l_{+f})
\gamma_{\mu}^{\bot} P_{T}(Q^{2}) S(l_{+i}) M_{j\beta}
\Delta_{\alpha\beta}^{1^{+}}(-l),
\end{eqnarray}
with $l_{\pm(i,f)}=l\pm P_{i,f}$, $M_{1\beta}=\gamma_{5}\gamma_{\beta}$, $M_{2\beta}=\gamma\hat{P}_{\beta}$, and $\gamma_{\mu}^{\perp} P_{T}(Q^{2})$ is the dressed-quark-photon vertex described in Sec.\,\ref{app:subsec:WTIs}.

\subsubsection{Photon on axial-vector diquark}
\label{subsubsec:agammaa}
The middle diagram in Fig.~\ref{fig:Transitioncurrent} depicts the photon scattering
elastically from an axial-vector diquark, with the dressed-quark as spectator.  Again, it can be expressed through the sum of two distinct terms
\begin{equation}
{\cal J}_{\mu,\alpha\beta}^{2}(K,Q) = {\cal J}_{\mu,\alpha\beta}^{2\{uu\}} +
{\cal J}_{\mu,\alpha\beta}^{2\{ud\}}\,,\quad
{\cal J}_{\mu\alpha}^{2}(K,Q) = {\cal J}_{\mu\alpha}^{2\{uu\}} +
{\cal J}_{\mu\alpha}^{2\{ud\}}\,.
\end{equation}
Capitalising on isospin symmetry, one finds
\begin{equation}
\begin{array}{ll}
\displaystyle {\cal J}_{\mu,\alpha\beta}^{2\{uu\}} = d^{\{uu\}} \, d^{\{uu\}} \,
e_{\{uu\}} \, {\cal I}_{\mu,\alpha\beta}^{2}, &
\displaystyle \quad
{\cal J}_{\mu,\alpha\beta}^{2\{ud\}} = d^{\{ud\}} \, d^{\{ud\}} \,
e_{\{ud\}} \, {\cal I}_{\mu,\alpha\beta}^{2}, \\
\displaystyle {\cal J}_{\mu\alpha}^{2\{uu\}} = d^{\{uu\}} \, a_{j}^{\{uu\}} \, e_{\{uu\}}
\, {\cal I}_{\mu\alpha}^{2j}, &
\displaystyle \quad {\cal J}_{\mu\alpha}^{2\{ud\}} = d^{\{ud\}} \, a_{j}^{\{ud\}} \,
e_{\{ud\}} \, {\cal I}_{\mu\alpha}^{2j},
\end{array}
\end{equation}
where $e_{\{uu\}}=4/3$, $e_{\{ud\}}=-1/3$ are the diquark-charges in units of the
positron charge and
\begin{eqnarray}
{\cal I}^{2}_{\mu,\alpha\beta} &=& \int \frac{d^{4}l}{(2\pi)^{4}} \, S(l) \,
\Delta_{\alpha\rho}^{1^{+}}(-l_{-f})  \Gamma_{\mu,\rho\sigma}^{1^{+}}(-l_{-f},-l_{-i})
\Delta_{\sigma\beta}^{1^{+}}(-l_{-i})\,,\\
{\cal I}^{2j}_{\mu\alpha} &=& \int \frac{d^{4}l}{(2\pi)^{4}} \, S(l) \,
\Delta_{\alpha\rho}^{1^{+}}(-l_{-f}) \Gamma_{\mu,\rho\sigma}^{1^{+}}(-l_{-f},-l_{-i})
\Delta_{\sigma\beta}^{1^{+}}(-l_{-i}) M_{j\beta}.
\end{eqnarray}

The photon--axial-vector-diquark elastic interaction was studied in Ref.\,\cite{Roberts:2011wy} and can be expressed as
\begin{equation}
\Gamma_{\mu,\rho\sigma}^{1^{+}}(k^{f} = K+Q/2,k^{i}=K-Q/2) =\sum_{j=1}^{3} T_{\mu,\rho\sigma}^{j}(K,Q) F_{j}^{1^{+}}(Q^{2}),
\end{equation}
where
\begin{subequations}
\begin{eqnarray}
T_{\mu,\rho\sigma}^{1}(K,Q) &=& 2 K_{\mu} {\cal T}_{\rho\alpha}^{k^{i}} {\cal
T}_{\alpha\sigma}^{k^{f}}, \\
T_{\mu,\rho\sigma}^{2}(K,Q) &=&
\left[Q_{\rho}-k_{\rho}^{i}\frac{Q^{2}}{2m_{qq_{1^{+}}}^{2}}\right] {\cal
T}_{\mu\sigma}^{k^{f}}
-\left[Q_{\sigma}+k_{\sigma}^{f}\frac{Q^{2}}{2m_{qq_{1^{+}}}^{2}}\right] {\cal
T}_{\mu\rho}^{k^{i}}, \\
T_{\mu,\rho\sigma}^{3}(K,Q) &=& \frac{K_{\mu}}{m_{qq_{1^{+}}}^{2}}
\left[Q_{\rho}-k_{\rho}^{i}\frac{Q^{2}}{2m_{qq_{1^{+}}}^{2}}\right]
\left[Q_{\sigma}+k_{\sigma}^{f}\frac{Q^{2}}{2m_{qq_{1^{+}}}^{2}}\right].
\end{eqnarray}
\end{subequations}
The electric, magnetic, and quadrupole form factors of the axial-vector diquark are
constructed as follows
\begin{equation}
\begin{array}{ll}
\displaystyle G_{E}^{1^{+}}(Q^{2}) = F_{1}^{1^{+}}(Q^{2}) + \frac{2}{3}\eta G_{Q}^{1^{+}}(Q^{2})\,, &
\displaystyle G_{M}^{1^{+}}(Q^{2}) = -F_{2}^{1^{+}}(Q^{2})\,,\\
\displaystyle
G_{Q}^{1^{+}}(Q^{2}) = F_{1}^{1^{+}}(Q^{2}) + F_{2}^{1^{+}}(Q^{2})
+ [1+\eta] F_{3}^{1^{+}}(Q^{2})\,, &
\end{array}
\end{equation}
where $\eta=Q^{2}/(4m_{qq_{1^{+}}}^{2})$.  These quantities were computed in Ref.\,\cite{Roberts:2011wy} and the following functions provide accurate interpolations on $Q^{2}\in[-m_{\rho}^{2},10]\,{\rm GeV}^{2}$:
\begin{equation}
\begin{split}
G_{E}^{1^{+}}(Q^{2}) &\stackrel{\rm interpolation}{=}
\frac{1.0-0.16Q^{2}}{1.0+1.17Q^{2}+0.012Q^{4}}, \\
G_{M}^{1^{+}}(Q^{2}) &\stackrel{\rm interpolation}{=}
\frac{2.13-0.19Q^{2}}{1.0+1.07Q^{2}-0.10Q^{4}}, \\
G_{Q}^{1^{+}}(Q^{2}) &\stackrel{\rm interpolation}{=}
-\frac{0.81-0.029Q^{2}}{1.0+1.11Q^{2}-0.054Q^{4}}. \\
\end{split}
\end{equation}

\subsubsection{Photon induced diquark transition}
\label{subsubsec:agammas}
The rightmost diagram in Fig.~\ref{fig:Transitioncurrent} depicts a dressed-quark spectator to a photon-induced transition between scalar and axial-vector diquark correlations. This diagram only contributes to the electromagnetic transition current: $\gamma+ N \to \Delta$.  It may be expressed
\begin{equation}
{\cal J}_{\mu\alpha}^{3}(K,Q) = {\cal J}_{\mu\alpha}^{3\{ud\}}
= d^{\{ud\}} \, s \, e_{\{ud\}} \, {\cal I}_{\mu\alpha}^{3},
\end{equation}
where the last step used isospin symmetry, $s$ is the scalar-diquark Faddeev component of the nucleon, Eq.\,\eqref{FAmplitudesContact}, and
\begin{equation}
{\cal I}_{\mu\alpha}^{3} = i \int \frac{d^{4}l}{(2\pi)^{4}} \,
S(l) \Delta_{\alpha\rho}^{1^{+}}(-l_{-f})
\Gamma^{10}_{\rho\mu}(-l_{-f},Q) \Delta^{0^{+}}(-l_{-i}).
\end{equation}
The photon-induced scalar$\leftrightarrow\,$axial-vector transition vertex is
\begin{equation}
\Gamma^{10}_{\rho\mu}(k_{2},k_{1}) = \Gamma^{01}_{\rho\mu}(-k_{2},k_{1}) =
\Gamma^{01}_{\mu\rho}(k_{1},k_{2}),
\end{equation}
where
\begin{equation}
\Gamma^{01}_{\mu\rho}(k_{1},k_{2}) =
\frac{g_{01}}{m_{qq_{1^{+}}}}\epsilon_{\mu\rho\alpha\beta}k_{1\alpha}k_{2\beta}G^{01}
(Q^{2}).
\end{equation}

The coupling constant and transition form factor were computed in Ref.\,\cite{Roberts:2011wy}, with the results $g_{01}=0.78$, and
\begin{equation}
G^{01}(Q^{2}) \stackrel{\rm interpolation}{=} \frac{1.0+0.10Q^{2}}{1.0+1.073Q^{2}}.
\end{equation}
providing an accurate interpolation of the form factor on $Q^{2}\in[-m_{\rho}^{2},10]\,{\rm GeV}^{2}$.

\subsection{Dressed-quark-core, isospin symmetry and the \mbox{\boldmath $\Delta$} currents}
\label{CurrentSymmetry}
Consider the $\Delta^{++}$.  There are two contributions to the current, which collapse as follows using isospin symmetry:
\begin{eqnarray}
\nonumber {\cal J}_{\mbox{$\Delta^{++}$}} & = &
{\cal J}_{\mu,\alpha\beta}^{1\{uu\}_{\Delta^{++}}}+{\cal J}_{\mu,\alpha\beta}^{2\{uu\}_{\Delta^{++}}}\\
\nonumber  &=& d_{\Delta^{++}}^{\{uu\}} \, d_{\Delta^{++}}^{\{uu\}} \, e_{u} \, {\cal I}_{\mu,\alpha\beta}^{1}
+ d_{\Delta^{++}}^{\{uu\}} \, d_{\Delta^{++}}^{\{uu\}} \, e_{\{uu\}} \, {\cal I}_{\mu,\alpha\beta}^{2}\\
&=& (d_{\Delta^{++}}^{\{uu\}})^2 e_u \left[ {\cal I}_{\mu,\alpha\beta}^{1} + 2 \, {\cal I}_{\mu,\alpha\beta}^{2}\right]\,.
\end{eqnarray}

In the case of the $\Delta^-$ one has
\begin{eqnarray}
\nonumber {\cal J}_{\mbox{$\Delta^{-}$}} & = &
{\cal J}_{\mu,\alpha\beta}^{1\{dd\}_{\Delta^{-}}}+{\cal J}_{\mu,\alpha\beta}^{2\{dd\}_{\Delta^{-}}}\\
\nonumber  &=& d_{\Delta^-}^{\{dd\}} \, d_{\Delta^-}^{\{dd\}} \, e_{d} \, {\cal I}_{\mu,\alpha\beta}^{1}
+ d_{\Delta^-}^{\{dd\}} \, d_{\Delta^-}^{\{dd\}} \, e_{\{dd\}} \, {\cal I}_{\mu,\alpha\beta}^{2}\\
&=& (d_{\Delta^-}^{\{dd\}})^2 e_d \left[ {\cal I}_{\mu,\alpha\beta}^{1} + 2 \, {\cal I}_{\mu,\alpha\beta}^{2}\right]\,.
\end{eqnarray}
Since $d_{\Delta^-}^{\{dd\}} = d_{\Delta^{++}}^{\{uu\}}$, the $\Delta^-$ form factors are precisely the same as those of the $\Delta^{++}$ apart from an overall electric-charge-related factor of $(-\frac{1}{2})$.

The current for the $\Delta^+$ exhibits a similar pattern:
\begin{eqnarray}
\nonumber {\cal J}_{\mbox{$\Delta^{+}$}} & = &
{\cal J}_{\mu,\alpha\beta}^{1\{uu\}_{\Delta^{+}}} + {\cal J}_{\mu,\alpha\beta}^{1\{ud\}_{\Delta^{+}}}
+ {\cal J}_{\mu,\alpha\beta}^{2\{uu\}_{\Delta^{+}}} + {\cal J}_{\mu,\alpha\beta}^{2\{ud\}_{\Delta^{+}}} \\
\nonumber
&=& d_{\Delta^+}^{\{uu\}} \, d_{\Delta^+}^{\{uu\}} \, e_d\, {\cal I}_{\mu,\alpha\beta}^{1}
+ d_{\Delta^+}^{\{ud\}} \, d_{\Delta^+}^{\{ud\}} \, e_u\, {\cal I}_{\mu,\alpha\beta}^{1}
+ d_{\Delta^+}^{\{uu\}} \, d_{\Delta^+}^{\{uu\}} \, e_{\{uu\}}\, {\cal I}_{\mu,\alpha\beta}^{2}
+ d_{\Delta^+}^{\{ud\}} \, d_{\Delta^+}^{\{ud\}} \, e_{\{ud\}}\, {\cal I}_{\mu,\alpha\beta}^{2}\\
\nonumber
& = & (d_{\Delta^+}^{\{uu\}})^2 \left([ e_d + 2 e_u]\, {\cal I}_{\mu,\alpha\beta}^{1}
+ [ e_{\{uu\}} + 2 e_{\{ud\}} ] {\cal I}_{\mu,\alpha\beta}^{2} \right)\\
\nonumber
&=& (d_{\Delta^+}^{\{uu\}})^2 \left[ {\cal I}_{\mu,\alpha\beta}^{1} + 2 \, {\cal I}_{\mu,\alpha\beta}^{2}\right]\\
&=& \, (d_{\Delta^{++}}^{\{uu\}})^2 |e_d| \, \left[ {\cal I}_{\mu,\alpha\beta}^{1} + 2 \, {\cal I}_{\mu,\alpha\beta}^{2}\right] ,
\end{eqnarray}
where we used $d_{\Delta^+}^{\{ud\}}=\surd 2 d_{\Delta^+}^{\{uu\}}$ in the third line and $d_{\Delta^+}^{\{uu\}} = d_{\Delta^{++}}^{\{uu\}}/\surd 3$ in the last, a result which originates in the Clebsch-Gordon isospin algebra.

Finally, the current for the $\Delta^0$ collapses completely:
\begin{eqnarray}
\nonumber {\cal J}_{\mbox{$\Delta^{0}$}} & = &
{\cal J}_{\mu,\alpha\beta}^{1\{uu\}_{\Delta^{0}}} + {\cal J}_{\mu,\alpha\beta}^{1\{ud\}_{\Delta^{0}}}
+ {\cal J}_{\mu,\alpha\beta}^{2\{uu\}_{\Delta^{0}}} + {\cal J}_{\mu,\alpha\beta}^{2\{ud\}_{\Delta^{0}}} \\
\nonumber
&=& d_{\Delta^0}^{\{dd\}} \, d_{\Delta^0}^{\{dd\}} \, e_u\, {\cal I}_{\mu,\alpha\beta}^{1}
+ d_{\Delta^0}^{\{ud\}} \, d_{\Delta^0}^{\{ud\}} \, e_d\, {\cal I}_{\mu,\alpha\beta}^{1}
+ d_{\Delta^0}^{\{dd\}} \, d_{\Delta^0}^{\{dd\}} \, e_{\{dd\}}\, {\cal I}_{\mu,\alpha\beta}^{2}
+ d_{\Delta^0}^{\{ud\}} \, d_{\Delta^0}^{\{ud\}} \, e_{\{ud\}}\, {\cal I}_{\mu,\alpha\beta}^{2}\\
\nonumber
& = & (d_{\Delta^0}^{\{uu\}})^2 \left([ e_u + 2 e_d]\, {\cal I}_{\mu,\alpha\beta}^{1}
+ [ e_{\{dd\}} + 2 e_{\{ud\}} ] {\cal I}_{\mu,\alpha\beta}^{2} \right)\\
&=& 0\,; \label{Delta0zero}
\end{eqnarray}
i.e., our formulation produces dressed-quark-core form factors for the $\Delta^0$ that are all identically zero.

\begin{figure}[t]
\begin{center}
\includegraphics[clip,width=0.35\textwidth]{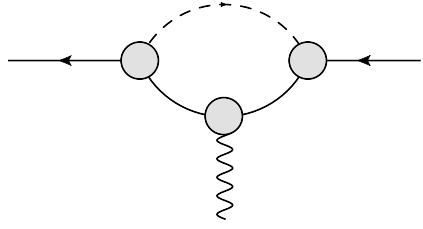}
\end{center}
\caption{\label{fig:PionLoop} Pion-loop correction to the quark-photon vertex: \emph{solid lines}, quark -- ${\cal S}(p)={\rm diag}[S_u,S_d]$; \emph{dashed-line}, pion; and \emph{wiggly-line}, photon.  The isospin structure of the quark-pion vertex is $\vec{\tau}\cdot \vec{\pi}$ and that of the quark photon vertex is $\hat Q={\rm diag}[e_u,e_d]$.  This momentum-dependent correction is readily seen to affect $u$- and $d$-quark couplings differently, Eq.\,\protect\eqref{SQS}, and hence leads to $\Delta^0$ form factors that are nonzero for $Q^2\neq 0$.}
\end{figure}

This result does not survive in the presence of mechanisms that lead to differences between the $Q^2$-dependence of $u$- and $d$-quark electromagnetic couplings.  One such process is associated with $\pi$-loop corrections to the dressed-quark-photon vertex \cite{Horikawa:2005dh}.  For example, with ${\cal S}={\rm diag}[S_u,S_d]$ and $\{\tau^i,i=1,2,3\}$ being the Pauli matrices, one may display the structure of the correction in Fig.\,\ref{fig:PionLoop} as follows:
\begin{equation}
\bigg(
\begin{array}{cc}
i \delta\Gamma_\mu^{u} & 0 \\
0 & i \delta\Gamma_\mu^{d}
\end{array}
\bigg)
\sim \tau^i \, {\cal S} \, e \hat Q i\gamma_\mu \, {\cal S} \, \tau^i
= \left(
\begin{array}{cc}
2 e_d S_d i\gamma_\mu S_d + e_u S_u i\gamma_\mu S_u & 0 \\
0 & e_d S_d i\gamma_\mu S_d + 2 e_u S_u i\gamma_\mu S_u
\end{array}\right) .\label{SQS}
\end{equation}
Since $2 e_d + e_u = 0$, this diagram alters the $u$- and $d$-quark couplings differently.  In particular, if one treats $S_u=S_d$, then this term only affects the $d$-quark's coupling.  In such circumstances,
${\cal J}_{\mu,\alpha\beta}^{1\{uu\}} \neq {\cal J}_{\mu,\alpha\beta}^{1\{ud\}}$ and hence Eq.\,\eqref{Delta0zero} is no longer true for $Q^2\neq 0$.  The diquark vertices are similarly affected.  Naturally, however, at $Q^2=0$ in a symmetry preserving treatment, all contributions to the vertex sum to produce an electric charge of zero for the $\Delta^0$.  Analogous considerations apply to the charged $\Delta$ states.

\subsection{Dressed-quark anomalous magnetic moment}
\label{app:subsec:anomalousmagnetic}
In the presence of dynamical chiral symmetry breaking, a dressed light-quark
possesses a large anomalous electromagnetic moment \cite{Bicudo:1998qb,Kochelev:1996pv,Chang:2010hb,Bashir:2011dp,Qin:2013mta}.  To illustrate the effect on form factors that one might expect from this phenomenon, we also present some results obtained with the dressed-quark--photon interaction modified as follows:
\begin{equation}
\Gamma_\mu^f(Q) =
\gamma_{\mu}^{\parallel} + \gamma_{\mu}^{\perp} P_{T}^f(Q^2) + \frac{\zeta_f}{2M_f} \sigma_{\mu\nu} Q_{\nu} \exp(-Q^{2}/4M_f^{2}) ,
\label{eq:DQAMM}
\end{equation}
where $f$ labels the quark flavour and $M_f$ is the dressed-quark mass.  The rate at which the anomalous moment term decays is taken from the distribution computed in
Ref.~\cite{Chang:2010hb}.

As we noted implicitly in App.\,\ref{CurrentSymmetry}, the two-flavour quark charge operator $\hat{Q} = \frac{1}{6} \boldmath{I} + \frac{1}{2} \tau_3$ has both isoscalar and isovector components.  Therefore the anomalous electromagnetic moment (AMM) of the dressed $u$- and $d$-quarks can differ.  This may again be illustrated by considering the contribution that pion loops can conceivably produce \cite{Horikawa:2005dh}: in this case it was found that $\zeta_u - \zeta_d \approx 1/2$; i.e., the isovector combination is large, and $\zeta_d + 2 \zeta_u \approx 0$.

It follows that when considering form factors involving the spin-$1/2$, isospin-$1/2$  neutron and proton, a reliable estimate of the effect produced by dressed-quark AMMs can be obtained by ignoring the flavour dependence in Eq.\,\eqref{eq:DQAMM} and using a common value of $\zeta =1/2$.  This is the procedure employed in Ref.\,\cite{Wilson:2011aa}.  The dressed-quark AMM thus implemented changes the form factors associated with the axial-vector diquarks; viz., with our standard parameter choice, Table~\ref{tab:CQM}, the following functions provide an accurate interpolation of the result on
$Q^{2}\in[-m_{\rho}^{2},10]\,{\rm GeV}^{2}$:
\begin{subequations}
\begin{eqnarray}
F_{1}^{1^{+}}(z) & \stackrel{\rm interpolation}{=} &\frac{1 + 0.98 Q^{2}}{1 + 2.75
Q^{2} + 1.26 Q^{4}}, \\
F_{2}^{1^{+}}(z) & \stackrel{\rm interpolation}{=} & -\frac{3.23 + 0.048 Q^{2}}{1 +
2.11 z + 0.0037 Q^{4}},\\
F_{3}^{1^{+}}(z) & \stackrel{\rm interpolation}{=} & \frac{1.19 + 0.33 Q^{2}}{1 + 1.38
Q^{2} + 4.62 Q^{4}}.
\end{eqnarray}
\end{subequations}
We use these formulae and Eq.\,\eqref{eq:DQAMM} in order to estimate the effect of dressed-quark AMMs on the $\gamma^\ast N\to \Delta$ transition form factors.

The $J=3/2$, $I=3/2$ $\Delta^+$ baryon, on the other hand, is predominantly a quark$\,+\,$axial-vector diquark in a relative $S$-wave, so that the spin-flavour wave function may be represented as $ \sqrt{2} \, u_\uparrow\{u_\uparrow d_\uparrow\} + d_\uparrow \{u_\uparrow u_\uparrow\}$ (see Sect.\,\ref{app:subsubsec:DKernel}).  Within a bound-state with this spin-flavour structure the dressed-quark AMM contribution must largely cancel.  This expectation is supported by an analysis of lattice-QCD results for $\Delta$-baryon form factors \cite{Cloet:2003jm}.  We therefore ignore the dressed-quark AMM when computing elastic form factors of the decuplet baryons.

\subsection{Current conservation}
As we indicated at the outset, one may compute the canonical normalisation constant for a charged-baryon Faddeev amplitude via the bound-states elastic electric charge form factor: its value at $Q^2=0$ must match that of the state's electric charge \cite{Oettel:1999gc}.
One must proceed carefully with that calculation, however.  Using the Ward-Green-Takahashi identities for the quark-photon and quark-diquark vertices, one can show that, at $Q^2=0$, Diagrams~1 and 2 in Fig.~\ref{fig:Transitioncurrent} must be equal.  Computationally, this is ensured by any $O(4)$- and translationally-invariant regularisation scheme. Whilst both are formally a property of our treatment of the contact-interaction, the latter is practically broken by the final step of introducing infrared and ultraviolet mass-scales in the proper-time regularisation of integrals. The effect is to produce a small mismatch between these diagrams at $Q^2=0$.

The weakness can be traced to quadratic divergences that arise through integrals such
as
\begin{equation}
\int\frac{d^4 \ell}{(2\pi)^4}
\frac{1}{[\ell^2+\omega]^2} \{(K\cdot \ell)^2,(Q\cdot \ell)^2, (K\cdot \ell) (Q\cdot
\ell)\} =
\int\frac{d^4 \ell}{(2\pi)^4}
 \frac{\ell^2}{[\ell^2+\omega]^2}\frac{1}{4}
\{K^2,Q^2 , K\cdot Q \}\,,
\label{eq:1on4}
\end{equation}
and analogous integrals with a quartic divergence.  As explained elsewhere (Ref.\,\cite{Chen:2012txa}, App.\,C), the weakness can be ameliorated via a simple expedient: in Eq.\,\eqref{eq:1on4}, replace $1/4 \to \theta = 1.874 (1/4)$ ($\to
\theta = 2.056 (1/4)$ in the case of the $\Omega$-baryon); and in results for those
integrals with a quartic divergence, replace the usual factor of $1/24$ by $\theta^2/24$.




\end{document}